\documentclass[12pt]{iopart}
\usepackage{amsfonts}
\usepackage{graphicx}
\usepackage{url}
\usepackage{cite}
\usepackage{caption}
\DeclareMathAlphabet{\pazocal}{OMS}{zplm}{m}{n}
\usepackage{amsmath}
\usepackage{xcolor}
\usepackage{calrsfs}
\usepackage{mathtools}
\DeclarePairedDelimiter{\ceil}{\lceil}{\rceil}

\begin{document}
\title[Popularity-similarity optimisation model beyond two dimensions]{Growing hyperbolic networks beyond two dimensions: the generalised popularity-similarity optimisation model}

\author{Bianka Kovács$^1$, Sámuel G. Balogh$^1$ and Gergely Palla$^{1,2,3}$}

\address{$^1$ Dept. of Biological Physics, Eötvös Loránd University, H-1117 Budapest, Pázmány P. stny. 1/A, Hungary}
\address{$^2$ MTA-ELTE Statistical and Biological Physics Research Group, H-1117 Budapest, Pázmány P. stny. 1/A, Hungary}
\address{$^3$ 
Health Services Management Training Centre, Semmelweis University,  H-1125, Kútvölgyi út 2, Budapest, Hungary}
\ead{pallag@hal.elte.hu}


\begin{abstract}

Hyperbolic network models have gained considerable attention in recent years, mainly due to their capability of explaining many peculiar features of real-world networks. One of the most widely known models of this type is the popularity-similarity optimisation (PSO) model, working in the native disk representation of the two-dimensional hyperbolic space and generating networks with small-world property, scale-free degree distribution, high clustering and strong community structure at the same time. With the motivation of better understanding hyperbolic random graphs, we hereby introduce the $d$PSO model, a generalisation of the PSO model to any arbitrary integer dimension $d>2$. The analysis of the obtained networks shows that their major structural properties can be affected by the dimension of the underlying hyperbolic space in a non-trivial way.
Our extended framework is not only interesting from a theoretical point of view but can also serve as a starting point for the generalisation of already existing two-dimensional hyperbolic embedding techniques.
\end{abstract}

\section{Introduction}
\label{sect:Intro}

Network theory has become an essential and ubiquitous tool for modelling various types of complex systems ranging from the level of interactions within cells to the level of the Internet, economic networks, and the society~\cite{Laci_revmod,Dorog_book}. In the past decades, a vast number of related studies reported a few universal features that most of the real networks seem to have in common, such as sparsity~\cite{sf_networks_are_sparse}, small-world property~\cite{Milgram_small_world,Kochen_book}, inhomogeneous degree distribution~\cite{Faloutsos,Laci_science}, high clustering coefficient~\cite{Watts-Strogatz} or community structure~\cite{Fortunato_coms,Fortunato_Hric_coms,Cherifi_coms}. Incorporating all, or at least some of these universal properties into a unified modelling framework is, however, a non-trivial issue and still presents a theoretical challenge of high relevance. Along this line, a variety of different network models have been proposed so far, including the celebrated Barabási--Albert (BA) model with preferential attachement~\cite{BA_model}, the hidden variables formalism~\cite{caldarelli_first_peaked,caldarelli_general_sf_recipe,boguna_general_hv,Garlaschelli_Entropy,gen_treshold_scirep} or models based on the mechanism of triadic closure, which has been specifically designed for explaining the high clustering of social networks~\cite{triadic_closure_holme,triadic_closure_toivonen}. Besides these examples, a further notable approach is given by hyperbolic network models that are capable of simultaneously explaining many observed network characteristics in a natural manner by assuming that nodes are embedded into a negatively curved hidden metric space ~\cite{hyperGeomBasics,PSO,EPSO_HyperMap,GPA_PSOsoftComms,nPSO,our_hyp_coms}. 

The random hyperbolic graph (RHG)~\cite{hyperGeomBasics}, for instance, is a static network model where nodes are placed at random on the hyperbolic disk of constant curvature $K=-\zeta^2$, and the connection probability between any pair of nodes is a decreasing function of their hyperbolic distance. A mathematically equivalent model is given by the $\mathbb{S}^1$~model~\cite{S1}, where the nodes are positioned on a circle and become connected according to a probability depending on the angular distance and a hidden variable drawn from a power-law distribution. By converting the hidden variables into radial coordinates we arrive to the hyperbolic $\mathbb{H}^2$ model~\cite{S1H2_Mercator} that is equivalent to the RHG model; hence, the RHG is often also referred to as the $\mathbb{S}^1/\mathbb{H}^2$ model.



In contrast to the RHG, in the popularity-similarity optimisation (PSO) model~\cite{PSO} the networks are not static but evolve over time via the continual appearance of new nodes on the hyperbolic plane. More precisely, new nodes are placed one by one in the native disk representation of the two-dimensional hyperbolic plane~\cite{hyperGeomBasics} with logarithmically increasing radial coordinates and uniformly random angular coordinates. Once a new node appears, it establishes connections to the previous ones with a probability depending on the hyperbolic distance in a similar way as in the RHG model. The tendency to connect to hyperbolically close nodes can be interpreted as an optimisation of a trade-off between the popularity (arising from the node birth time and reflected by the radial coordinate) of a possible candidate and its similarity (the angular distance abstracting the distance in an attribute space) compared to the newly arriving node. In vague terms, the degree of the nodes is determined by the radial coordinate, and owing to an outward shift of the nodes (referred to as the popularity fading, controlled by a parameter $\beta$), the degree distribution takes the scaling form of $\pazocal{P}(k)\sim k^{-\gamma}$ with a tuneable decay exponent $\gamma=1+\frac{1}{\beta}$. By changing the sharpness of the cutoff in the connection probability as a function of the hyperbolic distance with another parameter $T$ called temperature, the average clustering coefficient $\bar{c}$ of the resulting graphs can be adjusted as well. Although this model has been shown to be capable of generating networks that are small-world, highly clustered and scale-free at the same time, several other variants of the original PSO model have been suggested in order to explain further features of real-world graphs. Examples include the E-PSO model that inherently accounts for the creation of internal links, i.e. connections emerging between old nodes in the network~\cite{EPSO_HyperMap}, or alternatively, the deletion of already existing links~\cite{our_embedding}. Or, the nonuniform popularity-similarity optimisation (nPSO) model~\cite{nPSO,nPSO_2} that allows the generation of networks with an adjustable community structure by assuming a heterogeneous angular node distribution with multiple peaks is also worth mentioning. Nevertheless, it has also been revealed quite recently that both the RHG and the PSO models can generate networks that possess strong community structure despite lacking any explicitly built-in community generating mechanisms~\cite{commSector_hypEmbBasedOnComms_2016,commSector_linkPred,commSector_hypEmbBasedOnComms_2019,analogyBetweenHypEmbAndComms,commSector_commDetMethod,our_hyp_coms}.

In parallel with the developments of hyperbolic network models, another closely related field given by hyperbolic embedding techniques has also received great attention 
\cite{Boguna_Krioukov_Internet_2010,EPSO_HyperMap,Alanis-Lobato_LE_embedding,Alanis-Lobat_liekly_LE_emb,linkWeights_coalescentEmbedding,our_embedding,S1H2_Mercator}. Briefly, this tackles the problem of inferring the most plausible coordinates for the network nodes based on the topology of a given network. One of the first methods pointing in this direction was HyperMap~\cite{EPSO_HyperMap}, relying on a maximum likelihood estimation, where we assign hyperbolic coordinates to the nodes of the network by maximising the probability that the network was generated by the \mbox{E-PSO} model. Contrarily, in Refs.~\cite{Alanis-Lobato_LE_embedding,Alanis-Lobat_liekly_LE_emb} an embedding technique based on a nonlinear dimension reduction of the Laplacian matrix was introduced. Along similar lines, a whole set of embedding algorithms were studied in Ref.~\cite{linkWeights_coalescentEmbedding}, using different pre-weighted matrices encapsulating the network structure and multiple unsupervised dimension reduction techniques borrowed from machine learning. The rationale behind these approaches (usually coined as coalescent embeddings) is that when they are applied to hyperbolic networks, a common node aggregation pattern can be observed that is circularly or linearly ordered (angular coalescence) according to the original angular coordinates on the hyperbolic plane. An embedding algorithm that mixes the coalescent embedding with local angular optimisation based on likelihood maximisation was proposed in Ref.~\cite{our_embedding}. A further, very efficient embedding method is given by Mercator~\cite{S1H2_Mercator}, adopting the Laplacian eigenmaps approach with the coordinates optimised according to the RHG model.
 
Despite the excellent performance of the above embedding techniques, there is clearly a theoretical limitation behind most of them: they are defined on the hyperbolic disk, that is, in $d=2$ dimensions. Nevertheless, it has recently been revealed that higher-dimensional hyperbolic embeddings can outperform lower-dimensional ones in link prediction, for instance, in author collaboration networks~\cite{high_dim_embed_author_coll}. Moreover, in Ref.~\cite{Cannistraci_ASI} it has been also shown that the presence of additional dimensions can lead to a clearer separation between the communities of a network. Uncovering the role of the number of dimensions in hyperbolic embeddings is, therefore, of great interest that simultaneously provides a strong motivation for investigating appropriate higher-dimensional hyperbolic network models as well.

Along this line, the RHG model has recently been extended to $d>2$ dimensions~\cite{RHG_d_dim_mathematics,RHG_d_dim_krioukov}, placing the nodes in a $d$-dimensional hyperbolic ball. However, the extension of the PSO model to higher dimensions is still missing. Motivated by that, here we introduce the $d$PSO model, a generalisation of the original two-dimensional popularity-similarity optimisation model to any arbitrary integer dimension of $d\geq2$, which, we believe, provides a further substantial step towards a comprehensive theoretical characterisation of hyperbolic graphs.

Besides its theoretical relevance, our $d$PSO model opens up the possibility of systematically generalising already existing embedding techniques to higher dimensions, the necessity of which has explicitly been outlined e.g. for coalescent embeddings in Ref.~\cite{linkWeights_coalescentEmbedding}. Therein the authors claim that as a supplement to their findings, an additional interesting analysis could be to generate synthetic networks using for instance a three-dimensional PSO model and examine the accuracy of their methods by comparing the obtained embeddings to the original node arrangement. The present work contributes to this issue by thoroughly elaborating the PSO model for $d=3$ and higher dimensions. Since the higher number of dimensions of the underlying hyperbolic space allows a much richer characterisation of the nodes in general, here we conjecture that the suggested $d$-dimensional embeddings could provide further and deeper insights into the architecture of the hidden geometry behind the structure of complex networks.

In the present paper, we introduce the $d$PSO model as a natural generalisation of the well-known two-dimensional PSO model~\cite{PSO} to hyperbolic spaces of dimension $d>2$. We show analytically that the degree distribution of $d$PSO networks can be written as $\pazocal{P}(k)\sim k^{-\gamma}$ in the large $k$ regime, where the degree decay exponent $\gamma$ is directly related to the dimension $d$ and the popularity fading parameter $\beta$ as $\gamma=1+\frac{1}{(d-1)\beta}$. Besides the scale-free behaviour, the networks generated by the $d$PSO model can exhibit a large average clustering coefficient $\bar{c}$ and a strong community structure for a relatively wide range of the parameter settings if the number of dimensions of the underlying hyperbolic space is not extremely high. According to our results, $\bar{c}$ is controlled by an interesting interplay between the dimension $d$ and the temperature $T$, where for $T<\frac{1}{d-1}$ the average clustering coefficient is a decreasing function of $T$, whereas at temperatures near and above $T=\frac{1}{d-1}$, $\bar{c}$ becomes independent of $T$. A further noteworthy feature of the $d$PSO model is that in dimensions $d>2$ extremely skewed degree distributions with $\gamma<2$ become accessible, which can lead to networks displaying a number of exotic properties that are absent in scale-free networks with $2\leq \gamma$. The rich variety of networks that can be obtained in our proposed framework together with the capability of reproducing the fundamental properties of real networks in a natural way make the $d$PSO  model a very promising candidate upon which higher-dimensional hyperbolic embedding techniques may be developed in the future.

\section{Methods}
\label{sect:modelDescription}

The original popularity-similarity optimisation model places the network nodes in the native representation of the hyperbolic plane during the network generation. First, in Sect.~\ref{sect:nativeRepr} we describe the native representation of the $d$-dimensional hyperbolic space and the corresponding formula of the hyperbolic distance. Next, the network generation algorithm of the $d$PSO model is introduced by extending the PSO model to the hyperbolic space of any integer dimension $d\geq 2$.

\subsection{Native representation of the hyperbolic space}
\label{sect:nativeRepr}

The $d$-dimensional hyperbolic space of constant curvature $K<0$ is represented in the so-called native representation~\cite{hyperGeomBasics} by a $d$-dimensional ball of infinite radius in the Euclidean space (for which $K=0$). In this representation the Euclidean angles between hyperbolic lines are equal to their hyperbolic values, and the radial coordinate $r$ of a point (defined as its Euclidean distance from the centre of the ball) is equal to its hyperbolic distance from the ball centre. The hyperbolic distance between two points is measured along their connecting hyperbolic line, which is either the arc of the Euclidean circle going through the given points and intersecting the ball's boundary perpendicularly or -- if the ball centre falls on the Euclidean line connecting the two points in question -- the corresponding diameter of the ball. The hyperbolic distance $x$ between two points given by the Cartesian coordinate vectors $\underline{u}=(u_1,u_2,...,u_d)$ and $\underline{v}=(v_1,v_2,...,v_d)$ of norms $\|\underline{u}\|=\sqrt{\sum_{i=1}^d u_i^2}\equiv r_u$ and $\|\underline{v}\|=\sqrt{\sum_{i=1}^d v_i^2}\equiv r_v$ fulfills the hyperbolic law of cosines written as
\begin{equation}
    \mathrm{cosh}(\zeta x)=\mathrm{cosh}(\zeta r_u)\,\mathrm{cosh}(\zeta r_v)-\mathrm{sinh}(\zeta r_u)\,\mathrm{sinh}(\zeta r_v)\,\mathrm{cos}(\theta_{u,v}),
    \label{eq:hypDist}
\end{equation}
where $\zeta=\sqrt{-K}$, and $\theta_{u,v}=\mathrm{arccos}(\frac{\underline{u}\cdot\underline{v}}{\|\underline{u}\|\,\|\underline{v}\|})=\mathrm{arccos}(\frac{\sum_{i=1}^d u_i v_i}{r_u r_v})$ is the angle between the examined points. Note that in the case of $r_u=0$ simply $x=r_v$, and if $r_v=0$ then $x=r_u$. According to Ref.~\cite{hyperGeomBasics}, for sufficiently large $\zeta r_u$ and $\zeta r_v$ with an angular distance $\theta_{u,v}$ larger than $2\cdot\sqrt{e^{-2\zeta r_u}+e^{-2\zeta r_v}}$ but small enough to use the approximation $\sin(\theta_{u,v}/2)\approx\theta_{u,v}/2$, the hyperbolic distance can be approximated as
\begin{equation}
    x\approx r_u+r_v+\frac{2}{\zeta}\cdot\ln\left(\frac{\theta_{u,v}}{2}\right).
    \label{eq:hypDistApprox_main}
\end{equation}

\subsection{Description of the extended PSO model}
\label{sect:modelParametersAndSteps}

The PSO model~\cite{PSO} generates networks that have a scale-free degree distribution characterised by a degree decay exponent $\gamma$ that is determined by the radial arrangement of the network nodes. In terms of these, there are two natural possibilities for the extension of the two-dimensional PSO model to any integer number of dimensions $d\geq 2$ that both gives back the original PSO model at $d=2$: one can either make $\gamma$ independent of the number of dimensions by introducing a $d$-dependent multiplier in the radial coordinates of the low-temperature regime that yields highly clustered networks
, or not change this coordinate formula compared to the two-dimensional case and make the degree decay exponent $\gamma$ dependent on the value of $d$. 
In the present study, we chose the latter option since it offers the opportunity to expand the range of the achievable $\gamma$ values below 2 by increasing the number of dimensions. This choice is established in more detail in Sect.~\ref{sect:radCoordFactor} of the Supplementary Information.

In the $d$PSO model, the network nodes appear one by one in the above described native representation of the $d$-dimensional hyperbolic space and connect to previously appeared nodes with probabilities depending on the hyperbolic distances. The parameters of the model can be listed as follows:
\begin{itemize}
	\item The curvature $K\in\mathbb{R}^-$ of the hyperbolic space, controlled by $\zeta=\sqrt{-K}>0$. Changing the value of $\zeta$ corresponds to a simple rescaling of the hyperbolic distances; the usual custom is to set the value of $\zeta$ to $1$ (i.e. $K$ to $-1$).
	\item The dimension $2\leq d\in\mathbb{Z}^+$ of the hyperbolic space.
	\item The final number of nodes $N\in\mathbb{Z}^+$ in the network.
	\item The number of connections $m\in\mathbb{Z}^+$ established by each node after the $m$th one at its appearance. The average degree of the network is approximately $\bar{k}=2\cdot m$.
    \item The popularity fading parameter $\beta\in(0,1]$, controlling the outward drift of the nodes in the native ball. The exponent $\gamma$ of the power-law decaying tail of the degree distribution is related to the popularity fading parameter as 
    \begin{equation}
    \gamma=1+\frac{1}{(d-1)\cdot\beta}. 
    \label{eq:gamma}
    \end{equation}
    According to this relation between $\gamma$, $\beta$ and $d$, in the case of different dimensions different popularity fading parameters are needed to obtain the same degree decay exponent $\gamma$. Note that as the dimension $d$ increases, the achievable smallest degree decay exponent ($\gamma_{\mathrm{min}}=1+1/(d-1)$, yielded by $\beta=1$) decreases, meaning that in higher dimensions the attainable highest degree is larger than in hyperbolic spaces of smaller dimensions. (The details of the derivation of Eq.~(\ref{eq:gamma}) are given in Sect.~\ref{sect:degdist} and in Sect.~\ref{sect:degreeDist} of the Supplementary Information.)
    \item The temperature $0\leq T$, $T\neq\frac{1}{d-1}$, controlling the average clustering coefficient $\bar{c}$ of the network. As the temperature increases from $0$, the average clustering coefficient decreases, and settles to a more or less constant value at $T=\frac{1}{d-1}$. For $\beta\leq\frac{1}{d-1}$ (i.e., for $2\leq\gamma$), the clustering is asymptotically zero for any $\frac{1}{d-1}<T$, while for larger popularity fading parameters (i.e., for $\gamma<2$) the lowest possible $\bar{c}$ is an increasing function of $\beta$. 
\end{itemize}

During the random graph generation process, initially the network is empty, and at each time step $j=1,2,...,N$ a new node joins the network as follows:
\begin{enumerate}
    \item The new node $j$ appears at radial distance $r_{jj}$ from the origin with a position chosen uniformly at random on the surface of the corresponding $d$-dimensional ball, where
    \begin{enumerate}
        \item $r_{jj}=\frac{2}{\zeta}\ln{j}$ if $T<\frac{1}{d-1}$, and
        \item $r_{jj}=\frac{2T(d-1)}{\zeta}\ln{j}$ if $\frac{1}{d-1}<T$. 
    \end{enumerate}
    The change in the multiplying factor at $T=1/(d-1)$ was introduced to ensure that the formula of the degree decay exponent $\gamma$ remains the same (given by Eq.~(\ref{eq:gamma})) for all temperature settings. (Note that a similar adjustment of $r_{jj}$ was already introduced in the original PSO model of $d=2$~\cite{PSO}.)
    \item The radial coordinate of each previously (at time $i<j$) appeared node $i$ is increased according to the formula $r_{ij}=\beta r_{ii}+(1-\beta)r_{jj}$ in order to simulate popularity fading.
    \item The new node $j$ establishes connections with $m$ number of previously appeared nodes. Only single links are permitted. If the number of previously appeared nodes is not larger than $m$, then node $j$ connects to all of them. Otherwise (i.e., for $m+1<j$),
    \begin{enumerate}
        \item node $j$ connects to the $m$ hyperbolically closest nodes if $T=0$, and
        \item at temperatures $0<T$, any previous node $i=1,2,...,j-1$ gets connected to node $j$ with probability
            \begin{equation}
            p(x_{ij})=\frac{1}{1+e^{\frac{\zeta}{2T}(x_{ij}-R_j)}},
            \label{eq:PSO_con_prob}
            \end{equation}
        where the hyperbolic distance $x_{ij}$ between the node pair $i-j$ can be calculated based on Eq.~(\ref{eq:hypDist}) and the so-called cutoff distance $R_j$ can be obtained by solving the equation $m=\bar{k}_j$, where the expected number $\bar{k}_j$ of the realised connections of the new node $j$ at its arrival time $j$ can be written as 
            \begin{equation}
            \bar{k}_j=\eta(d)\cdot\int_1^j \int_0^{\pi} \frac{\sin^{d-2}{\theta_{ij}}}{1+\left(e^{\frac{\zeta}{2}\cdot(r_{ij}+r_{jj}-R_j)}\cdot\sin{\left(\frac{\theta_{ij}}{2}\right)}\right)^{\frac{1}{T}}} \,\mathrm{d}\theta_{ij}\,\mathrm{d}i
            \label{eq:cutOffDistEq}
            \end{equation}
using
            \begin{equation}
             \eta(d)=\left\lbrace \begin{array}{ll}
             \frac{\left(\frac{d}{2}-1\right)!\cdot \frac{d-2}{2}!\cdot 2^{d-2}}{(d-2)!\cdot\pi} & \mbox{if}\; d\mbox{ is even,} \\
             \frac{(d-1)!}{\left(\frac{d-1}{2}-1\right)!\cdot\frac{d-1}{2}!\cdot 2^{d-1}} & \mbox{if}\; d\mbox{ is odd.}
             \end{array} \right.
             \label{eq:etaFormula}
            \end{equation}
    \end{enumerate}
\end{enumerate}
The analytic study of the above model is given in the Supplementary Information: the expected form of the degree distribution is derived in Sect.~\ref{sect:degreeDist}, whereas the approximating formulae of the cutoff distance are presented in Sect.~\ref{sect:cutoffDistance}.

\section{Results}
\label{sect:results}

\begin{figure}[h!]
    \centering
    \captionsetup{width=\textwidth}
    \includegraphics[width=\textwidth]{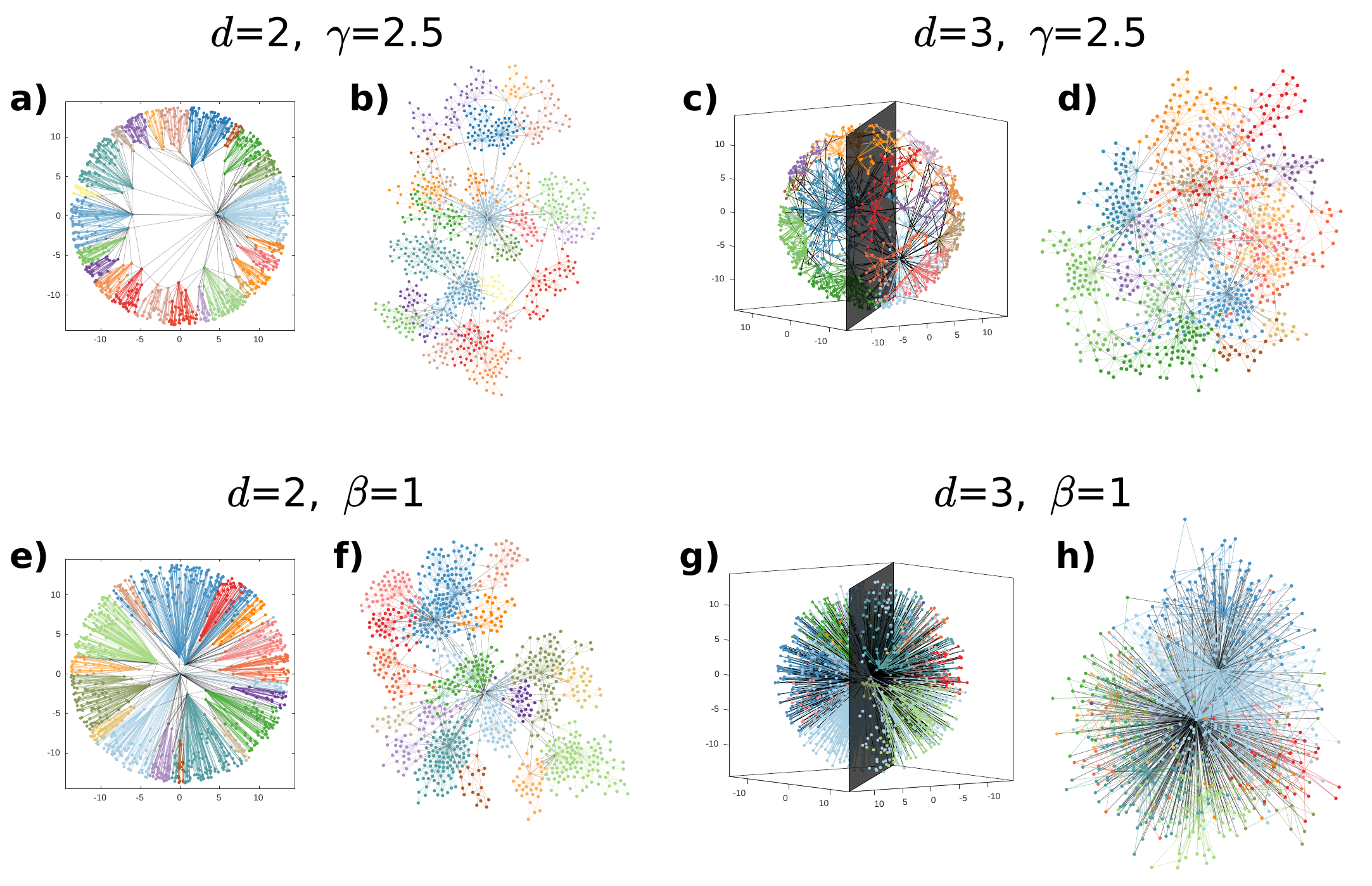}
    \caption{ {\bf Layouts of networks generated by the $d$PSO model in 2- and 3-dimensional hyperbolic spaces of curvature $K=-1$.} The colouring of the nodes and the links indicates communities found by the Louvain algorithm. In panels a), c), e) and g) we display the network in the native $d$-dimensional hyperbolic ball, whereas panels b), d), f) and h) show the standard Euclidean layout of the same graph for comparison. In the top row, we present networks of the same degree decay exponent $\gamma=2.5$, generated on the 2-dimensional hyperbolic plane, setting the popularity fading parameter $\beta$ to $2/3$ (panels a) and b)), and in the 3-dimensional hyperbolic space, setting the popularity fading parameter $\beta$ to $1/3$ (panels c) and d)). In the bottom row, we show networks of the same popularity fading parameter $\beta=1$, corresponding to the smallest degree decay exponent achievable in a given dimension, namely $\gamma=2.0$ in the 2-dimensional case (panels e) and f)) and $\gamma=1.5$ in the 3-dimensional case (panels g) and h)). For each network, we set the number of nodes $N$ to $1000$, the expected average degree $2m$ to $4$ and the temperature $T$ to $0$. The average clustering coefficient $\bar{c}$ of the displayed networks and the modularity $Q$ for their displayed partitions are the following: $\bar{c}=0.729$ and $Q=0.882$ for panels a) and b); $\bar{c}=0.644$ and $Q=0.845$ for panels c) and d); $\bar{c}=0.788$ and $Q=0.829$ for panels e) and f); while $\bar{c}=0.964$ and $Q=0.354$ for panels g) and h).
    }
    \label{fig:layoutExample}
\end{figure}

We generated networks with the above-described $d$PSO model using numerous parameter settings. As an illustration, in Fig.~\ref{fig:layoutExample} we show layouts of $d$PSO networks of size $N=1000$ both in the native representation of the $d$-dimensional hyperbolic space and according to a standard force-directed layout algorithm in the Euclidean plane. We analyzed the major structural properties of $d$PSO networks from various points of view. The results along with their detailed explanations are presented in the following subsections, focusing on a few fundamental network characteristics such as the degree distribution (Sect.~\ref{sect:degdist}), the average clustering coefficient (Sect.~\ref{sect:clust}) and the community structure (Sect.~\ref{sect:comms}). Further plots regarding simulation results are shown in Sect.~\ref{sect:simRes} of the Supplementary Information. Moreover, in Sect.~\ref{sect:nPSO} of the Supplementary Information we also study a three-dimensional extension of the nonuniform popularity-similarity optimisation model (nPSO)~\cite{nPSO,nPSO_2} that samples the angular coordinates of the network nodes from a multimodal distribution to allow control over the number and the size of the communities.

\subsection{Degree distribution}
\label{sect:degdist}
\begin{figure}[h!]
    \centering
    \captionsetup{width=1.0\textwidth}
    \includegraphics[width=1.0\textwidth]{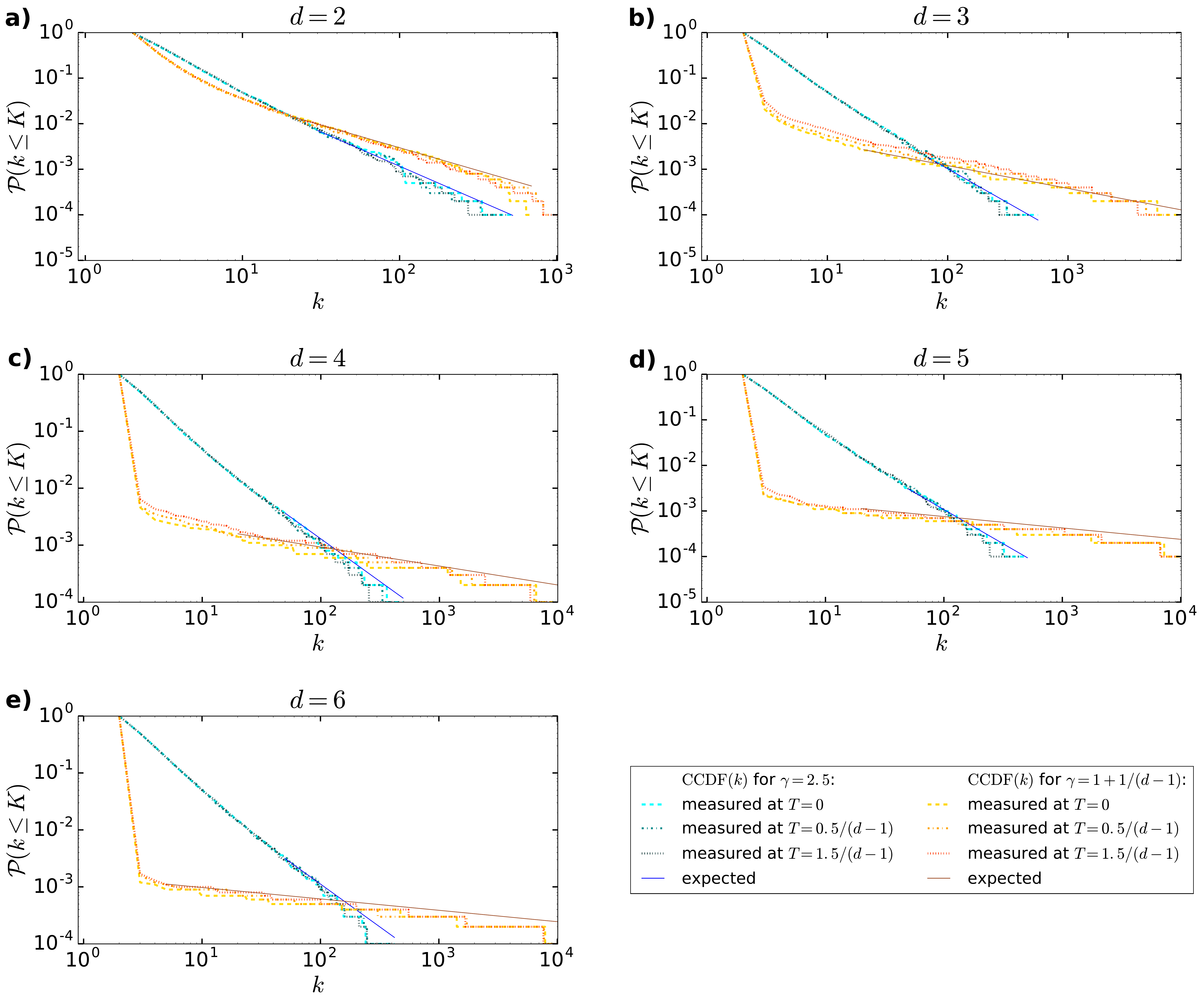}
    \caption{ {\bf Degree distribution of networks generated by the $d$PSO model at different parameter settings.} Each panel corresponds to a given dimension $d$. We display in all the panels the complementary cumulative distribution function (CCDF) of the node degrees for networks of two different degree decay exponents: $\gamma=2.5$ (blue) and $\gamma=1+1/(d-1)$ (orange). In both cases, we generated a network using the temperature setting $T=0$ (dashed lines), $T=0.5/(d-1)$ (dash-dotted lines) or $T=1.5/(d-1)$ (dotted lines). The curvature of the hyperbolic space, the number of nodes and the half of the expected average degree were the same for all networks, namely $K=-\zeta^2=-1$, $N=10,000$ and $m=2$. All the indicated simulation results match well the curve $\pazocal{P}(k\leq K) \sim k^{-(\gamma-1)}$ (shown by solid lines) that was expected based on the analytical calculations.
    }
    \label{fig:degreeDist_Tdependence}
\end{figure}

Our analytical calculations show that when generating networks according to the $d$PSO model presented in Sect.~\ref{sect:modelDescription}, we obtain degree distributions that follow a scaling form of
\begin{equation}
    \pazocal{P}(k)\sim k^{-\gamma},
    \label{eq:d_dim_deg_dist}
\end{equation}
where the degree decay exponent $\gamma$, given by Eq.~(\ref{eq:gamma}), depends on the popularity fading parameter $\beta$ and also the number of dimensions $d$, but is independent of the temperature $T$. 
This is a direct consequence of the fact that the probability for node $i$ (appearing at time $i$) to attract a link from node $j$ (appearing at time $j > i$) in the above model can be written as
\begin{equation}
    \Pi_{d\text{PSO}}(i,j)=m\cdot\frac{i^{-(d-1)\beta}}{\int^{j}_{1}\ell^{-(d-1)\beta}\mathrm{d}\ell},
    \label{eq:dpso_attraction}
\end{equation}
with $m$ denoting the number of connections established by node $j$ at its appearance. (The derivation of the above formula is given in Sect.~\ref{sect:degreeDist} of the Supplementary Information.) 
Indeed, by following the idea in Ref.~\cite{PSO}, it can be shown that Eq.~(\ref{eq:dpso_attraction}) is equivalent to imposing an extended preferential attachment rule (EPA) that defines the connection probability between an existing node $i$ with a degree $k_i(j)$ and the newly appearing node $j$ as
\begin{equation}
    \Pi_{\text{EPA}}[k_i(j)]=m\frac{k_i(j)-m+A}{(m+A)j},
    \label{eq:EPA}
\end{equation}
where $A=(\gamma-2)m$ is a parameter called initial attractiveness. According to Ref.~\cite{PA_degreeDistr}, in networks generated by this EPA rule, the expected degree of node $i$ becomes
\begin{equation}
    \overline{k_i(j)}=m+A\left[\left(\frac{i}{j}\right)^{-\alpha}-1\right]
    \label{eq:EPA_ki}
\end{equation}
at time $j$, and the degree distribution develops into a scale-free form of $\pazocal{P}(k)\sim k^{-\gamma}$, where the exponents $\alpha$ and $\gamma$ are connected by
$\alpha(\gamma)=\frac{1}{\gamma-1}$. We can relate Eqs.~(\ref{eq:EPA}) and (\ref{eq:EPA_ki}) to Eq.~(\ref{eq:dpso_attraction}) by identifying $\alpha$ as $\alpha=(d-1)\beta$, verifying that
\begin{equation}
   \Pi_{d\text{PSO}}(i,j)=\Pi_{\text{EPA}}[\overline{k_i(j)}].
\end{equation}
Finally, since $\gamma$ can be expressed from $\alpha(\gamma)$ as $\gamma=1+\frac{1}{\alpha}$, we arrive at the formula in Eq.~(\ref{eq:gamma}) for the decay exponent of the degree distribution in Eq.~(\ref{eq:d_dim_deg_dist}). 
Thus, networks that grow according to the rules outlined in Sect.~\ref{sect:modelDescription} inherit the scale-free property from the original PSO model, albeit using the same $\beta$ parameter 
leads to a smaller $\gamma$ exponent for greater values of $d$. Note that by setting $d$ to $2$, one can easily recover the results of the original PSO model in Ref.~\cite{PSO}, where the degree decay exponent has been found to be $\gamma=1+\frac{1}{\beta}$. The analytical results are in perfect agreement with the numerical simulations, as indicated by Fig.~\ref{fig:degreeDist_Tdependence}, where we display the complementary cumulative distribution function (CCDF) of the node degrees for several networks obtained from the $d$PSO model with different combinations of the $d$ and the $\beta$ parameters at different values of the temperature $T$. 

\subsection{Clustering coefficient}
\label{sect:clust}

\begin{figure}[h!]
    \centering
    \captionsetup{width=1.0\textwidth}
    \includegraphics[width=1.0\textwidth]{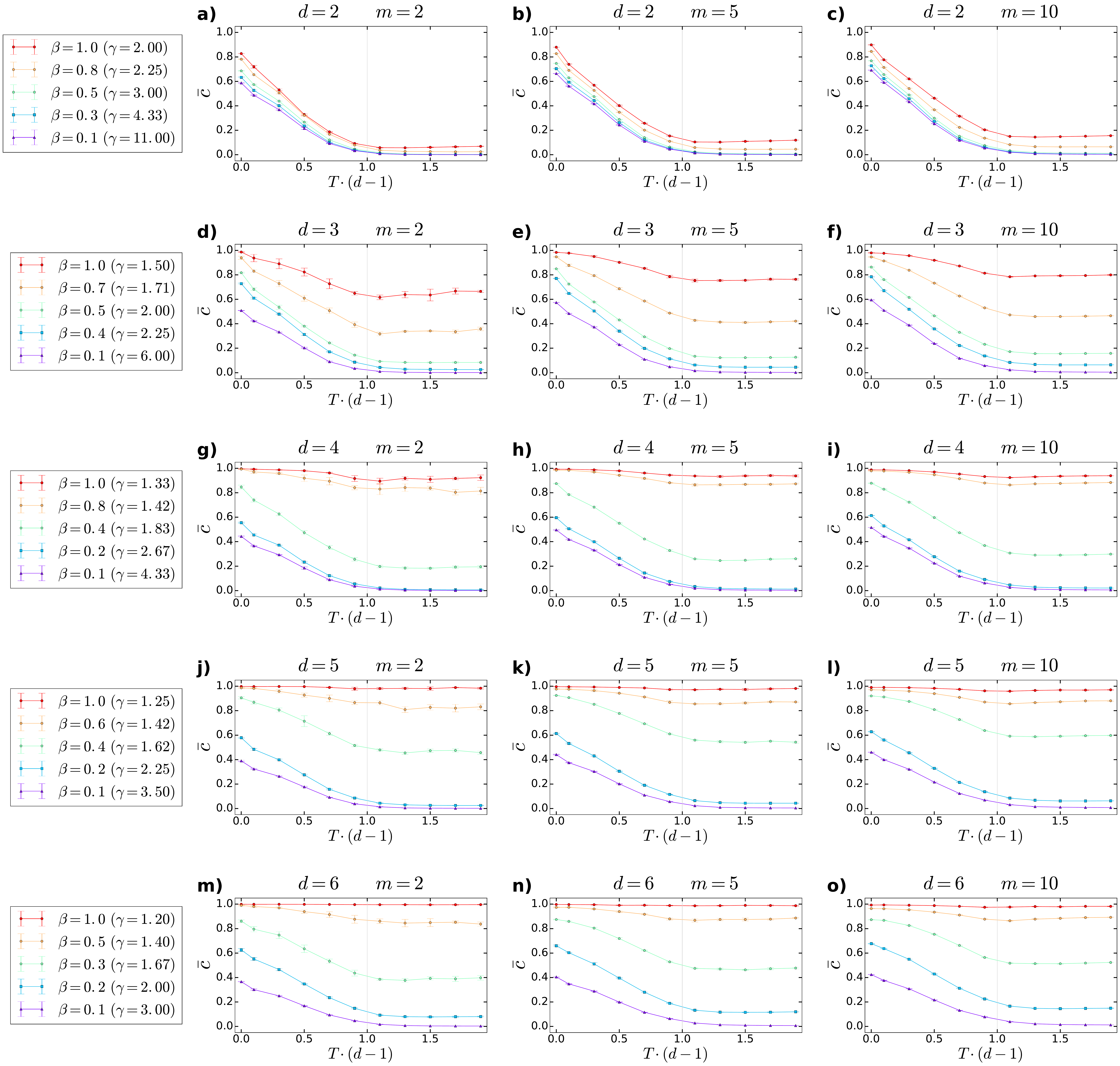}
    \caption{ {\bf Average clustering coefficient $\bar{c}$ of $d$PSO networks as a function of the rescaled temperature $T\cdot(d-1)$.} Each row of panels was created using a given dimension $d$, and each column of subplots presents the results obtained with a given value of the expected average degree $2m$, as written in the panel titles. The different curves of each panel correspond to different values of the popularity fading parameter $\beta$ (yielding different degree decay exponents $\gamma$), listed for each dimension in the leftmost panel of the corresponding row. The displayed data points were obtained by averaging over 5 $d$PSO networks generated independently with a given set of model parameters, setting the number of nodes to $10,000$ and the curvature of the hyperbolic space to $-1$ in each case. The error bars show the standard deviations measured among the 5 networks. The grey vertical lines indicate the critical point $T_{\mathrm{c}}=1/(d-1)$. 
    }
    \label{fig:clustCoeff_Tdependence}
\end{figure}

Based on simulations, we have found that the average clustering coefficient $\bar{c}$ of $d$PSO networks displays an unusually rich behaviour depending on the specific choices of the model parameters. First, in Fig.~\ref{fig:clustCoeff_Tdependence} we show the measured $\bar{c}$ as a function of the rescaled temperature $T \cdot(d-1)$ for different settings of the further model parameters. A rather simple observation regarding this figure is that $\bar{c}$ is an increasing function of $m$. This is reasonable in the light of the role of this model parameter outlined in Sect.~\ref{sect:modelDescription}: $m$ is related to the expected average degree of the $d$PSO networks as $\bar{k}=2\cdot m$, meaning that higher values of $m$ correspond to higher average degrees. 

Besides, according to Fig.~\ref{fig:clustCoeff_Tdependence}, the average clustering coefficient is always maximal at $T=0$, i.e. when newly appearing nodes connect only to the hyperbolically closest existing nodes. If, however, the temperature increases, the cutoff in the connection probability becomes milder. This implies that rather distant nodes also become likely to be connected and consequently, the value of $\bar{c}$ decreases. 
Nevertheless, above a certain point $T_{\mathrm{c}}=\frac{1}{d-1}$, hereinafter referred to as the critical temperature, the average clustering coefficient remains more or less constant. This can be attributed to the fact that at the critical temperature we change the formula of the radial coordinates from $r_{jj}=\frac{2}{\zeta}\ln{j}$ to $r_{jj}=\frac{2T(d-1)}{\zeta}\ln{j}$, meaning that the radial coordinates become an increasing function of the temperature above $T_{\mathrm{c}}$. According to the approximating formula given by Eq.~(\ref{eq:hypDistApprox_main}), the larger radial coordinates obviously yield larger hyperbolic distances between all node pairs. Therefore, despite the continuous slowing down of the decay 
in the connection probability as a function of the hyperbolic distance, above $T_{\mathrm{c}}=\frac{1}{d-1}$ the increase in the temperature does not increment further the number of nodes that are reachable for a newly coming node, and thus, the average clustering coefficient becomes settled to a constant value. 

Finally, it can be clearly seen in Fig.~\ref{fig:clustCoeff_Tdependence} that the average clustering coefficient is an increasing function of the popularity fading parameter $\beta$. This is especially striking in the higher-dimensional cases, where the range of the degree decay exponents that are achievable 
extends to lower values. 
When $\gamma=1+\frac{1}{(d-1)\beta}$ is decreased, the degree distribution of the emerging network decays more slowly and, consequently, the largest occurring degrees 
increase. This is realised by the increase in the preference of the newly coming nodes for connecting to the early-appeared ones
, overshadowing the attractiveness of the angular neighbours located at larger radii. 

Now let us concentrate on the subgraph of nodes that appeared before a given time point. Since each node has to create $m$ links at its appearance, the ratio between the connected and the non-connected node pairs in this subgraph increases towards earlier times (i.e., as the number of nodes in the subgraph decreases). Therefore, if the degree decay exponent $\gamma$ of the network is decreased, 
the set of the primarily attractive inner nodes tightens and the new nodes tend to connect into a more densely connected group of nodes, which increases the number of triangles and, accordingly, the average clustering coefficient. It is important to note that at extremely small values of $\gamma$, the high density of the connections among the most popular inner nodes moderates the effect of the temperature increase on $\bar{c}$ by limiting the probability to connect to nodes that are not connected to each other. Due to this, the networks of small enough degree decay exponents are highly clustered not only at small temperatures, but in the high-temperature regime as well. 

\begin{figure}[h!]
    \centering
    \captionsetup{width=1.0\textwidth}
    \includegraphics[width=1.0\textwidth]{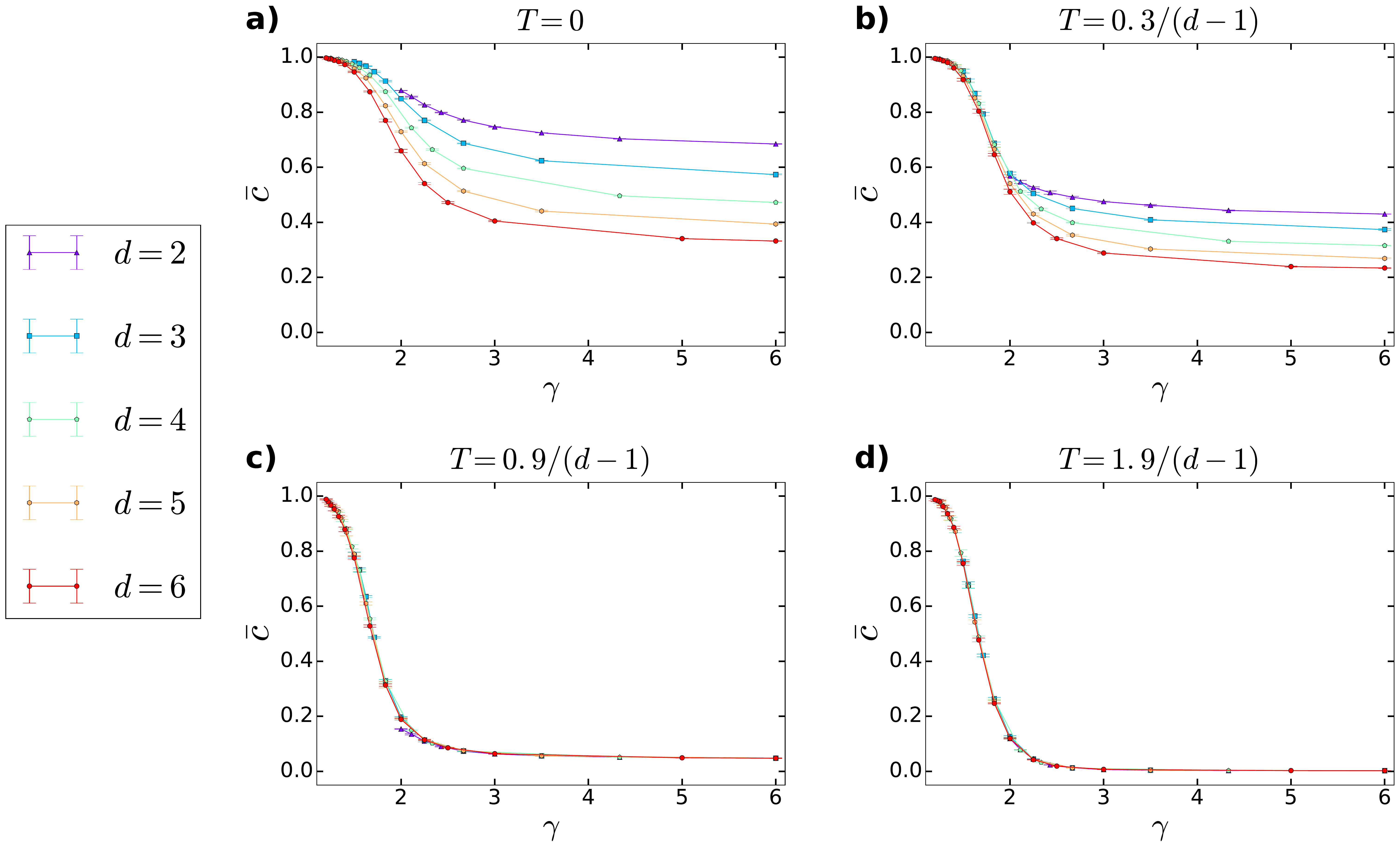}
    \caption{ {\bf Average clustering coefficient $\bar{c}$ measured in $d$-dimensional PSO networks as a function of the degree decay exponent $\gamma$ at different values of the temperature $T$ and the dimension $d$.} We plotted the average clustering coefficient averaged over 5 networks for each parameter setting, with the error bars indicating the standard deviations among the 5 networks. The size of the networks was $N=10,000$, the expected average degree was $\bar{k}=2m=10$, and each network was generated in a hyperbolic space of curvature $K=-1$. The curves of different colours correspond to different values of the dimension $d$ of the hyperbolic space, listed in the legend. The popularity fading parameter was always set to $\beta=\frac{1}{(d-1)\cdot(\gamma-1)}$. Each panel was created using a given value of the temperature $T$, specified in the panel title. 
    }
    \label{fig:clustCoeff_gammaDependence}
\end{figure}

To provide further insight into how the model parameters affect the triangle formation in $d$PSO networks, in Fig.~\ref{fig:clustCoeff_gammaDependence} we show the average clustering coefficient as a function of the degree decay exponent $\gamma$ 
at different settings of the temperature $T$ and the dimension $d$. This figure again proves that the critical temperature $T_{\mathrm{c}}=\frac{1}{d-1}$ separates two distinct regimes, where $\bar{c}$ shows a fundamentally different nature.

Below the critical temperature $T_{\mathrm{c}}=\frac{1}{d-1}$ (Figs.~\ref{fig:clustCoeff_gammaDependence}a and b), the value of $\bar{c}$ measured at a given degree decay exponent $\gamma$ is affected by the number of dimensions of the underlying hyperbolic space: $d$PSO networks created in higher-dimensional spaces using smaller popularity fading parameters display smaller values of $\bar{c}$ compared to networks that have the same degree decay exponent $\gamma$ but were generated in lower-dimensional hyperbolic spaces with higher popularity fading parameters. This can be explained roughly by considering that at not too high temperatures, the newly appearing nodes tend to connect to already existing nodes that are relatively similar to them, where high similarity means small angular distance~\cite{PSO}. If the number of dimensions or, equivalently, the number of independent angular coordinates characterising the node attributes is increased, the number of possible coordinate combinations that define the set of positions of similar attributes for a given new node also increases. 
Therefore, any two nodes that can be considered to be similar from a third node's point of view are less and less likely to have angular coordinates that are relatively 
close to each other as well with the increase in the number of dimensions. 
Consequently, the selected nodes to which the new node connects tend to share fewer links in higher dimensions; thus, the number of triangles in the emerging network is reduced.

As we approach the critical point from below, the role of the number of dimensions in local triangle formation gradually weakens. Near to and above $T = T_{\mathrm{c}}$ (Figs.~\ref{fig:clustCoeff_gammaDependence}c and d), where connections between rather distant nodes are likely to occur too, the value of $\bar{c}$ is independent of the individual values of the $d$ and the $\beta$ parameters that produce the same degree decay exponent $\gamma$. Here, it seems that the randomising influence of the high temperature on link formation is so strong that the similar effect of the large number of dimensions is negligible compared to it. 

\subsection{Finding and evaluating communities}
\label{sect:comms}

Communities are very important structural units in complex networks at the "mesoscopic" scale, without a widely accepted unique definition, but usually associated with subgraphs with a larger internal and a smaller external link density. 
The automated extraction of communities based solely on the network topology is a challenging problem, with an immense number of different solutions proposed in the literature~\cite{Fortunato_coms,Fortunato_Hric_coms,Cherifi_coms}. An interesting related feature of two-dimensional hyperbolic networks is that they also contain communities for the major part of the parameter space, in spite of the lack of any explicit built-in community formation mechanism in the graph generation algorithms~\cite{our_hyp_coms}. Motivated by this and the overall importance of communities in network science, here we examine also the community structure of $d$-dimensional PSO networks. 

Along this line, we apply three independent and well-established community finding algorithms to locate the modules, namely the asynchronous label propagation algorithm~\cite{alabprop,alabprop_code}, the Infomap method~\cite{Infomap,Infomap_code} and the Louvain algorithm~\cite{Louvain,Louvain_code}. The basic idea of asynchronous label propagation is to simulate the diffusion of community labels along the examined network, where the regular updating of the labels based on the neighbouring nodes brings a rapid consensus among the members of a dense group on a unique label. In contrast, the Infomap algorithm provides an information-theoretic approach for finding communities, taking advantage of the fact that communities can actively help in achieving the most parsimonious description of the trajectory of an infinitely long random walk on the network. The algorithm itself searches for the minimum of the so-called map equation, which expresses the code length for an average movement in the above-mentioned random walk process. Finally, the Louvain method performs a fast and efficient heuristic maximisation of modularity, which corresponds to the most widely used quality measure for communities~\cite{Newman_modularity_original,modularity_code}, expressed in general as 
\begin{equation}
    Q = \frac{1}{2E}\sum_{i=1}^N\sum_{j=1}^N\left[A_{ij} -P_{ij}\right]\delta_{c_i,c_j},
\end{equation}
where $N$ is the number of nodes in the network, $A_{ij}$ denotes an element of the adjacency matrix ($A_{ij}\equiv A_{ji}=1$ if $i$ is connected to $j$, and otherwise $A_{ij}\equiv A_{ji}=0$), $P_{ij}$ gives the connection probability between nodes $i$ and $j$ in a random null model, $E$ stands for the total number of links in the network, $c_i$ is the community to which node $i$ belongs and the Kronecker delta $\delta_{c_i,c_j}$ ensures that non-zero contribution can come only from node pairs of the same community. A natural choice for the null model is given by the configuration model, yielding $P_{ij}=\frac{k_ik_j}{2E}$. In the Louvain approach, $Q$ is optimised in a hierarchical manner, where after finding the local maximum at a given organisation level of the network, in the next step we move up to the next level by aggregating the current communities into single nodes. 

As demonstrated by Fig.~\ref{fig:layoutExample}, a community in a hyperbolic network arises from some inner nodes that serve as community cores and the outer nodes of the corresponding angular sector that are held together by their common preference toward the same attractive centres. As it was detailed in Ref.~\cite{our_hyp_coms} in the case of the two-dimensional PSO model, the emergence of a strong community structure can be achieved under two conditions: the existence of inner nodes that are distant from each other enough to provide well-separated attractive centres for the different angular regions, and the localisation of the connections. The distance between the inner community cores can be increased by accelerating their outward drift that simulates the popularity fading via decreasing the popularity fading parameter $\beta$. The strong localisation of the connections can be ensured primarily by setting the temperature $T$ to a small value and thus making the cutoff in the connection probability sharp as a function of the hyperbolic distance. However, it has to be also taken into consideration that if the ratio between the number $N$ of nodes and the number $m$ of connections established by each node at its appearance is smaller, then, to create all the $m$ number of links, the nodes are forced more often to connect even to farther nodes. As a consequence, a small temperature in itself is not always enough to make the connections localised, but it has to be complemented with a relatively large $N/m$ ratio in order to make the connections more strongly determined by the hyperbolic distances and create hereby a more clear separation between the angular regions with respect to the links, thus supporting community formation. 

\begin{figure}[h!]
    \centering
    \captionsetup{width=1.0\textwidth}
    \includegraphics[width=1.0\textwidth]{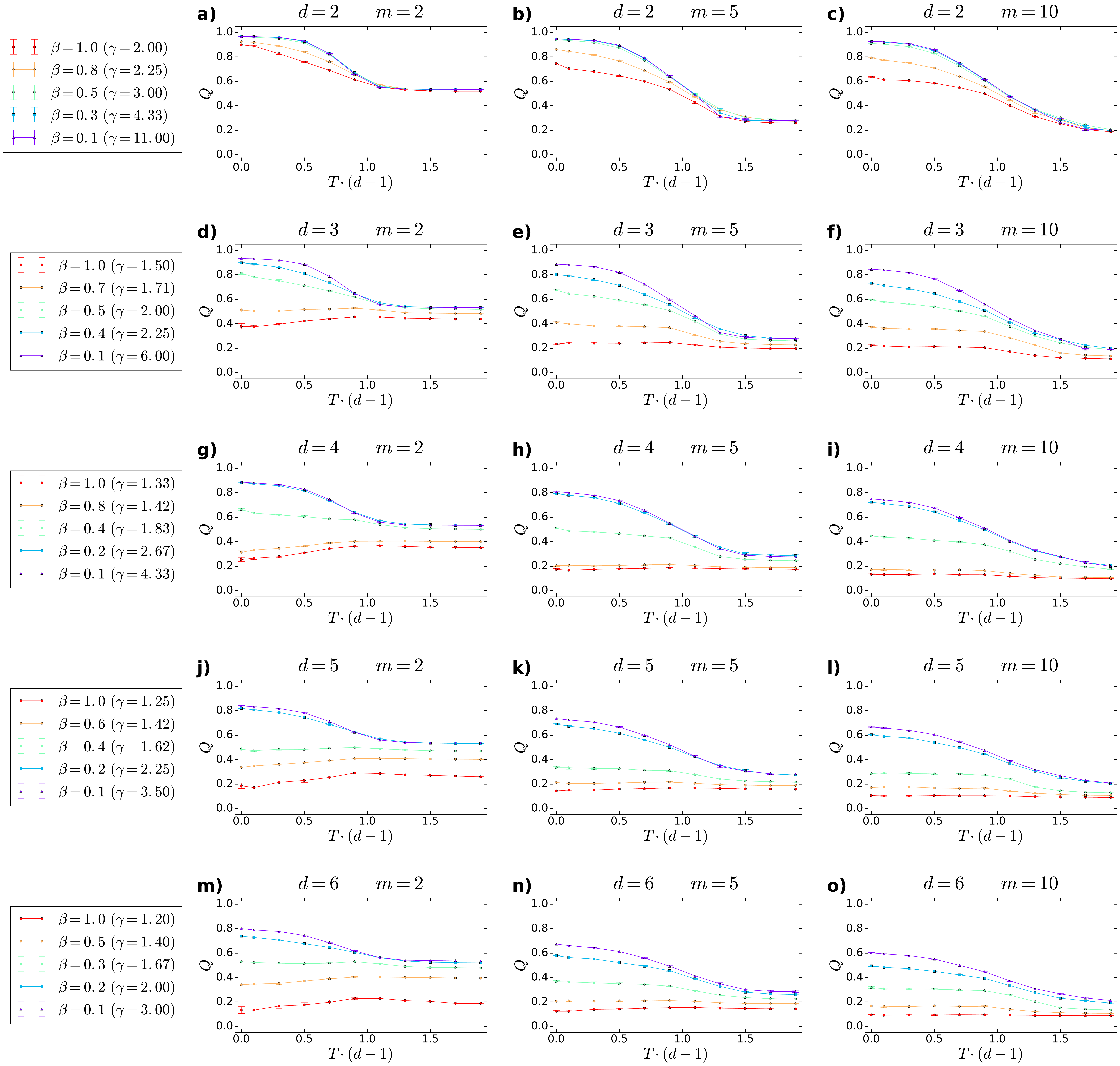}
    \caption{ {\bf The highest modularity $Q$ achieved among the communities obtained by the asynchronous label propagation, the Louvain and the Infomap algorithms in $d$PSO networks, as a function of the rescaled temperature $T\cdot(d-1)$.} The dimension $d$ is constant across the panel rows, whereas the expected average degree $2m$ is constant across the panel columns, as indicated by the panel titles. 
    The different curves in a given subplot correspond to different values of the popularity fading parameter $\beta$ (yielding different degree decay exponents $\gamma$), listed for each dimension in the leftmost panel of the corresponding row. The displayed data points were obtained by averaging over 5 $d$PSO networks of $N=10,000$ nodes at curvature $K=-\zeta^2=-1$, 
    the error bars indicate the standard deviations. 
    }
    \label{fig:modularity_Tdependence}
\end{figure}

\begin{figure}[h!]
    \centering
    \captionsetup{width=1.0\textwidth}
    \includegraphics[width=1.0\textwidth]{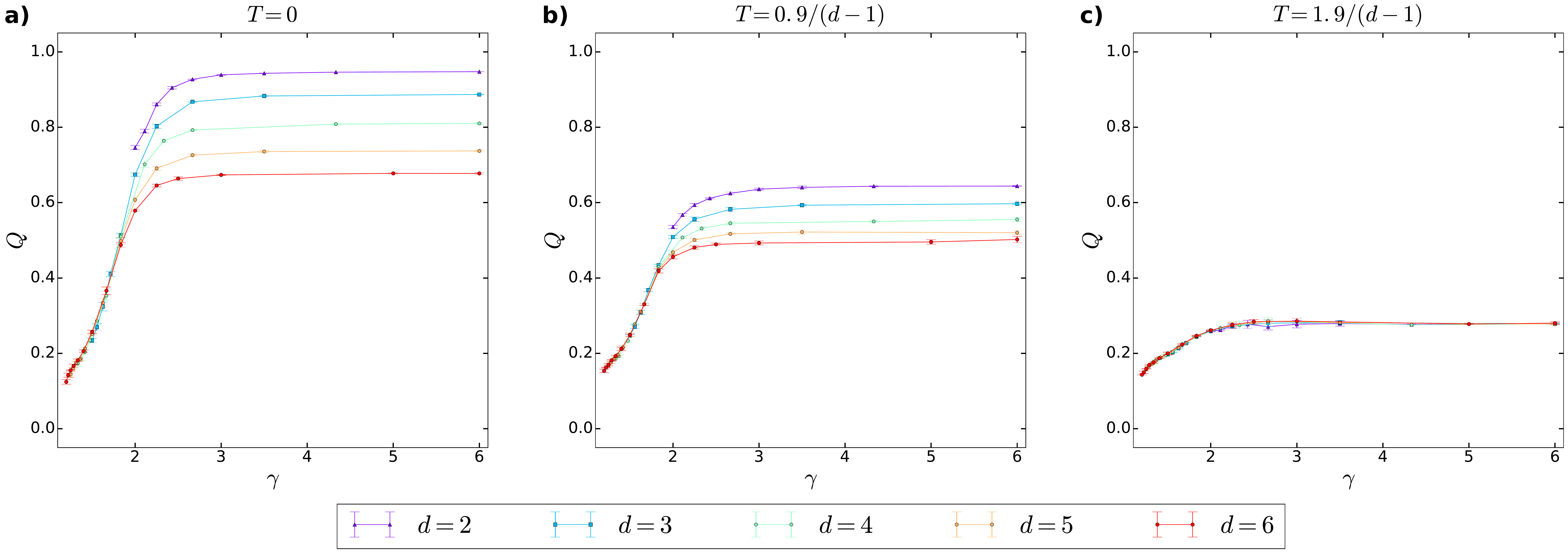}
    \caption{ 
    {\bf The highest modularity $Q$ achieved between the asynchronous label propagation, the Louvain and the Infomap algorithms in $d$PSO networks as a function of the degree decay exponent $\gamma$ at different values of the temperature $T$ and the dimension $d$.} The panels refer to different values of the temperature $T$, given in the title of the subplots. The curves of different colours correspond to different number of dimensions $d$, as listed below the panels. The popularity fading parameter was calculated as $\beta=\frac{1}{(d-1)\cdot(\gamma-1)}$. We always set the curvature $K=-\zeta^2$ of the hyperbolic space to $-1$, the network size $N$ to $10,000$ and the expected average degree $2m$ to $10$. We searched for communities once with all three community detection methods on 5 $d$PSO networks and plotted the obtained highest modularity averaged over the 5 networks for each parameter setting, with the error bar indicating the standard deviation among the 5 networks.
    }
    \label{fig:modularity_gammaDependence}
\end{figure}

In Fig.~\ref{fig:modularity_Tdependence}, we show the highest modularity $Q$ obtained among the applied three community finding methods as a function of the rescaled temperature $T\cdot(d-1)$ for $d$PSO networks generated at different parameter settings. In each panel, the bundle of curves (showing $Q$ for networks with different degree decay exponents) seems to form a fork-like pattern, in which the curves are more distant from each other in the low-temperature regime and approach each other at high values of $T\cdot(d-1)$, where $Q$ becomes more or less constant and independent from the rescaled temperature. 
In Fig.~\ref{fig:modularity_gammaDependence}, we show the achieved highest modularity $Q$ as a function of the degree decay exponent $\gamma$ for different number of dimensions at a low temperature (Fig.~\ref{fig:modularity_gammaDependence}a), a moderate temperature (Fig.~\ref{fig:modularity_gammaDependence}b) and a high temperature (Fig.~\ref{fig:modularity_gammaDependence}c). For all of the examined temperatures, $Q$ starts at low values and shows first a strong, then a mild increase toward the higher values of $\gamma$, i.e. as the largest node degrees decrease. 
The effect of the temperature can be observed by comparing the three panels, where we can see that the constant value to which $Q$ settles in the large $\gamma$ regime is higher if $T$ is lower, as expected. Furthermore, according to Fig.~\ref{fig:modularity_gammaDependence}c, when the rescaled temperature $T\cdot(d-1)$ is high enough, the $Q(\gamma)$ curves seem to collapse onto a universal curve for all the examined dimensions, while at smaller temperatures (Fig.~\ref{fig:modularity_gammaDependence}a and b), the modularity measured at a given degree decay exponent $\gamma$ is a decreasing function of the dimension $d$, similarly to what has been seen on the local scale for the average clustering coefficient in Fig.~\ref{fig:clustCoeff_gammaDependence}.

\section{Discussion}
\label{sect:disc}

The PSO model~\cite{PSO} is arguably one of the most successful hyperbolic network models since it offers a quite natural way to reproduce the major structural properties of real-world networks. The original formulation of this approach is, however, given only for the two-dimensional hyperbolic disk; therefore, a question arising naturally in this context is how it can reasonably be extended to higher-dimensional hyperbolic spaces. In the present paper, we studied this issue in detail and introduced the $d$PSO model as a $d$-dimensional generalisation of the original PSO model.

The obtained analytical results show that the scale-free property of $d$PSO networks is inherited from the original PSO model, meaning that the tail of the degree distribution decays as a power-law, $\displaystyle{\lim_{k \to \infty}} \pazocal{P}(k)\sim k^{-\gamma}$, for any number of dimensions $d\geq 2$. The exponent $\gamma$ is determined by the popularity fading parameter $\beta$ and the number of dimensions $d$ via the formula $\gamma=1+\frac{1}{(d-1)\cdot\beta}$, which can be basically explained as follows. 
The solid angle subtended by the entire surface of the $d$-dimensional ball representing the $d$-dimensional hyperbolic space is an increasing function of the number of dimensions, meaning that in the case of a higher $d$, the uniform distribution of the same number of nodes on the surface results in larger angular distances between the nearest neighbours. Therefore, at a given popularity fading parameter (controlling the attractiveness of the early-appeared nodes arising from their relatively low radial coordinate),
in higher dimensions we see an increased propensity of the newly coming nodes to connect to the innermost nodes of any angular position instead of the angular neighbours located at larger radii. Due to this, the same popularity fading parameter $\beta$ yields larger maximum degree, and thus, smaller degree decay exponent $\gamma$ for larger values of $d$. To compensate the decrease in the attractiveness of the angular neighbours caused by the increase in the number of dimensions and keep the degree decay exponent $\gamma$ at a given value, one has to reduce the distinguished attractiveness of the inner nodes and limit the angular range in which they are preferred. For this, the advantage of the inner nodes in their radial position has to be decreased by shifting them more outwards, which can be achieved by setting the popularity fading parameter $\beta$ to a smaller value and enhancing thereby the process of popularity fading.

It is worth emphasising that the smallest attainable degree decay exponent (obtained at $\beta=1$) is $2$ in the original PSO model~\cite{PSO}, whereas in $d$PSO networks 
\begin{equation}
\gamma_{\text{min}}(d)=1+\frac{1}{d-1},   
\end{equation} 
which can be decreased below the two-dimensional limit $\gamma_{\text{min}}(2)=2$ by simply increasing the number of dimensions $d$ above $2$. Note that the analysis of scale-free networks with a degree decay exponent smaller than two is in itself an interesting topic since such networks exhibit many exotic features that can not be observed in scale-free networks with $2\leq\gamma$, including the divergence of the average degree or the presence of macroscopical hubs being connected to a finite fraction of the nodes even in the thermodynamic limit $N\to\infty$~\cite{bianconi_exp_less_than_two,dorogov_exp_one}.

In terms of $d$PSO networks, we have found that such extremely skewed degree distributions lead to further unexpected results. As indicated in~Fig.~\ref{fig:clustCoeff_gammaDependence} and Fig.~\ref{fig:modularity_gammaDependence}, $d$PSO networks of $\gamma<2$ are characterised by high average clustering coefficient $\bar{c}$, but remarkably at the same time relatively small modularity $Q$ at basically any temperature $T$. Intuitively, this behaviour can be understood if we consider that on the one hand, since the largest hubs are formed from the first few nodes of the network generation process, they are densely connected to each other. Thus, in the presence of extremely large hubs to which most of the nodes connect at their appearance, triangles are formed with large probability, resulting in a large $\bar{c}$. On the other hand, these large hubs make the partitioning of the network into disjunct communities with high modularity practically impossible since the community of any such hub has a macroscopic number of links pointing outside of the given community, resulting in low $Q$ values. 

In addition, we found that the number of dimensions $d$ along with the temperature $T$ play a joint role in controlling the average clustering coefficient $\bar{c}$ of $d$PSO networks as well. More precisely, as it can be seen in Fig.~\ref{fig:clustCoeff_Tdependence}, we can distinguish two phases separated by the critical point $T_{\text{c}}=\frac{1}{d-1}$, where the average clustering coefficient of the networks behaves in a fundamentally different way. At temperatures $T<T_{\text{c}}$, $\bar{c}$ is a decreasing function of the temperature; however, at temperatures near and above $T_{\text{c}}$, $\bar{c}$ tends to become independent from $T$. Besides, based on Fig.~\ref{fig:clustCoeff_gammaDependence}, at low temperatures we can state that the average clustering coefficient measured at a given degree decay exponent $\gamma$ decreases with the number of dimensions $d$, in perfect accordance with the results found in Ref.~\cite{boguna_multiscale_unfolding} for the RHG model. Meanwhile, near the critical temperature $\bar{c}$ begins to show a universal decay with $\gamma$, irrespectively of the separate values of the popularity fading parameter $\beta$ and the dimension $d$. Despite the obvious differences between the $d$-dimensional PSO and RHG models, remarkably, a similar separation of phases with the same critical temperature has also been observed in the $d$-dimensional RHG model and discussed in a slightly different context in Ref.~\cite{RHG_d_dim_krioukov}. (Note however that the analogous phases of the two models are different in certain aspects.) 

As it has already been reported in Ref.~\cite{our_hyp_coms}, the original, two-dimensional PSO model~\cite{PSO} is capable of generating networks with strong communities for a wide range of the parameter settings, despite the fact that it does not include any explicitly built-in community structure generating mechanism. Based on the high modularity values measured on $d$PSO networks (Figs.~\ref{fig:modularity_Tdependence} and \ref{fig:modularity_gammaDependence}), here we conclude that the emergence of a strong community structure is not a peculiar feature of the two-dimensional case, but it can be observed even if the number of dimensions is larger than $2$, provided that the degree decay exponent $\gamma$ is not too small and the temperature $T$ is not too high. Nevertheless, at such settings of $\gamma$ and $T$, the modularity $Q$ obtained at a given $\gamma$ decreases as $d$ is increased, similarly to the average clustering coefficient $\bar{c}$.

In conclusion, motivated by the fact that there is evidently no particular reason to assume that the underlying hyperbolic space 
of complex networks 
is certainly two-dimensional, here we proposed a 
generalisation of the two-dimensional popularity-similarity optimisation model of network growth, 
namely the $d$PSO model, in which the dimension $d$ acts as an additional degree of freedom. This increase in the number of freely variable model parameters allows a much richer characterisation of the networks, and therefore, enables us to better adjust the properties of the generated networks to that of the empirically observed data. In connection with this, we found that there exists a relatively broad range of parameter settings where $d$PSO networks are suitable for capturing many essential characteristics of real-world networks such as the scale-free property, the high value of the average clustering coefficient or the strong community structure. Namely, $d$PSO networks generated in lower-dimensional hyperbolic spaces using not too high temperatures and not too low degree decay exponents simultaneously exhibit all the above-mentioned features of real-world networks. 
Nevertheless, as the number of dimensions $d$ increases, both the maximal average clustering coefficient and the maximal modularity that can be obtained at a given degree decay exponent $\gamma$ (by setting the temperature $T$ to $0$) decreases. 
This implies that $d$PSO networks generated in high-dimensional spaces (e.g., at $d>10$) can be characterised by neither a clustering nor a community structure of similar strength that can be observed in various real-world examples; hence, a reasonable upper limit can be found on the number of dimensions of the hyperbolic space that underlies real networks. By following similar considerations regarding the clustering coefficient of RHG networks, in Ref.~\cite{boguna_network_geometry} the authors claim that in terms of both the modelling and the embedding techniques, the most suitable choice for the number of dimensions of the underlying hyperbolic space is $d=2$. Here, we complement this result by showing that in general, the $d$PSO model with $d$ slightly above $2$ also performs excellently on the modelling ground (still yielding relatively high values of the average clustering coefficient and the modularity), which, along with the recent success of $d>2$ hyperbolic embedding techniques for instance in link prediction~\cite{high_dim_embed_author_coll} or the separation of communities~\cite{Cannistraci_ASI}, confirmes the relevance of low-dimensional hyperbolic spaces (e.g. with $d=3$ or $d=4$) in the theory of complex networks. In Ref.~\cite{linkWeights_coalescentEmbedding}, several different methods have been already provided for assigning angular coordinates to the network nodes in hyperbolic spaces of arbitrary curvature $K=-\zeta^2$ and number of dimensions $d$. Using the angular positions yielded by one of these methods, in order to create an embedding that corresponds to our $d$PSO model, the radial coordinate of the node having the $\ell$th ($\ell=1,2,...,N$) largest degree (with ties in the order of node degrees broken arbitrarily) has to be calculated as $r_{\ell N}=\beta\cdot(2/\zeta)\cdot\ln{\ell}+(1-\beta)\cdot(2/\zeta)\cdot\ln{N}$, where the popularity fading parameter $\beta$ is determined by the degree decay exponent $\gamma$ and the number of dimensions $d$ as $\beta=\frac{1}{(d-1)\cdot(\gamma-1)}$.


\section*{Code availability}
The code used for generating networks with the $d$-dimensional PSO model will be available at https://github.com/BianKov/dPSO upon publication.

\section*{Acknowledgments}
The research was partially supported by the Hungarian National Research, Development and Innovation Office (grant no. K 128780, NVKP\_16-1-2016-0004),
by the European Union’s Horizon 2020 research and innovation programme, VEO (grant agreement No. 874735) and the Thematic Excellence Programme (Tématerületi Kiválósági Program, 2020-4.1.1.-TKP2020) of the Ministry for Innovation and Technology in Hungary, within the framework of the Digital Biomarker thematic programme of the Semmelweis University.

\section*{Author contributions statement}
B.K. and G.P. developed the concept of the study, B.K. constructed the generative steps of the $d$PSO, the $f$PSO and the three-dimensional nPSO model, B.K. carried out the analytical calculations for the degree distribution of $d$PSO networks in the $T=0$ case, B.K. and S.G.B. performed the analytical calculations for the degree distribution of $d$PSO networks in the $0<T$ case, B.K. performed the numerical analyses and prepared the figures, B.K., S.G.B. and G.P. analysed and interpreted the results, G.P., S.G.B. and B.K. wrote the paper. All authors reviewed the manuscript.

\section*{Competing Interests}
The authors declare no competing interests.

\section*{References}
\bibliography{references}

\begin{thebibliography}{10}
\bibitem{alabprop_code}
We used the Python function `asyn\_lpa\_communities', an implementation of the
  asynchronous label propagation algorithm available in the
  `networkx.algorithms.community.label\_propagation' package.;.
\bibitem{modularity_code}
We calculated the modularity values with the Python function `modularity'
  available in the `networkx.algorithms.community.quality' package.;.
\bibitem{AMI_code}
We calculated the adjusted mutual information values with the Python function
  `adjusted\_mutual\_info\_score' available in the `sklearn.metrics.cluster'
  package.;.
\end{thebibliography}

\clearpage
\setcounter{section}{0}
\renewcommand{\thefigure}{S\arabic{figure}}
\renewcommand{\thetable}{S\arabic{table}}
\renewcommand{\theequation}{S\arabic{equation}}
\renewcommand{\thesection}{S\arabic{section}}

\title[Supplementary Material]{Popularity-similarity optimisation model beyond two dimensions -- Supplementary Material}

%
%

\section{Degree distribution of $d$PSO networks}
\label{sect:degreeDist}
\setcounter{figure}{0}
\setcounter{table}{0}
\setcounter{equation}{0}
\renewcommand{\thefigure}{S1.\arabic{figure}}
\renewcommand{\thetable}{S1.\arabic{table}}
\renewcommand{\theequation}{S1.\arabic{equation}}

It has been shown in Ref.~\cite{PSO} that in the two-dimensional PSO model of expected average degree $\bar{k}\approx 2\cdot m$ and popularity fading parameter $\beta$, the probability that node $i$ (appearing at time $i$) and node $j$ (appearing at time $j>i$) connect to each other can be written as
\begin{equation}
    \Pi(i,j)=m\cdot\frac{i^{-\beta}}{\int_1^j i^{-\beta}\,\mathrm{d}i}.
    \label{eq:connProbIn2D}
\end{equation}
The preferential attachment model in Ref.~\cite{PA_degreeDistr} yields the same connection probability with $\beta=1/(\gamma-1)$ when the number of connections created at the appearance of a new node is set to $m$ and the degree distribution $\pazocal{P}(K=k)$ is proportional to $k^{-\gamma}$.
In this section we show that in the $d$-dimensional popularity-similarity optimisation model
\begin{equation}
    \Pi(i,j)=m\cdot\frac{i^{-(d-1)\cdot\beta}}{\int_1^j i^{-(d-1)\cdot\beta}\,\mathrm{d}i},
    \label{eq:connProb_statement}
\end{equation}
which -- according to the analogy with the preferential attachment model -- means that the degree distribution of the networks generated by the $d$PSO model takes the form of $\pazocal{P}(K=k) \sim k^{-\gamma}$ with $\gamma=1+\frac{1}{(d-1)\cdot\beta}$. Sect.~\ref{sect:DetConnDegreeDist} deals with the case of $T=0$, where the new node always connects to the $m$ hyperbolically closest nodes, while Sect.~\ref{sect:ProbConnDegreeDist} describes settings where the temperature is strictly larger than zero ($0<T$), enabling hyperbolically father nodes to become connected as well.

\subsection{Connection probability in the case of deterministic connection}
\label{sect:DetConnDegreeDist}
\setcounter{figure}{0}
\setcounter{table}{0}
\setcounter{equation}{0}
\renewcommand{\thefigure}{S1.1.\arabic{figure}}
\renewcommand{\thetable}{S1.1.\arabic{table}}
\renewcommand{\theequation}{S1.1.\arabic{equation}}

At temperature $T=0$, each node $j=1,2,...,N$ connects at its appearance to the $m$ hyperbolically closest nodes or, in other words, to all the previously appeared nodes that lie from node $j$ within a certain hyperbolic distance $R_j$. Thus, the probability $\Pi(i,j)$ of the emergence of a link between nodes $i$ and $j$ (with $i<j$) equals to the probability that the hyperbolic distance of node $i$ from node $j$ at the appearance of the latter is not larger than $R_j$, i.e.
\begin{equation}
    \Pi(i,j)=P(x_{ij}(j)\leq R_j),
    \label{eq:determinConnProb_basic}
\end{equation}
where $R_j$ is determined by the equation
\begin{equation}
    m=\int_1^j \Pi(i,j)\,\mathrm{d}i
    \label{eq:equationForR_determin}
\end{equation}
expressing that the expected number of the previously appeared nodes that connect to node $j$ must be $m$. The hyperbolic distance between nodes $i$ and $j$ at time $j$ can be approximated as
\begin{equation}
    x_{ij}(j)\approx r_{ij}+r_{jj}+\frac{2}{\zeta}\cdot\ln\left(\frac{\theta_{ij}}{2}\right),
    \label{eq:hypDistApprox}
\end{equation}
assuming that $\zeta r_{ij}$ and $\zeta r_{jj}$ are sufficiently large (which is true for most of the network nodes if the total number of nodes $N$ is large enough) and therefore, $2\cdot\sqrt{e^{-2\zeta r_{ij}}+e^{-2\zeta r_{jj}}}<\theta_{ij}$, but in the meantime the angular distance between the possibly connecting nodes $i$ and $j$ is small enough to use the approximation $\sin(\theta_{ij}/2)\approx\theta_{ij}/2$~\cite{hyperGeomBasics}. Using Eq.~(\ref{eq:hypDistApprox}), the connection probability of nodes $i$ and $j$ can be written for $T=0$ as
\begin{equation}
    \Pi(i,j)=P(x_{ij}(j)\leq R_j)=P\left(\theta_{ij}\leq 2\cdot e^{-\frac{\zeta}{2}\cdot(r_{ij}+r_{jj}-R_j)}\right),
    \label{eq:determinConnProb_basicWithDistApprox}
\end{equation}
i.e., we are searching for the probability that the angular distance between nodes $i$ and $j$ is not larger than a given value, namely $\theta_{ij}^{\mathrm{max}}=2\cdot e^{-\frac{\zeta}{2}\cdot(r_{ij}+r_{jj}-R_j)}$. This can be rephrased as the probability that the angular position of node $i$ falls in the spherical sector characterised by an apex angle $2\cdot\theta_{ij}^{\mathrm{max}}$ and an axis going through node $j$.

Since the angular position of the network nodes is chosen uniformly at random on the surface of a $d$-dimensional ball, the probability that the angular location of a given node falls in a certain spherical sector can be written simply as the fraction of the solid angle subtended by the sector in question and the solid angle subtended by the complete $d$-dimensional ball, independently of the direction of the axis of the examined spherical sector. Consequently, the probability that the angular distance measured between a new node and a previously appeared node is not larger than a given value $\theta^{\mathrm{max}}$ equals to the solid angle subtended by a spherical sector of apex angle $2\cdot\theta^{\mathrm{max}}$ divided by the solid angle subtended by the complete ball. Thus, 
\begin{equation}
    \Pi(i,j)=\frac{\Omega_d\left(2\cdot e^{-\frac{\zeta}{2}\cdot(r_{ij}+r_{jj}-R_j)}\right)}{\Omega_d^{\mathrm{total}}}=\frac{\Omega_{d-1}^{\mathrm{total}}}{\Omega_d^{\mathrm{total}}}\cdot\int_0^{2\cdot e^{-\frac{\zeta}{2}\cdot(r_{ij}+r_{jj}-R_j)}}\sin^{d-2}{\phi}\,\mathrm{d}\phi,
    \label{eq:determinConnProb_withSolidAngles}
\end{equation}
where $\Omega_d^{\mathrm{total}}$ denotes the solid angle subtended by the complete surface of a $d$-dimensional ball, namely
\begin{equation}
 \Omega_d^{\mathrm{total}}=\left\lbrace \begin{array}{ll}
 \frac{2\cdot\pi^{\frac{d}{2}}}{\left(\frac{d}{2}-1\right)!} & \mbox{if}\; d\mbox{ is even,} \\ \\
 \frac{\frac{d-1}{2}!\cdot2^d\cdot\pi^{\frac{d-1}{2}}}{(d-1)!} & \mbox{if}\; d\mbox{ is odd,}
 \end{array} \right.
 \label{eq:solidAngle_total}
\end{equation}
and in the last step we used that the solid angle subtended by a $d$-dimensional spherical sector of an apex angle $2\psi$ can be calculated as
\begin{equation}
    \Omega_d(\psi) = \Omega_{d-1}^{\mathrm{total}}\cdot\int_0^{\psi}\sin^{d-2}{\phi}\,\mathrm{d}\phi.
    \label{eq:solidAngle_sphericalSector}
\end{equation}
In the case of large networks, the maximum angular distance $\theta_{ij}^{\mathrm{max}}=2\cdot e^{-\frac{\zeta}{2}\cdot(r_{ij}+r_{jj}-R_j)}$ that still enables the link formation is small enough for most of the nodes to assume that in the range of the integration in Eq.~(\ref{eq:determinConnProb_withSolidAngles}), $\sin^{d-2}{\phi}\approx\phi^{d-2}$. Based on this,
\begin{align}
    \Pi(i,j)\approx&\frac{\Omega_{d-1}^{\mathrm{total}}}{\Omega_d^{\mathrm{total}}}\cdot\int_0^{2\cdot e^{-\frac{\zeta}{2}\cdot(r_{ij}+r_{jj}-R_j)}}\phi^{d-2}\,\mathrm{d}\phi = \frac{\Omega_{d-1}^{\mathrm{total}}}{\Omega_d^{\mathrm{total}}}\cdot\left[\frac{\phi^{d-1}}{d-1}\right]^{2\cdot e^{-\frac{\zeta}{2}\cdot(r_{ij}+r_{jj}-R_j)}}_0 =
    \nonumber \\ & =\frac{\Omega_{d-1}^{\mathrm{total}}}{\Omega_d^{\mathrm{total}}}\cdot\frac{2^{d-1}\cdot e^{-\frac{\zeta\cdot(d-1)}{2}\cdot(r_{ij}+r_{jj}-R_j)}}{d-1}.
    \label{eq:determinConnProb_withSolidAngles_smallAngles}
\end{align}
Introducing the notation
\begin{equation}
 \eta(d)=\frac{\Omega_{d-1}^{\mathrm{total}}}{\Omega_d^{\mathrm{total}}}=\left\lbrace \begin{array}{ll}
 \frac{\left(\frac{d}{2}-1\right)!\cdot \frac{d-2}{2}!\cdot 2^{d-2}}{(d-2)!\cdot\pi} & \mbox{if}\; d\mbox{ is even,} \\ \\
 \frac{(d-1)!}{\left(\frac{d-1}{2}-1\right)!\cdot\frac{d-1}{2}!\cdot 2^{d-1}} & \mbox{if}\; d\mbox{ is odd,}
 \end{array} \right.
 \label{eq:etaFormula1}
\end{equation}
we arrive at the formula
\begin{equation}
    \Pi(i,j)=\eta(d)\cdot\frac{2^{d-1}}{d-1}\cdot e^{-\frac{\zeta\cdot(d-1)}{2}\cdot(r_{ij}+r_{jj}-R_j)}
    \label{eq:determinConnProb_finalButWithR}
\end{equation}
of the probability that nodes $i$ and $j$ get connected during the network growth.

According to the model definition, here the initial radial coordinate of the $j$th node is $r_{jj}=(2/\zeta)\ln{j}$ and the radial coordinate of node $i$ at time $j$ is $r_{ij}=\beta\cdot r_{ii}+(1-\beta)\cdot r_{jj}=$ $=\beta\cdot(2/\zeta)\ln{i}+(1-\beta)\cdot(2/\zeta)\ln{j}$, while $R_j$ can be expressed from the equation
\begin{equation}
    m=\eta(d)\cdot\frac{2^{d-1}}{d-1}\cdot e^{-\frac{\zeta\cdot(d-1)}{2}\cdot(2-\beta)\cdot r_{jj}}\cdot e^{\frac{\zeta\cdot(d-1)}{2}\cdot R_j}\cdot\int_1^j e^{-\frac{\zeta\cdot(d-1)}{2}\cdot\beta\cdot r_{ii}}\,\mathrm{d}i
    \label{eq:eqForRj}
\end{equation}
obtained by substituting
~(\ref{eq:determinConnProb_finalButWithR}) into Eq.~(\ref{eq:equationForR_determin}). After some rearrangement, we 
can express $R_j$ as
\begin{equation}
    R_j = \frac{2}{\zeta\cdot(d-1)}\cdot \ln\left(\frac{(d-1)\cdot m}{\eta(d)\cdot2^{d-1}\cdot e^{-\frac{\zeta\cdot(d-1)}{2}\cdot(2-\beta)\cdot r_{jj}}\cdot \int_1^j e^{-\frac{\zeta\cdot(d-1)}{2}\cdot\beta\cdot r_{ii}}\,\mathrm{d}i}\right).
    \label{eq:RjForT=0}
\end{equation}
Substituting this expression back into ~(\ref{eq:determinConnProb_finalButWithR}) yields
\begin{equation}
    \Pi(i,j)=m\cdot \frac{e^{-\frac{\zeta\cdot(d-1)}{2}\cdot\beta\cdot r_{ii}}}{\int_1^j e^{-\frac{\zeta\cdot(d-1)}{2}\cdot\beta\cdot r_{ii}}\,\mathrm{d}i} = m\cdot \frac{i^{-(d-1)\cdot\beta}}{\int_1^j i^{-(d-1)\cdot\beta}\,\mathrm{d}i}.
    \label{eq:determinConnProb_final}
\end{equation}
Hereby we proved that the probability that nodes $i$ and $j$ connect to each other indeed can be written in the form of Eq.~(\ref{eq:connProb_statement}) at $T=0$.


\subsection{Connection probability in the case of $0<T$}
\label{sect:ProbConnDegreeDist}
\setcounter{figure}{0}
\setcounter{table}{0}
\setcounter{equation}{0}
\renewcommand{\thefigure}{S1.2.\arabic{figure}}
\renewcommand{\thetable}{S1.2.\arabic{table}}
\renewcommand{\theequation}{S1.2.\arabic{equation}}

At temperature $0<T$, the newly appearing node $j$ ($j=1,2,...,N$) repeatedly makes attempts to create connections with the previously appeared nodes (indexed by $i=1,2,...,j-1$) until the emergence of $m$ number of links. Thus, using the probability $P(i,j)$ that the new node $j$ connects in a given connection attempt to node $i$ and the probability $P(j)=\int_1^j P(i,j) \,\mathrm{d}i$ that the new node $j$ connects in a given connection attempt to any of the already existing nodes, the probability that node $i$ becomes connected to node $j$ can be written as
\begin{equation}
\Pi(i,j) = m\cdot\frac{P(i,j)}{P(j)}.
\label{eq:Pi_ij_asFractionOfProbs}
\end{equation}

In one connection attempt, the new node $j$ chooses randomly one of the previously appeared nodes, each with probability $1/(j-1)$, and if there is still no connection between the new node and the selected one, then the link formation occurs with probability $p(x)=1/\left[1+e^{\zeta(x-R_j)/(2T)}\right]$, given that the current hyperbolic distance between the two nodes in question equals to $x$. Notice that in the case of large networks, most of the nodes appear at times $j\gg m$, when the probability that the new node selects such a random node to which it is already connected is insignificant and can be ignored to ease the analysis. However, it has to be taken into consideration that -- contrary to the radial coordinates -- the angular coordinates of the nodes are not strictly determined by the node identifiers, but are random variables; therefore, the hyperbolic distance between two nodes is also a random variable, and the probability that a given link creation attempt succeeds can be formulated as $\int_0^{\infty} p(x)\cdot P(x_{ij}(j)=x)\,\mathrm{d}x$. This can be rewritten as $\int_0^{\pi} p(x)\cdot P(\theta_{ij}=\phi)\,\mathrm{d}\phi$, since for given nodes with given radial coordinates the only source of randomness in the hyperbolic distance is the angular distance $\theta_{ij}$ between the nodes. As described in Sect.~\ref{sect:DetConnDegreeDist}, due to the uniformity of the angular node arrangement, the probability that the angular distance between two nodes falls in the range $[\phi,\phi+\mathrm{d}\phi)$ can be calculated by dividing the solid angle subtended by the volume enclosed between two coaxial spherical sectors of apex angles $2\cdot\phi$ and $2\cdot(\phi+\mathrm{d}\phi)$ by the solid angle subtended by the complete $d$-dimensional ball, i.e.,
\begin{equation}
P(\theta_{ij}=\phi)\,\mathrm{d}\phi = P(\phi\leq\theta_{ij}<\phi+\mathrm{d}\phi) = \frac{\Omega_d(\phi+\mathrm{d}\phi)-\Omega_d(\phi)}{\Omega_d^{\mathrm{total}}}=\frac{\frac{\mathrm{d}\Omega_d}{\mathrm{d}\phi} \,\mathrm{d}\phi}{\Omega_d^{\mathrm{total}}}.
\label{eq:P_angDist}
\end{equation}
Using Eq.~(\ref{eq:solidAngle_sphericalSector}) and that
\begin{equation}
\frac{\mathrm{d}}{\mathrm{d}y}\int_{f(y)}^{g(y)} h(z)\mathrm{d}z = \frac{\mathrm{d}g}{\mathrm{d}y}\cdot h(g(y))-\frac{\mathrm{d}f}{\mathrm{d}y}\cdot h(f(y)),
\label{eq:derivationOfIntegral}
\end{equation}
we arrive at the formula
\begin{equation}
P(\theta_{ij}=\phi)\,\mathrm{d}\phi = \frac{\Omega_{d-1}^{\mathrm{total}}\cdot\sin^{d-2}{\phi} \,\mathrm{d}\phi}{\Omega_d^{\mathrm{total}}},
\label{eq:P_angDist2}
\end{equation}
which, using the notation introduced in Eq.~(\ref{eq:etaFormula1}), takes the form of
\begin{equation}
P(\theta_{ij}=\phi)\,\mathrm{d}\phi = \eta(d)\cdot \sin^{d-2}{\phi} \,\mathrm{d}\phi.
\label{eq:P_angDist3}
\end{equation}
All things considered, the probability that node $i$ attracts a link from node $j$ in a given connection attempt can be formulated as
\begin{equation}
P(i,j) = \frac{\eta(d)}{j-1}\cdot\int_0^{\pi} \frac{\sin^{d-2}{\phi}}{1+e^{\frac{\zeta(x-R_j)}{2T}}} \,\mathrm{d}\phi.
\label{eq:P_ij}
\end{equation}

Using that at the arrival of node $j$ its hyperbolic distance from node $i$ can be written as~\cite{hyperGeomBasics}
\begin{equation}
    x_{ij}(j)\approx r_{ij}+r_{jj}+\frac{2}{\zeta}\cdot\ln\left(\sin{\left(\frac{\theta_{ij}}{2}\right)}\right),
    \label{eq:hypDistApprox_sin}
\end{equation}
we arrive at the formula
\begin{equation}
P(i,j) = \frac{\eta(d)}{j-1}\cdot\int_0^{\pi} \frac{\sin^{d-2}{\phi}}{1+\left(e^{\frac{\zeta}{2}\cdot(r_{ij}+r_{jj}-R_j)}\cdot\sin{\left(\frac{\phi}{2}\right)}\right)^{\frac{1}{T}}} \,\mathrm{d}\phi.
\label{eq:P_ij_radCoord}
\end{equation}

If the temperature $T$ is small enough, then in the case of sufficiently large networks we can assume for most of the node pairs that 
the main contribution of the above integral comes from the range of small angular distances. This implies that the approximation $\sin{\phi}\approx\phi$ can be used, and after that, changing the upper limit of the integral from $\pi$ to infinity practically does not affect the value of the integral. Using these two assumptions, the connection probability becomes
\begin{align}
&P(i,j) \approx \frac{\eta(d)}{j-1}\cdot\int_0^{\infty} \frac{\phi^{d-2}}{1+\left(e^{\frac{\zeta}{2}\cdot(r_{ij}+r_{jj}-R_j)}\cdot\frac{\phi}{2}\right)^{\frac{1}{T}}} \,\mathrm{d}\phi =\nonumber \\ &= \frac{\eta(d)}{j-1}\cdot 2^{d-1}\cdot T\cdot\Gamma\left((d-1)\cdot T\right)\cdot\Gamma\left(1-(d-1)\cdot T\right)\cdot\frac{1}{\left(e^{\frac{\zeta}{2}\cdot(r_{ij}+r_{jj}-R_j)}\right)^{d-1}}.
\label{eq:P_ij_lowT_gammaFunction}
\end{align}
Finally, the substitution of Euler's reflection formula $\Gamma(z)\cdot\Gamma(1-z)=\pi/\sin(z\cdot\pi)$ and the radial coordinate formulas ${r_{jj}=(2/\zeta)\ln{j}}$ and $r_{ij}=\beta\cdot r_{ii}+(1-\beta)\cdot r_{jj}=$ $=\beta\cdot(2/\zeta)\ln{i}+(1-\beta)\cdot(2/\zeta)\ln{j}$ yields
\begin{align}
P(i,j) &\approx \frac{\eta(d)\cdot\pi \cdot 2^{d-1}\cdot T}{(j-1)\cdot\sin((d-1)\cdot T\cdot\pi)}\cdot\frac{1}{\left(e^{\frac{\zeta}{2}\cdot(r_{ij}+r_{jj}-R_j)}\right)^{d-1}} =\nonumber \\ &= \frac{\eta(d)\cdot\pi \cdot 2^{d-1}\cdot T}{(j-1)\cdot\sin((d-1)\cdot T\cdot\pi)}\cdot j^{(d-1)\cdot(\beta-2)}\cdot i^{-(d-1)\cdot\beta}\cdot e^{\frac{\zeta\cdot(d-1)}{2}\cdot R_j},
\label{eq:P_ij_lowT_final}
\end{align}
which gives back for $d=2$ the result of Ref.~\cite{PSO} in the $0<T<1$ case.

Note that since $\Gamma(1-(d-1)\cdot T)$ is not defined for $T=1/(d-1)$ and becomes negative as $T$ exceeds $1/(d-1)$, the approximation in Eq.~(\ref{eq:P_ij_lowT_gammaFunction}) can be valid only in the case of $T<1/(d-1)$. However, at least in the case of sufficiently large networks, 
one can neglect the $1$ in the denominator of the formula~\ref{eq:P_ij_radCoord} for most of the node pairs at higher temperatures, yielding
\begin{equation}
P(i,j) \approx \frac{\eta(d)}{j-1}\cdot\int_0^{\pi}\frac{\sin^{d-2}{\phi}}{\sin^{\frac{1}{T}}{\left(\frac{\phi}{2}\right)}} \,\mathrm{d}\phi\cdot\frac{1}{\left(e^{\frac{\zeta}{2}\cdot(r_{ij}+r_{jj}-R_j)}\right)^{\frac{1}{T}}}.
\label{eq:P_ij_highT_radCoord}
\end{equation}
This approximation can hold only for $1/(d-1)<T$, as otherwise the integral $\int_0^{\pi}\sin^{d-2}(\phi)\cdot\sin^{-1/T}(\phi/2) \,\,\mathrm{d}\phi$ is divergent since $\lim_{\phi\to 0} \sin^{d-2}(\phi)\cdot\sin^{-1/T}(\phi/2)=$ ${=2^{1/T}\cdot\phi^{d-2-1/T}}$ increases slower than $\phi^{-1}$ towards $\phi\to0$ only if $1/(d-1)<T$. To ensure that the term of the connection probability $P(i,j)$ depending on the smaller node index $i$ remains the same for $1/(d-1)<T$ as in the case of $0\leq T<1/(d-1)$, the initial radial coordinate of each node $\ell\geq 1$ must be set to $r_{\ell\ell}=(2T(d-1)/\zeta)\cdot\ln{\ell}$ instead of $r_{\ell\ell}=(2/\zeta)\cdot\ln{\ell}$. This means that in Eq.~(\ref{eq:P_ij_highT_radCoord}) $r_{jj}=(2T(d-1)/\zeta)\cdot\ln{j}$ and $r_{ij}=\beta\cdot r_{ii}+(1-\beta)\cdot r_{jj}=$ $=\beta\cdot(2T(d-1)/\zeta)\ln{i}+(1-\beta)\cdot(2T(d-1)/\zeta)\ln{j}$. Thus,
\begin{equation}
P(i,j) \approx \frac{\eta(d)}{j-1}\cdot \int_0^{\pi}\frac{\sin^{d-2}\phi}{\sin^{\frac{1}{T}}\left(\frac{\phi}{2}\right)} \,\mathrm{d}\phi\, \cdot j^{(d-1)\cdot(\beta-2)}\cdot i^{-(d-1)\cdot\beta}\cdot e^{\frac{\zeta}{2\cdot T}\cdot R_j},
\label{eq:P_ij_highT_final}
\end{equation}
which, using also the approximation $\sin{(\phi/2)}\approx\phi/2$, can be written for $d=2$ as
\begin{align}
P(i,j) \approx& \frac{1/\pi}{j-1}\cdot 2^{\frac{1}{T}}\cdot \int_0^{\pi}\phi^{-\frac{1}{T}} \,\mathrm{d}\phi\, \cdot j^{\beta-2}\cdot i^{-\beta}\cdot e^{\frac{\zeta}{2\cdot T}\cdot R_j} = \frac{1/\pi}{j-1}\cdot 2^{\frac{1}{T}}\cdot \frac{\pi^{1-1/T}}{1-1/T} \cdot j^{\beta-2}\cdot i^{-\beta}\cdot e^{\frac{\zeta}{2\cdot T}\cdot R_j} = \nonumber \\ & =\left(\frac{2}{\pi}\right)^{\frac{1}{T}}\cdot\frac{T}{(j-1)\cdot(T-1)}\cdot j^{\beta-2}\cdot i^{-\beta}\cdot e^{\frac{\zeta}{2\cdot T}\cdot R_j},
\label{eq:P_ij_highT_2D}
\end{align}
corresponding to the result of Ref.~\cite{PSO} in the case of $1<T$.


Based on Eqs.~(\ref{eq:P_ij_lowT_final}) and (\ref{eq:P_ij_highT_final}), the probability that node $j$ connects to any of the previously appeared nodes in a given link formation attempt is
\begin{equation}
P(j) = \int_1^j P(i,j) \,\mathrm{d}i = C\cdot\int_1^j i^{-(d-1)\cdot\beta}\,\mathrm{d}i = C\cdot\frac{j^{1-(d-1)\cdot\beta}-1}{1-(d-1)\cdot\beta}
\label{eq:P_j}
\end{equation}
with
\begin{equation}
 C=\frac{P(i,j)}{i^{-(d-1)\cdot\beta}}=\left\lbrace \begin{array}{ll}
 \frac{\eta(d)\cdot\pi \cdot 2^{d-1}\cdot T}{(j-1)\cdot\sin((d-1)\cdot T\cdot\pi)}\cdot j^{(d-1)\cdot(\beta-2)}\cdot e^{\frac{\zeta\cdot(d-1)}{2}\cdot R_j} & \mbox{if}\; T<\frac{1}{d-1}, \\ \\
 \frac{\eta(d)}{j-1}\cdot \int_0^{\pi}\frac{\sin^{d-2}{\phi}}{\sin^{\frac{1}{T}}{\left(\frac{\phi}{2}\right)}} \,\mathrm{d}\phi\, \cdot j^{(d-1)\cdot(\beta-2)}\cdot e^{\frac{\zeta}{2\cdot T}\cdot R_j} & \mbox{if}\; \frac{1}{d-1}<T
 \end{array} \right.
 \label{eq:P_j_multiplyingFactors}
\end{equation}
denoting the term in $P(i,j)$ that is independent of the node index $i$.

Finally, using Eqs.~(\ref{eq:Pi_ij_asFractionOfProbs}), (\ref{eq:P_j}) and (\ref{eq:P_j_multiplyingFactors}), the probability that nodes $i$ and $j$ connect to each other can be written as
\begin{equation}
\Pi(i,j) = m\cdot\frac{P(i,j)}{P(j)} = m\cdot\frac{C\cdot i^{-(d-1)\cdot\beta}}{C\cdot \int_1^j i^{-(d-1)\cdot\beta}\,\mathrm{d}i} = m\cdot\frac{i^{-(d-1)\cdot\beta}}{\int_1^j i^{-(d-1)\cdot\beta}\,\mathrm{d}i},
\label{eq:connProb_Tnot0}
\end{equation}
which is the same formula as the one given by Eq.~(\ref{eq:connProb_statement}) that we wanted to prove.

\subsection{The role of the multiplying factor in the initial radial coordinates and the $f$PSO model}
\label{sect:radCoordFactor}
\setcounter{figure}{0}
\setcounter{table}{0}
\setcounter{equation}{0}
\renewcommand{\thefigure}{S1.3.\arabic{figure}}
\renewcommand{\thetable}{S1.3.\arabic{table}}
\renewcommand{\theequation}{S1.3.\arabic{equation}}

According to Refs.~\cite{PA_degreeDistr,PSO}, in the properly parametrised preferential attachment model where the connection probability of nodes $i$ (appearing at time $i$) and $j$ (appearing at time $j>i$) can be written as $\Pi(i,j) = m\cdot\frac{i^{-q}}{\int_1^j\ell^{-q}\mathrm{d}\ell}$ (as described in Eq.~(\ref{eq:connProbIn2D})), the degree decay exponent takes the form of $\gamma = 1+\frac{1}{q}$. The results of the previous sections show that in the $d$PSO model
\begin{equation}
 \Pi(i,j)=\left\lbrace \begin{array}{ll}
 m\cdot\frac{e^{-\frac{\zeta\cdot(d-1)}{2}\cdot\beta\cdot r_{ii}}}{\int_1^j e^{-\frac{\zeta\cdot(d-1)}{2}\cdot\beta\cdot r_{\ell\ell}}\,\mathrm{d}\ell} & \mbox{if}\; 0\leq T<\frac{1}{d-1}, \\ \\
 m\cdot\frac{e^{-\frac{\zeta}{2\cdot T}\cdot\beta\cdot r_{ii}}}{\int_1^j e^{-\frac{\zeta}{2\cdot T}\cdot\beta\cdot r_{\ell\ell}}\,\mathrm{d}\ell} & \mbox{if}\; \frac{1}{d-1}<T.
 \end{array} \right.
 \label{eq:connProbFormulaWithRadCoord}
\end{equation}
Thus, setting the initial radial coordinate of each node $\ell\geq 1$ to $r_{\ell\ell}=f\cdot\ln{\ell}$ yields
\begin{equation}
 \Pi(i,j)=\left\lbrace \begin{array}{ll}
 m\cdot\frac{i^{-f\cdot\frac{\zeta\cdot(d-1)}{2}\cdot\beta}}{\int_1^j \ell^{-f\cdot\frac{\zeta\cdot(d-1)}{2}\cdot\beta}\,\mathrm{d}\ell} & \mbox{if}\; 0\leq T<\frac{1}{d-1}, \\ \\
 m\cdot\frac{i^{-f\cdot\frac{\zeta}{2\cdot T}\cdot\beta}}{\int_1^j \ell^{-f\cdot\frac{\zeta}{2\cdot T}\cdot\beta}\,\mathrm{d}\ell} & \mbox{if}\; \frac{1}{d-1}<T,
 \end{array} \right.
 \label{eq:connProbFormulaWithq}
\end{equation}i.e.,
\begin{equation}
 \gamma=\left\lbrace \begin{array}{ll}
 1+\frac{1}{f\cdot\frac{\zeta\cdot(d-1)}{2}\cdot\beta} & \mbox{if}\; 0\leq T<\frac{1}{d-1}, \\ \\
 1+\frac{1}{f\cdot\frac{\zeta}{2T}\cdot\beta} & \mbox{if}\; \frac{1}{d-1}<T.
 \end{array} \right.
 \label{eq:gammaFormulaGeneral}
\end{equation}
Since at $T<1/(d-1)$ we used  for any dimension $d$ the same formula 
\begin{equation}
r_{\ell\ell}=\frac{2}{\zeta}\cdot\ln{\ell}
\label{eq:ourRadCoord_belowTc}
\end{equation}
that was introduced in the original, two-dimensional PSO model~\cite{PSO}, below the critical temperature ${T_{\mathrm{c}}=1/(d-1)}$ the degree decay exponent became dependent on the number of dimensions $d$ besides the popularity fading parameter $\beta$ and, at the same time, independent of all the other model parameters ($\gamma=1+\frac{1}{(d-1)\cdot\beta}$). Then, in order to ensure that the formula of the degree decay exponent remains the same above the critical temperature 
(just as in the original PSO model), we defined the initial radial coordinates as
\begin{equation}
r_{\ell\ell}=\frac{2T(d-1)}{\zeta}\cdot\ln{\ell}
\label{eq:ourRadCoord_aboveTc}
\end{equation}
at temperatures $1/(d-1)<T$. Note that using the same initial radial coordinates as in the $T<T_{\mathrm{c}}$ case would yield $\gamma=1+T/\beta$ above the critical temperature.

Nevertheless, by redefining $r_{\ell\ell}$ as
\begin{equation}
 r_{\ell\ell}=\left\lbrace \begin{array}{ll}
 \frac{2}{\zeta\cdot(d-1)}\cdot\ln{\ell} & \mbox{if}\; 0\leq T<\frac{1}{d-1}, \\ \\
 \frac{2T}{\zeta}\cdot\ln{\ell} & \mbox{if}\; \frac{1}{d-1}<T,
 \end{array} \right.
 \label{eq:radCoordsForDimIndependentGamma}
\end{equation}
we can obtain a degree decay exponent formula that coincides with the well-known expression $\gamma=1+1/\beta$ of the two-dimensional PSO model~\cite{PSO} for any number of dimensions, and we again recover the original radial coordinate formulas
\begin{equation}
 r_{\ell\ell}=\left\lbrace \begin{array}{ll}
 \frac{2}{\zeta}\cdot\ln{\ell} & \mbox{if}\; 0\leq T<\frac{1}{d-1}, \\ \\
 \frac{2T}{\zeta}\cdot\ln{\ell} & \mbox{if}\; \frac{1}{d-1}<T
 \end{array} \right.
 \label{eq:2dRadCoords}
\end{equation}
for $d=2$.  
However, Eq.~(\ref{eq:radCoordsForDimIndependentGamma}) does not provide the possibility to achieve a degree decay exponent below 2, since in this case the smallest possible value of $\gamma$ (obtained at $\beta=1$) is
\begin{equation}
\gamma_{\mathrm{min}}(d)=\min{\left(1+\frac{1}{\beta}\right)}=2,
\label{eq:gammaMin_dimIndependent}
\end{equation}
independently of the number of dimensions. In contrast, if the initial radial coordinates are defined by Eqs.~(\ref{eq:ourRadCoord_belowTc}) and (\ref{eq:ourRadCoord_aboveTc}), then by increasing the number $d$ of dimensions, the lower limit of the degree decay exponent can be decreased as
\begin{equation}
\gamma_{\mathrm{min}}(d)=\min{\left(1+\frac{1}{(d-1)\cdot\beta}\right)}=1+\frac{1}{(d-1)}.
\label{eq:gammaMin_dimDependent}
\end{equation}

With respect to the clustering and the community structure, the behaviour of the above-highlighted two variations of the $d$PSO model (differing in the initial radial coordinates given by either Eqs.~(\ref{eq:ourRadCoord_belowTc}) and (\ref{eq:ourRadCoord_aboveTc}) or Eq.~(\ref{eq:radCoordsForDimIndependentGamma})) is the same, since for given values of $\zeta$, $d$, $N$, $m$, $\gamma$ and $T$, all the connection probabilities emerging during the network growth are equal. Looking at the approximating formula of the hyperbolic distance~\cite{hyperGeomBasics} written up for a new node $j$ and a previously appeared node $i$ as
\begin{equation}
    x_{ij}(j)\approx r_{ij}+r_{jj}+\frac{2}{\zeta}\cdot\ln\left(\sin{\left(\frac{\theta_{ij}}{2}\right)}\right) = \beta\cdot r_{ii}+(1-\beta)\cdot r_{jj}+r_{jj}+\frac{2}{\zeta}\cdot\ln\left(\sin{\left(\frac{\theta_{ij}}{2}\right)}\right),
    \label{eq:hypDistApprox_reasoning}
\end{equation}
one can observe that the differences in the attractiveness of the already existing nodes arise from the term $\beta\cdot r_{ii}$ and from the angular term. However, in a network characterised by a degree decay exponent $\gamma$, the term $\beta\cdot r_{ii}$ is equal to $\frac{2}{\zeta\cdot(d-1)\cdot(\gamma-1)}\cdot\ln{i}$ for both model variants. Thus, the difference in the
initial radial coordinate formulas appears only in the term $(2-\beta)\cdot r_{jj}$ that is independent of $i$. 
And since the cutoff distance $R_j$ is always set to that value at which the expected number of connections will be $m$, together with the equal change of the (radial) distances between the new node and all of its possible neighbours, the connection probability function
\begin{equation}
    p(x_{ij})=\frac{1}{1+e^{\frac{\zeta(x_{ij}-R_j)}{2T}}}
    \label{eq:connProbWithHypDist}
\end{equation}
becomes shifted as well, 
leading to the emergence of a connection between node $j$ and any previous node $i$ eventually with the same probability in the case of using Eq.~(\ref{eq:radCoordsForDimIndependentGamma}) as in the case of defining the initial radial coordinates by Eqs.~(\ref{eq:ourRadCoord_belowTc}) and (\ref{eq:ourRadCoord_aboveTc}). 

According to the above, in the $2\leq\gamma$ regime 
these two model variants are equivalent. Consequently, although both definition of the initial radial coordinates are applicable, because of the wider range of achievable degree decay exponents, we stick to using Eqs.~(\ref{eq:ourRadCoord_belowTc}) and (\ref{eq:ourRadCoord_aboveTc}).

Let us now take a look at the effects of using the most general choice of the initial radial coordinates $r_{\ell\ell}=f\cdot\ln{\ell}$ in the two-dimensional case, which we will refer to as the $f$PSO model. 
According to Eq.~(\ref{eq:gammaFormulaGeneral}), the initial radial coordinates $r_{\ell\ell}=f\cdot\ln{\ell}$ with $2/(\zeta\cdot(d-1))<f\cdot\beta$ for $T<1/(d-1)$ and $2T/\zeta<f\cdot\beta$ for $1/(d-1)<T$ yield $\gamma<2$ even in the two-dimensional hyperbolic space (i.e., at $d=2$). Note that $\gamma$ is a decreasing function of the multiplying factor $f$, meaning that the $f$ factors that yield relatively small degree decay exponents are high enough to ensure that in the hyperbolic law of cosines
\begin{equation}
    \mathrm{cosh}(\zeta x_{ij}(j))=\mathrm{cosh}(\zeta r_{ij})\,\mathrm{cosh}(\zeta r_{jj})-\mathrm{sinh}(\zeta r_{ij})\,\mathrm{sinh}(\zeta r_{jj})\,\mathrm{cos}(\theta_{ij})
    \label{eq:hypLawOfCosines}
\end{equation}
the terms $\zeta r_{ij}$ and $\zeta r_{jj}$ are sufficiently large and the hyperbolic distance can be written for most of the node pairs as
\begin{equation}
    x_{ij}(j)\approx r_{ij}+r_{jj}+\frac{2}{\zeta}\cdot\ln\left(\sin{\left(\frac{\theta_{ij}}{2}\right)}\right),
    \label{eq:hypDistApprox_fPSO}
\end{equation}
which is an essential approximation in the derivation of the scale-free degree distribution. Nevertheless, to obtain higher values of $\gamma$, it is better to decrease only the popularity fading parameter $\beta$ and not the multiplying factor $f$, since otherwise the approximation in Eq.~(\ref{eq:hypDistApprox_fPSO}) becomes invalid due to the smallness of the radial coordinates.

To generate networks on the hyperbolic plane with initial radial coordinates $r_{\ell\ell}=f\cdot\ln{\ell}$, the cutoff distance $R_j$ has to be calculated for $T\neq 0$. According to Eqs.~(\ref{eq:P_ij_lowT_final}) and (\ref{eq:P_ij_highT_radCoord}), for $d=2$
\begin{equation}
 \hspace*{-0.8cm}
 P(i,j)\approx\left\lbrace \begin{array}{ll}
 \frac{2\cdot T}{(j-1)\cdot\sin(T\cdot\pi)}\cdot e^{-\frac{\zeta}{2}\cdot(r_{ij}+r_{jj}-R_j)} & \mbox{if}\; T<\frac{1}{d-1}, \\ \\
 \frac{1}{(j-1)\cdot\pi}\cdot \int_0^{\pi}\frac{1}{(\phi/2)^{1/T}} \,\mathrm{d}\phi \cdot e^{-\frac{\zeta}{2T}\cdot(r_{ij}+r_{jj}-R_j)}=\frac{2^{1/T}\cdot T}{(j-1)\cdot\pi^{1/T}\cdot(T-1)}\cdot e^{-\frac{\zeta}{2T}\cdot(r_{ij}+r_{jj}-R_j)} & \mbox{if}\; \frac{1}{d-1}<T.
 \end{array} \right.
 \label{eq:PijCases_withr}
\end{equation}
Note that here, in accordance with Ref.~\cite{PSO}, $\sin{(\phi/2)}$ was approximated with $\phi/2$ above the critical temperature $T_{\mathrm{c}}=1/(d-1)$ too. Using the initial radial coordinate formula $r_{\ell\ell}=f\cdot\ln{\ell}$ and that $r_{ij}=\beta\cdot r_{ii}+(1-\beta)\cdot r_{jj}$, we arrive at
\begin{equation}
 P(i,j)\approx\left\lbrace \begin{array}{ll}
 \frac{2\cdot T}{(j-1)\cdot\sin(T\cdot\pi)}\cdot i^{-\frac{\zeta\cdot f\cdot\beta}{2}}\cdot j^{-\frac{\zeta\cdot f\cdot(2-\beta)}{2}} \cdot e^{\frac{\zeta}{2}\cdot R_j} & \mbox{if}\; T<\frac{1}{d-1}, \\ \\
 \frac{2^{1/T}\cdot T}{(j-1)\cdot\pi^{1/T}\cdot(T-1)}\cdot i^{-\frac{\zeta\cdot f\cdot\beta}{2T}}\cdot j^{-\frac{\zeta\cdot f\cdot(2-\beta)}{2T}} \cdot e^{\frac{\zeta}{2T}\cdot R_j} & \mbox{if}\; \frac{1}{d-1}<T.
 \end{array} \right.
 \label{eq:PijCases_withID}
\end{equation}
The cutoff distance $R_j$ of the connection probability at the appearance of node $j$ can be expressed from the equation
\begin{equation}
 \hspace*{-2cm}
 \resizebox{1.3\hsize}{!}{$
 m=(j-1)\cdot\int_1^j P(i,j)\,\mathrm{d}i=\left\lbrace \begin{array}{ll}
 \frac{2\cdot T}{\sin(T\cdot\pi)}\cdot \frac{j^{1-\zeta\cdot f\cdot\beta/2}-1}{1-\zeta\cdot f\cdot\beta/2}\cdot j^{-\frac{\zeta\cdot f\cdot(2-\beta)}{2}} \cdot e^{\frac{\zeta}{2}\cdot R_j} = \frac{2\cdot T}{\sin(T\cdot\pi)}\cdot \frac{j^{1-\zeta\cdot f}-j^{\zeta\cdot f\cdot(\beta-2)/2}}{1-\zeta\cdot f\cdot\beta/2} \cdot e^{\frac{\zeta}{2}\cdot R_j} & \mbox{if}\; T<\frac{1}{d-1}, \\ \\
 \frac{2^{1/T}\cdot T}{\pi^{1/T}\cdot(T-1)}\cdot
 \frac{j^{1-\zeta\cdot f\cdot\beta/(2T)}-1}{1-\zeta\cdot f\cdot\beta/(2T)}\cdot j^{-\frac{\zeta\cdot f\cdot(2-\beta)}{2T}} \cdot e^{\frac{\zeta}{2T}\cdot R_j} = \frac{2^{1/T}\cdot T}{\pi^{1/T}\cdot(T-1)}\cdot
 \frac{j^{1-\zeta\cdot f/T}-j^{\zeta\cdot f\cdot(\beta-2)/(2T)}}{1-\zeta\cdot f\cdot\beta/(2T)} \cdot e^{\frac{\zeta}{2T}\cdot R_j} & \mbox{if}\; \frac{1}{d-1}<T
 \end{array} \right.
 $}
 \label{eq:cutoffEq}
\end{equation}
as
\begin{equation}
 R_j=\left\lbrace \begin{array}{ll}
 \frac{2}{\zeta}\cdot\ln{\left(\frac{m\cdot\sin{(T\cdot\pi)}\cdot(1-\zeta\cdot f\cdot\beta/2)}{2\cdot T\cdot\left(j^{1-\zeta\cdot f}-j^{\zeta\cdot f\cdot(\beta-2)/2}\right)}\right)} & \mbox{if}\; T<\frac{1}{d-1}, \\
 \frac{2T}{\zeta}\cdot\ln{\left(\frac{m\cdot\pi^{1/T}\cdot(T-1)\cdot(1-\zeta\cdot f\cdot\beta/(2T))}{2^{1/T}\cdot T\cdot\left(j^{1-\zeta\cdot f/T}-j^{\zeta\cdot f\cdot(\beta-2)/(2T)}\right)}\right)} & \mbox{if}\; \frac{1}{d-1}<T.
 \end{array} \right.
 \label{eq:cutoff_q}
\end{equation}
Our implementation of this two-dimensional $f$PSO model is available from Ref.~\cite{our_code}.

\begin{figure}[h!]
    \centering
    \captionsetup{width=1.0\textwidth}
    \includegraphics[width=1.0\textwidth]{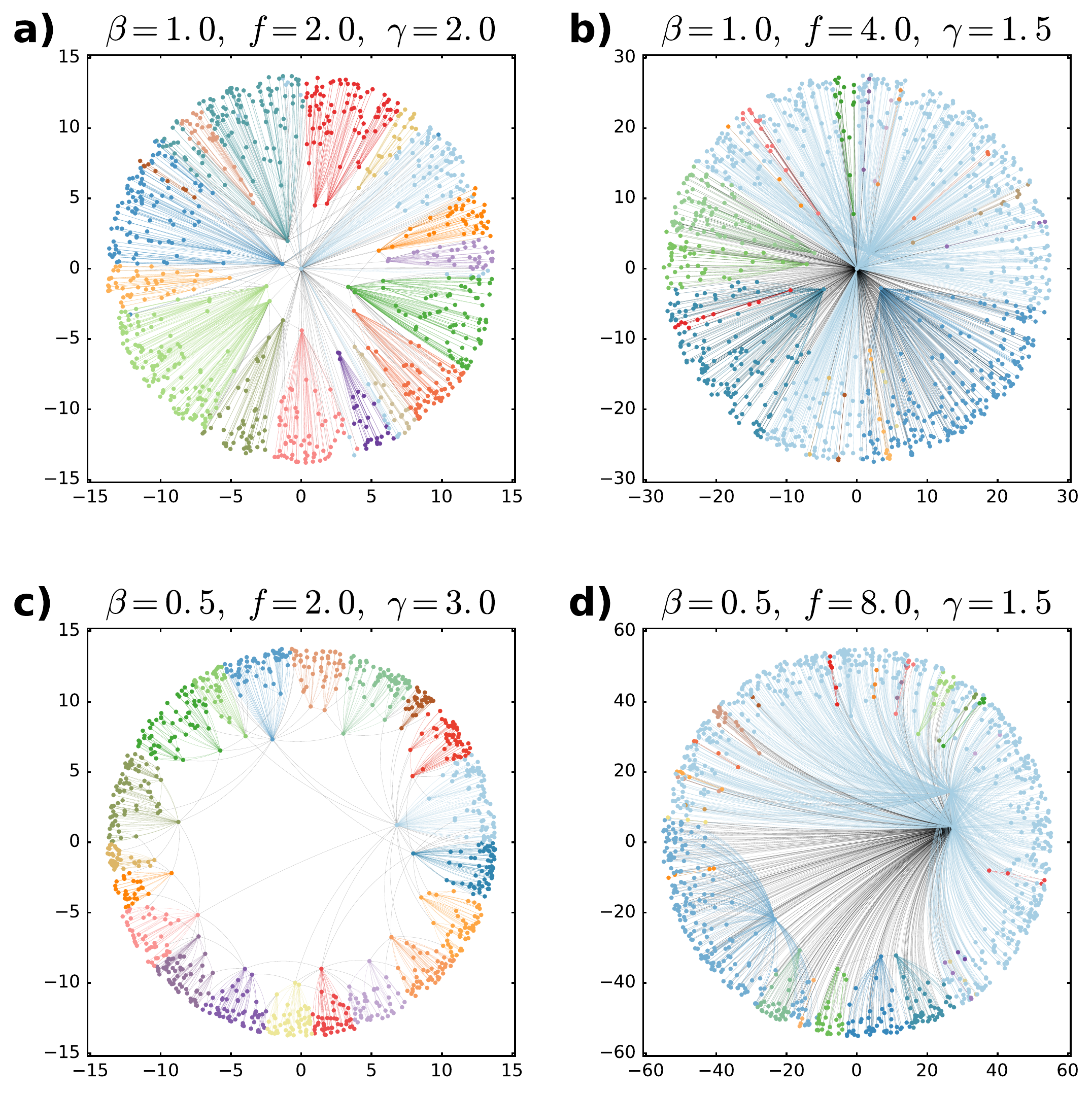}
    \caption{ {\bf Native layouts of networks generated by the \textit{f}PSO model on the hyperbolic plane.} The curvature of the hyperbolic plane, the number of nodes, the half of the expected average degree and the temperature were the same for all networks, namely $K=-\zeta^2=-1$, $N=1000$, $m=2$ and $T=0$. Since $T<1/(d-1)=1$, the degree decay exponent $\gamma$ can be expressed with the popularity fading parameter $\beta$ and the multiplying factor $f$ of the initial radial coordinates as $\gamma=1+\frac{2}{f\cdot\zeta\cdot(d-1)\cdot\beta}$ (see Eq.~(\ref{eq:gammaFormulaGeneral})). Note that for the here-applied $d=2$ and $\zeta=1$ settings, $f=2$ gives back the original PSO model. The colouring of the nodes and the links indicates communities found by the Louvain algorithm.}
    \label{fig:layouts_fPSO}
\end{figure}

\begin{figure}[h!]
    \centering
    \captionsetup{width=1.0\textwidth}
    \includegraphics[width=1.0\textwidth]{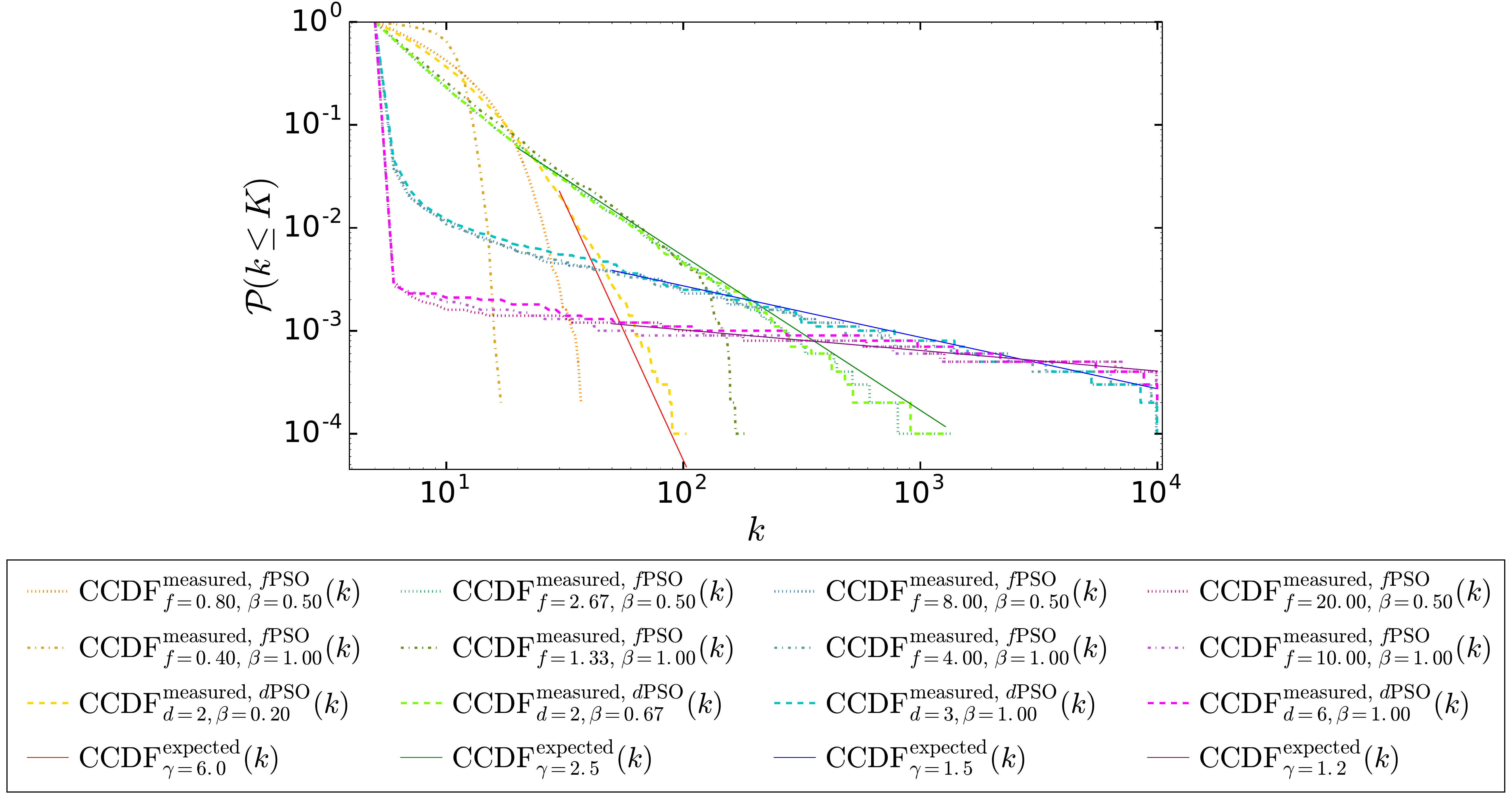}
    \caption{ {\bf Complementary cumulative distribution function (CCDF) of the node degrees for networks generated by the \textit{f}PSO model and the \textit{d}PSO model.} We examined two degree decay exponents that can not be obtained from the original PSO model: $\gamma=1.5$ and $\gamma=1.2$. Both of them are achievable with several $\beta-f$ settings in the two-dimensional $f$PSO model (where $\gamma=1+\frac{2}{f\cdot\zeta\cdot\beta}$ at $T<1$), just as with the $d$PSO model of high enough number of dimensions $d$ (where $\gamma=1+\frac{1}{(d-1)\cdot\beta}$ at $T<\frac{1}{d-1}$). However, with the decrease of the multiplying factor $f$ of the initial radial coordinates in the $f$PSO model, the approximation of the hyperbolic distance formula given by Eq.~(\ref{eq:hypDistApprox_fPSO}) becomes invalid. As a result, the $f$PSO model of $\beta=1$ already failed to generate the degree distribution characterised by the exponent $\gamma=2.5$, and both the $\beta=1$ and the $\beta=0.5$ settings of the $f$PSO model yielded considerable deviations from the expected curve of $\gamma=6$, while the degree distribution of the original PSO model (corresponding to the $d$PSO model with $d=2$ or the two-dimensional $f$PSO model with $f=2/\zeta$) obtained at $\beta=1/(\gamma-1)$ behaves as expected even for these higher values of $\gamma$. The curvature of the hyperbolic space, the number of nodes, the half of the expected average degree and the temperature were the same for all networks, namely $K=-\zeta^2=-1$ (using $\zeta=1$), $N=10,000$, $m=5$ and $T=0$. One network was generated with each parameter setting.}
    \label{fig:degreeDist_fPSO}
\end{figure}

\begin{figure}[h!]
    \centering
    \captionsetup{width=1.0\textwidth}
    \includegraphics[width=1.0\textwidth]{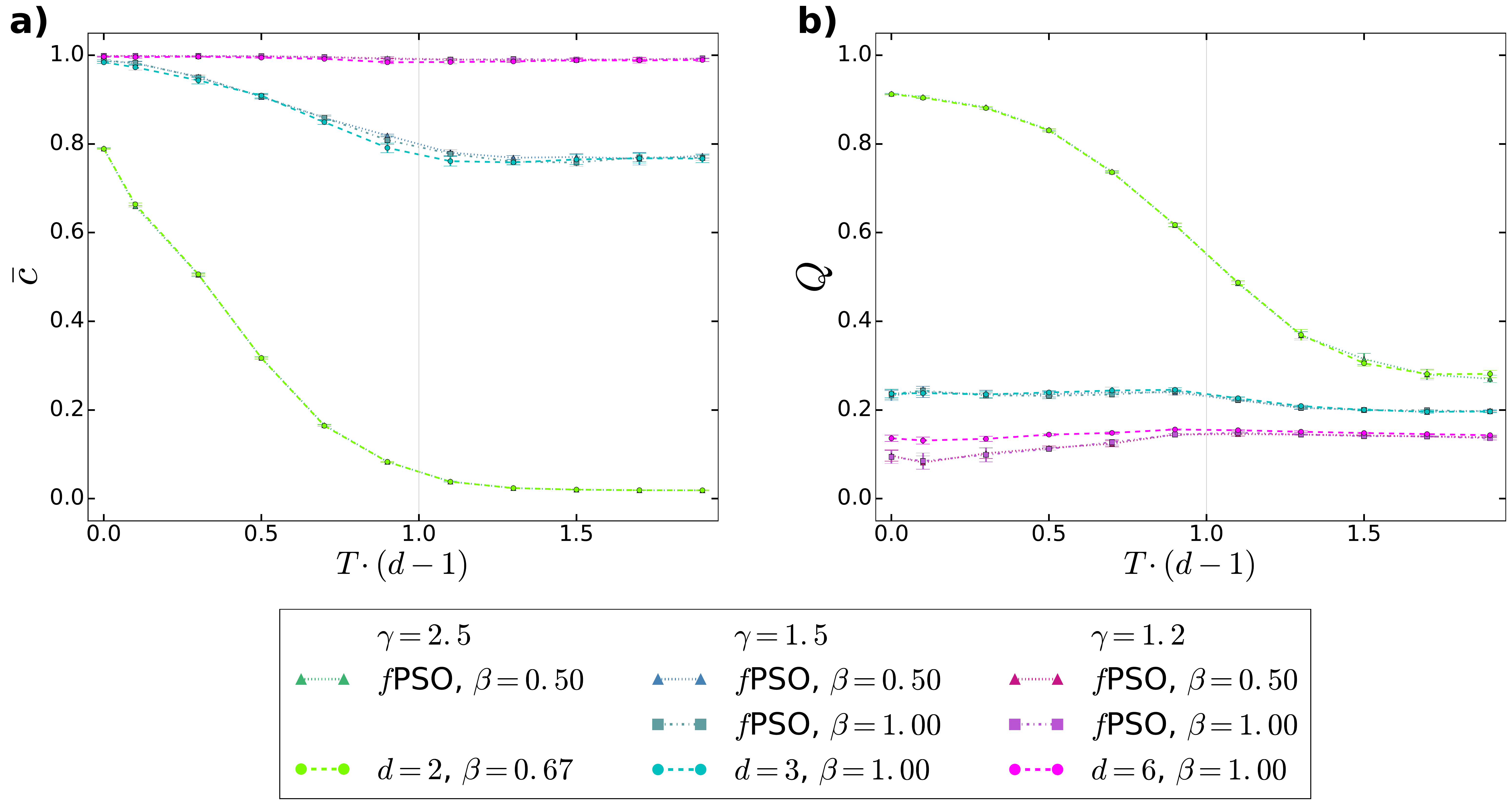}
    \caption{ {\bf Average clustering coefficient $\bar{c}$ and modularity $Q$ of the community structure detected by the Louvain algorithm in networks generated by the two-dimensional \textit{f}PSO model and the \textit{d}PSO model as a function of the rescaled temperature $T\cdot(d-1)$.} In order to create a fair comparison with the corresponding $d$PSO networks, for the $f$PSO model we show here only the results of such parameter settings that, according to Fig.~\ref{fig:degreeDist_fPSO}, actually yield the expected degree distribution, i.e. where the multiplying factor $f$ of the initial radial coordinates is not too small. Each displayed data point was obtained by averaging over 5 networks generated independently with a given set of model parameters, setting the curvature of the hyperbolic space to $-1$ (using $\zeta=1$), the number of nodes to $10,000$ and the half of the expected average degree to $5$ in each case. The error bars show the standard deviations measured among the 5 networks. The grey vertical lines indicate the critical point $T_{\mathrm{c}}=1/(d-1)$.}
    \label{fig:cQvsT_fPSO}
\end{figure}

\begin{figure}[h!]
    \centering
    \captionsetup{width=1.0\textwidth}
    \includegraphics[width=1.0\textwidth]{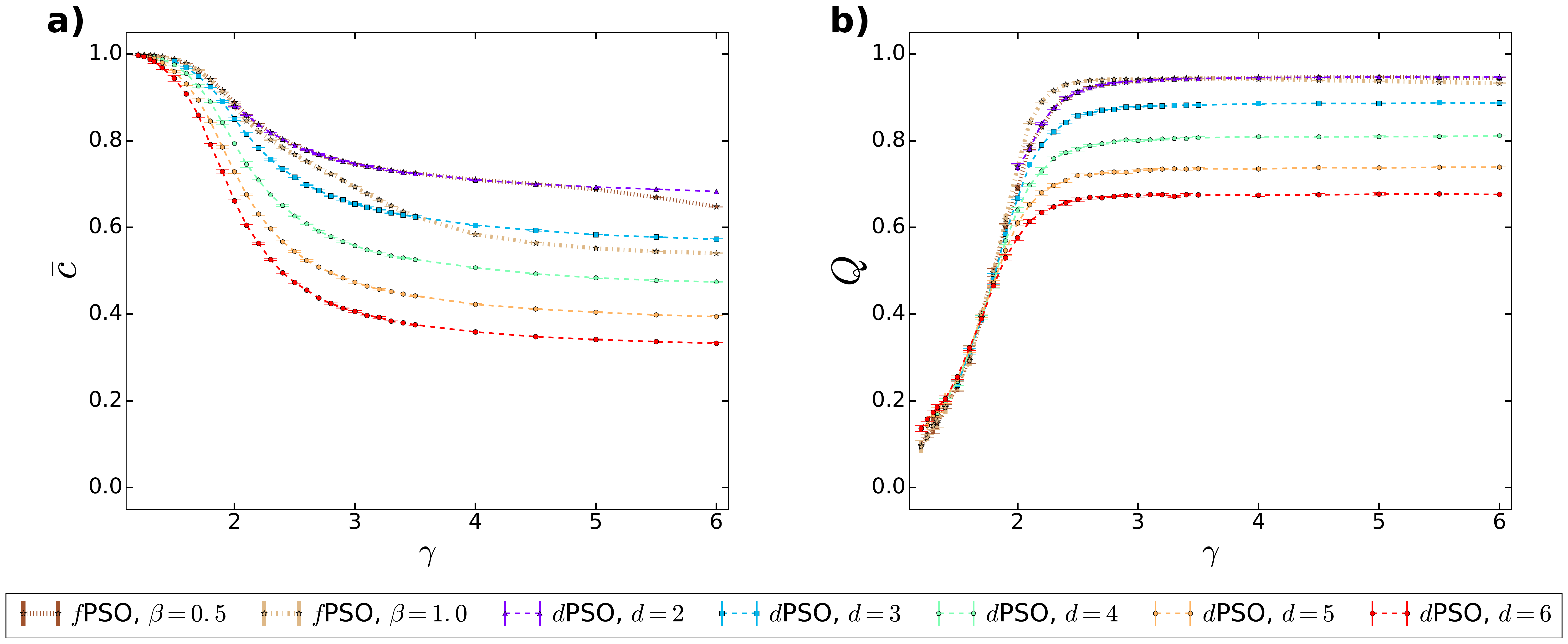}
    \caption{ {\bf Average clustering coefficient $\bar{c}$ and modularity $Q$ of the community structure detected by the Louvain algorithm in networks generated by the two-dimensional \textit{f}PSO model and the \textit{d}PSO model as a function of the degree decay exponent $\gamma$.} The depicted data points correspond to the values averaged over 5 networks generated with the same parameter settings, while the error bars show the standard deviations measured among the 5 networks. The curvature of the hyperbolic space, the number of nodes, the half of the expected average degree and the temperature were the same for all networks, namely $K=-\zeta^2=-1$, $N=10,000$, $m=5$ and $T=0$. Note that for the $f$PSO model of a given popularity fading parameter $\beta$, the increase in the expected degree decay exponent $\gamma$ corresponds to a decrease in the multiplying factor $f$ of the initial radial coordinates, which ruins the approximation in Eq.~(\ref{eq:hypDistApprox_fPSO}) and hereby increases the deviation between the degree distribution of the $f$PSO networks and the $d$PSO networks of the same expected degree decay exponent.}
    \label{fig:cQvsGamma_fPSO}
\end{figure}

First, in Fig.~\ref{fig:layouts_fPSO}, we show examples for $f$PSO network layouts generated in the native representation of the hyperbolic plane. Then, Fig.~\ref{fig:degreeDist_fPSO} compares the behaviour of the $f$PSO model and the $d$PSO model with regard to the degree distribution. Finally, Figs.~\ref{fig:cQvsT_fPSO} and \ref{fig:cQvsGamma_fPSO} present the changes in the average clustering coefficient $\bar{c}$ and the modularity $Q$~\cite{Newman_modularity_original,modularity_code} of the community structure detected by the Louvain algorithm~\cite{Louvain,Louvain_code} in $f$PSO and $d$PSO networks as a function of the temperature $T$ and the degree decay exponent $\gamma$. According to these plots, in that parameter regime where the degree distribution of the $f$PSO networks is well-described by the theory built on the approximating formula of the hyperbolic distance given by Eq.~(\ref{eq:hypDistApprox_fPSO}), i.e. when the multiplying factor $f$ of the initial radial coordinates is large enough, there is no substantial difference 
between a two-dimensional $f$PSO network and a $d$PSO network of the same degree decay exponent and the possible 
lowest number of dimensions with regard to the strength of the clustering and the community structure. 
However, while it is easy to determine the lowest $d$ at which a given small degree decay exponent $\gamma$ is achievable ($d_{\mathrm{min}}=\ceil[\big]{1+\frac{1}{\beta_{\mathrm{max}}\cdot(\gamma-1)}}=\ceil[\big]{1+\frac{1}{\gamma-1}}$), it is rather burdensome to determine the exact limit of the factor $f$ that separates the regions in which the hyperbolic distance is well or poorly approximable in practice. Therefore, instead of using the two-dimensional $f$PSO model and adjusting the $\beta-f$ parameter pair, we prefer the $d$PSO model where $\gamma$ can be controlled via the $\beta-d$ setting. 

\subsubsection{Formulas of the radial coordinates and the popularity fading parameter in the case of $d$-dimensional embeddings}
\label{sect:embeddingVersions}
\setcounter{figure}{0}
\setcounter{table}{0}
\setcounter{equation}{0}
\renewcommand{\thefigure}{S1.3.1.\arabic{figure}}
\renewcommand{\thetable}{S1.3.1.\arabic{table}}
\renewcommand{\theequation}{S1.3.1.\arabic{equation}}

As it is described above, there is more than one possible choice regarding the multiplying factor of the initial radial coordinates in the extension of the two-dimensional PSO model to any integer number of dimensions $d\geq 2$. Although of the detailed two approaches given by
\begin{equation}
 r_{\ell\ell}=\left\lbrace \begin{array}{ll}
 \frac{2}{\zeta\cdot(d-1)}\cdot\ln{\ell} & \mbox{if}\; 0\leq T<\frac{1}{d-1}, \\ \\
 \frac{2T}{\zeta}\cdot\ln{\ell} & \mbox{if}\; \frac{1}{d-1}<T
 \end{array} \right.
 \label{eq:radCoordsForDimIndependentGamma_2}
\end{equation}
and
\begin{equation}
 r_{\ell\ell}=\left\lbrace \begin{array}{ll}
 \frac{2}{\zeta}\cdot\ln{\ell} & \mbox{if}\; 0\leq T<\frac{1}{d-1}, \\ \\
 \frac{2T(d-1)}{\zeta}\cdot\ln{\ell} & \mbox{if}\; \frac{1}{d-1}<T
 \end{array} \right.
 \label{eq:radCoordCases}
\end{equation}
only the latter can be used to obtain degree decay exponents below $2$, both of these approaches may be suitable for the generation, and thus also the hyperbolic embedding of networks with $2\leq\gamma$. The two model variants provide two different ways for the radial arrangement of a network with size $N$ and degree decay exponent $\gamma$ in a $d$-dimensional hyperbolic space of curvature $K=-\zeta^2$. In both cases, unless the average clustering coefficient of the network to be embedded is very close to $0$, it can be assumed that the temperature $T$ that corresponds to the network is smaller than the critical value $T_{\mathrm{c}}=1/(d-1)$, and therefore the radial coordinate formulas of the $0\leq T<1/(d-1)$ case can be used. Accordingly, if the popularity fading parameter $\beta$ is determined as 
\begin{equation}
    \beta=\frac{1}{\gamma-1},
    \label{eq:betaInEmbedding1}
\end{equation}
then the radial coordinate of the node having the $\ell$th ($\ell=1,2,...,N$) largest degree (with ties in the order of node degrees broken arbitrarily) can be formulated as 
\begin{equation}
    r_{\ell N}=\beta\cdot r_{\ell\ell}+(1-\beta)\cdot r_{NN}= \beta\cdot\frac{2}{\zeta\cdot(d-1)}\cdot\ln{\ell}+(1-\beta)\cdot\frac{2}{\zeta\cdot(d-1)}\cdot\ln{N},
    \label{eq:radCoordInEmbedding1}
\end{equation}
while if the formula 
\begin{equation}
    \beta=\frac{1}{(d-1)\cdot(\gamma-1)}
    \label{eq:betaInEmbedding2}
\end{equation}
is used, then the radial coordinate in question can be calculated as
\begin{equation}
    r_{\ell N}=\beta\cdot r_{\ell\ell}+(1-\beta)\cdot r_{NN}=\beta\cdot\frac{2}{\zeta}\cdot\ln{\ell}+(1-\beta)\cdot\frac{2}{\zeta}\cdot\ln{N}.
    \label{eq:radCoordInEmbedding2}
\end{equation}
Note that angular coordinates can be assigned to the network nodes independently from the radial positions using e.g. a method proposed in Ref.~\cite{linkWeights_coalescentEmbedding}.

\clearpage
\section{Cutoff distance of the connection probability in the $d$PSO model}
\label{sect:cutoffDistance}
\setcounter{figure}{0}
\setcounter{table}{0}
\setcounter{equation}{0}
\renewcommand{\thefigure}{S2.\arabic{figure}}
\renewcommand{\thetable}{S2.\arabic{table}}
\renewcommand{\theequation}{S2.\arabic{equation}}

The cutoff distance $R_j$ of the connection probability $p(x)=1/[1+e^{\zeta(x-R_j)/(2T)}]$ applied at time $j$ is set to the value ensuring that the expected number of nodes connecting to node $j$ at its arrival is equal to $m$. As derived in Sect.~\ref{sect:DetConnDegreeDist}, for $T=0$ the cutoff distance can be calculated as 
   \begin{equation}
    R_j = \frac{2}{\zeta\cdot(d-1)}\cdot \ln\left(\frac{(d-1)\cdot m}{\eta(d)\cdot2^{d-1}\cdot e^{-\frac{\zeta\cdot(d-1)}{2}\cdot(2-\beta)\cdot r_{jj}}\cdot \int_1^j e^{-\frac{\zeta\cdot(d-1)}{2}\cdot\beta\cdot r_{ii}}\,\mathrm{d}i}\right),
    \label{eq:detCutoffDist1}
    \end{equation}
or, using that $r_{\ell\ell}=(2/\zeta)\cdot\ln{\ell}$, as
    \begin{multline}
    R_j = \frac{2}{\zeta\cdot(d-1)}\cdot \ln\left(\frac{(d-1)\cdot m}{\eta(d)\cdot2^{d-1}\cdot j^{-(d-1)\cdot(2-\beta)}\cdot \int_1^j i^{-(d-1)\cdot\beta}\,\mathrm{d}i}\right) = \\
    =\frac{2}{\zeta\cdot(d-1)}\cdot \ln\left(\frac{(d-1)\cdot m}{\eta(d)\cdot2^{d-1}\cdot j^{-(d-1)\cdot(2-\beta)}\cdot \frac{j^{1-(d-1)\cdot\beta}-1}{1-(d-1)\cdot\beta}}\right) = \\
    =\frac{2}{\zeta\cdot(d-1)}\cdot \ln\left(\frac{(d-1)\cdot m\cdot(1-(d-1)\cdot\beta)}{\eta(d)\cdot2^{d-1}\cdot\left(j^{3-2\cdot d}-j^{(d-1)\cdot(\beta-2)}\right)}\right).
    \label{eq:detCutoffDist2}
    \end{multline}
Note that for $d=2$ this formula is the same as the result
    \begin{equation}
    R_j = r_{jj} - \frac{2}{\zeta}\cdot\ln\left(\frac{2\cdot \left(1-e^{-\frac{\zeta}{2}(1-\beta)\cdot r_{jj}}\right)}{\pi\cdot m\cdot(1-\beta)}\right)
    \label{eq:detCutoffDistTwoDim_originalForm}
    \end{equation}
of Ref.~\cite{PSO}, which can be written as 
    \begin{equation}
    R_j = \frac{2}{\zeta}\cdot\ln\left(\frac{\pi\cdot m\cdot(1-\beta)}{2\cdot\left(j^{-1}-j^{\beta-2}\right)}\right)
    \label{eq:detCutoffDistTwoDim}
    \end{equation}
after the substitution of $r_{jj}=(2/\zeta)\cdot\ln{j}$. 

In the case of $0<T$, the cutoff distance at time $j$ is defined by the equation
    \begin{equation}
    m = (j-1)\cdot P(j),
    \label{eq:probCutOffDist_simpleGeneralEq}
    \end{equation}
where $j-1$ is the number of already existing nodes at the appearance of node $j$ and $P(j)$ is the probability that node $j$ connects to any of the previously appeared nodes in a given link formation attempt. According to Sect.~\ref{sect:ProbConnDegreeDist},
\begin{equation}
 P(j) = \int_1^j P(i,j) \,\mathrm{d}i = \int_1^j \frac{\eta(d)}{j-1}\cdot\int_0^{\pi} \frac{\sin^{d-2}{\phi}}{1+\left(e^{\frac{\zeta}{2}\cdot(r_{ij}+r_{jj}-R_j)}\cdot\sin{\left(\frac{\phi}{2}\right)}\right)^{\frac{1}{T}}} \,\mathrm{d}\phi\,\mathrm{d}i
 \label{eq:P_j_formula_summarised_exact}
\end{equation}
with
\begin{equation}
 \eta(d)=\left\lbrace \begin{array}{ll}
 \frac{\left(\frac{d}{2}-1\right)!\cdot \frac{d-2}{2}!\cdot 2^{d-2}}{(d-2)!\cdot\pi} & \mbox{if}\; d\mbox{ is even,} \\ \\
 \frac{(d-1)!}{\left(\frac{d-1}{2}-1\right)!\cdot\frac{d-1}{2}!\cdot 2^{d-1}} & \mbox{if}\; d\mbox{ is odd.}
 \end{array} \right.
 \label{eq:etaFormula3}
\end{equation}
Knowing also that the radial coordinate of node $\ell$ at time $j$ is
\begin{equation}
 \hspace*{-1.0cm}
 r_{\ell j}=\beta\cdot r_{\ell\ell}+(1-\beta)\cdot r_{jj}=\left\lbrace \begin{array}{ll}
 \beta\cdot\frac{2}{\zeta}\cdot\ln{\ell} + (1-\beta)\cdot\frac{2}{\zeta}\cdot\ln{j} & \mbox{if}\; 0\leq T<\frac{1}{d-1}, \\ \\
 \beta\cdot\frac{2T(d-1)}{\zeta}\cdot\ln{\ell} + (1-\beta)\cdot\frac{2T(d-1)}{\zeta}\cdot\ln{j} & \mbox{if}\; \frac{1}{d-1}<T,
 \end{array} \right.
 \label{eq:r_lj_formula_summarised}
\end{equation}
we arrive at the equation
    \begin{equation}
    m=
    \left\lbrace \begin{array}{ll}
    \eta(d)\cdot\int_1^j \int_0^\pi \frac{\sin^{d-2}{\phi}}{1+\sin^{\frac{1}{T}}{(\phi/2)}\cdot i^{\frac{\beta}{T}}\cdot j^{\frac{2-\beta}{T}}\cdot e^{-\frac{\zeta}{2T}\cdot R_j}} \,\mathrm{d}\phi\,\mathrm{d}i & \mbox{if}\; 0<T<\frac{1}{d-1}, \\ \\
    \eta(d)\cdot\int_1^j \int_0^\pi \frac{\sin^{d-2}{\phi}}{1+\sin^{\frac{1}{T}}{(\phi/2)}\cdot i^{\beta\cdot(d-1)}\cdot j^{(2-\beta)\cdot(d-1)}\cdot e^{-\frac{\zeta}{2T}\cdot R_j}} \,\mathrm{d}\phi\,\mathrm{d}i & \mbox{if}\; \frac{1}{d-1}<T,
    \end{array} \right.
    \label{eq:cutOffDistNumEq}
    \end{equation}
which can be solved numerically to obtain the cutoff distance $R_j$.

However, in the case of sufficiently large networks, for most of the nodes formula~\ref{eq:P_j_formula_summarised_exact} can be approximated as
\begin{equation}
 P(j)\approx\left\lbrace \begin{array}{ll}
 \frac{\eta(d)\cdot\pi \cdot 2^{d-1}\cdot T}{(j-1)\cdot\sin((d-1)\cdot T\cdot\pi)}\cdot \frac{j^{3-2\cdot d}- j^{(d-1)\cdot(\beta-2)}}{1-(d-1)\cdot\beta}\cdot e^{\frac{\zeta\cdot(d-1)}{2}\cdot R_j} & \mbox{if}\; 0<T<\frac{1}{d-1}, \\ \\
 \frac{\eta(d)}{j-1}\cdot \int_0^{\pi}\frac{\sin^{d-2}{\phi}}{\sin^{\frac{1}{T}}{(\phi/2)}} \,\mathrm{d}\phi\, \cdot \frac{j^{3-2\cdot d}-j^{(d-1)\cdot (\beta-2)}}{1-(d-1)\cdot\beta}\cdot e^{\frac{\zeta}{2\cdot T}\cdot R_j} & \mbox{if}\; \frac{1}{d-1}<T.
 \end{array} \right.
 \label{eq:P_j_formula_summarised_analytic}
\end{equation}
Substituting this in Eq.~(\ref{eq:probCutOffDist_simpleGeneralEq}), after some rearrangement of the terms one can write up the cutoff distance of the connection probability at the appearance of node $j$ as 
\begin{equation}
 R_j\approx\left\lbrace \begin{array}{ll}
 \frac{2}{\zeta\cdot(d-1)}\cdot\ln\left(\frac{m\cdot\sin((d-1)\cdot T\cdot\pi)\cdot(1-(d-1)\cdot\beta)}{\eta(d)\cdot\pi\cdot 2^{d-1}\cdot T\cdot\left(j^{3-2\cdot d}- j^{(d-1)\cdot(\beta-2)}\right)}\right) & \mbox{if}\; 0<T<\frac{1}{d-1}, \\ \\
 \frac{2\cdot T}{\zeta}\cdot \ln\left(\frac{m\cdot(1-(d-1)\cdot\beta)}{\eta(d)\cdot \int_0^{\pi}\sin^{d-2}(\phi)\cdot\sin^{-1/T}({\phi/2})\,\,\mathrm{d}\phi\,\cdot\left(j^{3-2\cdot d}- j^{(d-1)\cdot(\beta-2)}\right)}\right) & \mbox{if}\; \frac{1}{d-1}<T,
 \end{array} \right.
 \label{eq:R_j_probFormulas_approx}
\end{equation}
where $\int_0^{\pi}\sin^{d-2}(\phi)\cdot\sin^{-1/T}({\phi/2})\,\,\mathrm{d}\phi$ can be calculated numerically. Note that since $\lim_{T\to 0} \sin((d-1)\cdot T\cdot\pi)/(T\cdot\pi)=$ $=d-1$, for $T\rightarrow 0$ the approximated cutoff distance formula of the $0<T<1/(d-1)$ case becomes Eq.~(\ref{eq:detCutoffDist2}) as expected. 
For $d=2$, the approximating formula of the $0<T<1/(d-1)$ case gives back
    \begin{equation}
    R_j = r_{jj} - \frac{2}{\zeta}\cdot\ln\left(\frac{2\cdot T\cdot\left(1-e^{-\frac{\zeta}{2}(1-\beta)\cdot r_{jj}}\right)}{\sin(T\cdot\pi)\cdot m\cdot(1-\beta)}\right) = \frac{2}{\zeta}\cdot\ln\left(\frac{m\cdot\sin(T\cdot\pi)\cdot(1-\beta)}{2\cdot T\cdot\left(j^{-1}- j^{\beta-2}\right)}\right)
    \label{eq:probCutoffDistTwoDim_below1}
    \end{equation}
derived in Ref.~\cite{PSO} for $0<T<1$ using $r_{jj}=(2/\zeta)\cdot\ln{j}$. Furthermore, if one uses the approximation $\sin{(\phi/2)}\approx\phi/2$ also in the $1/(d-1)<T$ case as in Ref.~\cite{PSO}, then Eq.~(\ref{eq:R_j_probFormulas_approx}) yields for $d=2$ and $1<T$
    \begin{equation}
    R_j = \frac{2\cdot T}{\zeta}\cdot \ln\left(\frac{m\cdot(1-\beta)}{\frac{1}{\pi}\cdot 2^{1/T}\cdot \int_0^{\pi}\phi^{-1/T}\,\,\mathrm{d}\phi\,\cdot\left(j^{-1}- j^{\beta-2}\right)}\right),
    \label{eq:cutoffDistIndDim_Tabove1_d2}
    \end{equation}
which corresponds to
    \begin{equation}
    R_j = r_{jj}-\frac{2\cdot T}{\zeta}\cdot\ln\left(\left(\frac{2}{\pi}\right)^{\frac{1}{T}}\cdot\frac{T}{T-1}\cdot\frac{1-e^{-\frac{\zeta}{2\cdot T}\cdot(1-\beta)\cdot r_{jj}}}{m\cdot(1-\beta)}\right) = \frac{2\cdot T}{\zeta}\cdot\ln\left(\frac{\pi^{1/T}\cdot(T-1)\cdot m\cdot(1-\beta)}{2^{1/T}\cdot T\cdot(j^{-1}-j^{\beta-2})}\right)
    \label{eq:probCutoffDistTwoDim_above1}
    \end{equation}
derived in Ref.~\cite{PSO} for $1<T$ using $r_{jj}=(2T/\zeta)\cdot\ln{j}$. 

Finally, it is important to clarify how Eqs.~(\ref{eq:detCutoffDist2}) and (\ref{eq:R_j_probFormulas_approx}) behave in the $\beta\rightarrow1/(d-1)$ limit. As 
\begin{equation}
\lim_{\beta\to \frac{1}{d-1}} \frac{1-(d-1)\cdot\beta}{j^{3-2\cdot d}-j^{(d-1)\cdot(\beta-2)}} = \frac{j^{2\cdot d-3}}{\ln{j}},
\label{eq:betaLim}
\end{equation}
in the $\beta=1/(d-1)$ case the cutoff distance takes the form of
\begin{equation}
 R_j\approx\left\lbrace \begin{array}{ll}
 \frac{2}{\zeta\cdot(d-1)}\cdot \ln\left(\frac{(d-1)\cdot m\cdot j^{2\cdot d-3}}{\eta(d)\cdot2^{d-1}\cdot\ln{j}}\right) & \mbox{if}\; T=0, \\ \\
 \frac{2}{\zeta\cdot(d-1)}\cdot\ln\left(\frac{m\cdot\sin((d-1)\cdot T\cdot\pi)\cdot j^{2\cdot d-3}}{\eta(d)\cdot\pi\cdot 2^{d-1}\cdot T\cdot\ln{j}}\right) & \mbox{if}\; 0<T<\frac{1}{d-1}, \\ \\
 \frac{2\cdot T}{\zeta}\cdot \ln\left(\frac{m\cdot j^{2\cdot d-3}}{\eta(d)\cdot \int_0^{\pi}\sin^{d-2}(\phi)\cdot\sin^{-1/T}(\phi/2)\,\,\mathrm{d}\phi\,\cdot\ln{j}}\right) & \mbox{if}\; \frac{1}{d-1}<T.
 \end{array} \right.
 \label{eq:R_j_probFormulas_approx_beta1}
\end{equation}

\begin{figure}
    \centering
    \captionsetup{width=1.0\textwidth}
    \includegraphics[width=1.0\textwidth]{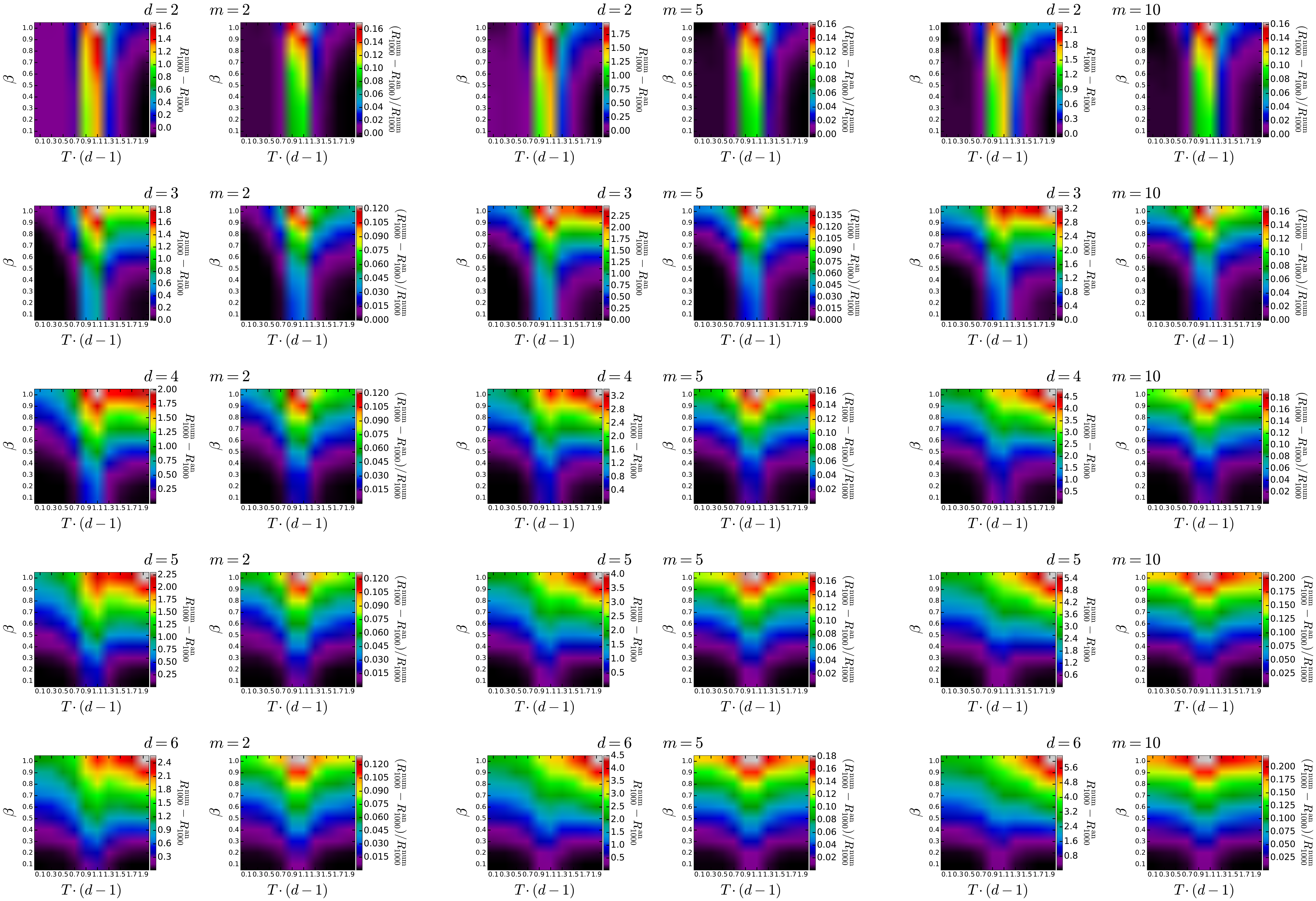}
    \caption{ {\bf Absolute and relative difference between the results of the two methods proposed for calculating the cutoff distance of the connection probability at $0<T$.} We calculated the cutoff distance for the node that arrives at the 1000th time step using Eq.~(\ref{eq:cutOffDistNumEq}), or Eqs.~(\ref{eq:R_j_probFormulas_approx}) and (\ref{eq:R_j_probFormulas_approx_beta1}). The result of the first approach is denoted by $R_{1000}^{\mathrm{num}}$, since it is based on the numerical solution of Eq.~(\ref{eq:cutOffDistNumEq}), while $R_{1000}^{\mathrm{an}}$ stands for the cutoff distance obtained from the latter approach that is, at least for $T<1/(d-1)$, an analytical calculation. Each pair of subplots depicts the effect of changing the popularity fading parameter $\beta$ and the rescaled temperature $T\cdot(d-1)$, with the number of dimensions $d$ and the half $m$ of the expected average degree $\bar{k}$ given in the title of the subplot pair. The curvature $K$ of the hyperbolic space was set to $-1$ in each case, i.e. we always used $\zeta=1$. 
    }
    \label{fig:numAnCutoffDistDiff}
\end{figure}

In our simulations, in order to reduce the computational time, we always calculated the cutoff distances at $0<T$ based on the approximating formulas given by Eqs.~(\ref{eq:R_j_probFormulas_approx}) and (\ref{eq:R_j_probFormulas_approx_beta1}) instead of solving numerically Eq.~(\ref{eq:cutOffDistNumEq}). According to Fig.~\ref{fig:numAnCutoffDistDiff}, the difference between the results of the approximating formulas and the numerical equation solution measured for the 1000th node is already acceptable at most of the examined parameter settings. Therefore, in the case of the studied networks of size $N=10,000$ we can assume for most of the network nodes that the approximated cutoff distance was close enough to the value that could have been obtained by solving numerically Eq.~(\ref{eq:cutOffDistNumEq}).

\clearpage
\section{Simulation results for $d$PSO networks}
\label{sect:simRes}
\setcounter{figure}{0}
\setcounter{table}{0}
\setcounter{equation}{0}
\renewcommand{\thefigure}{S3.\arabic{figure}}
\renewcommand{\thetable}{S3.\arabic{table}}
\renewcommand{\theequation}{S3.\arabic{equation}}

\begin{figure}[h!]
    \centering
    \captionsetup{width=1.0\textwidth}
    \includegraphics[width=1.0\textwidth]{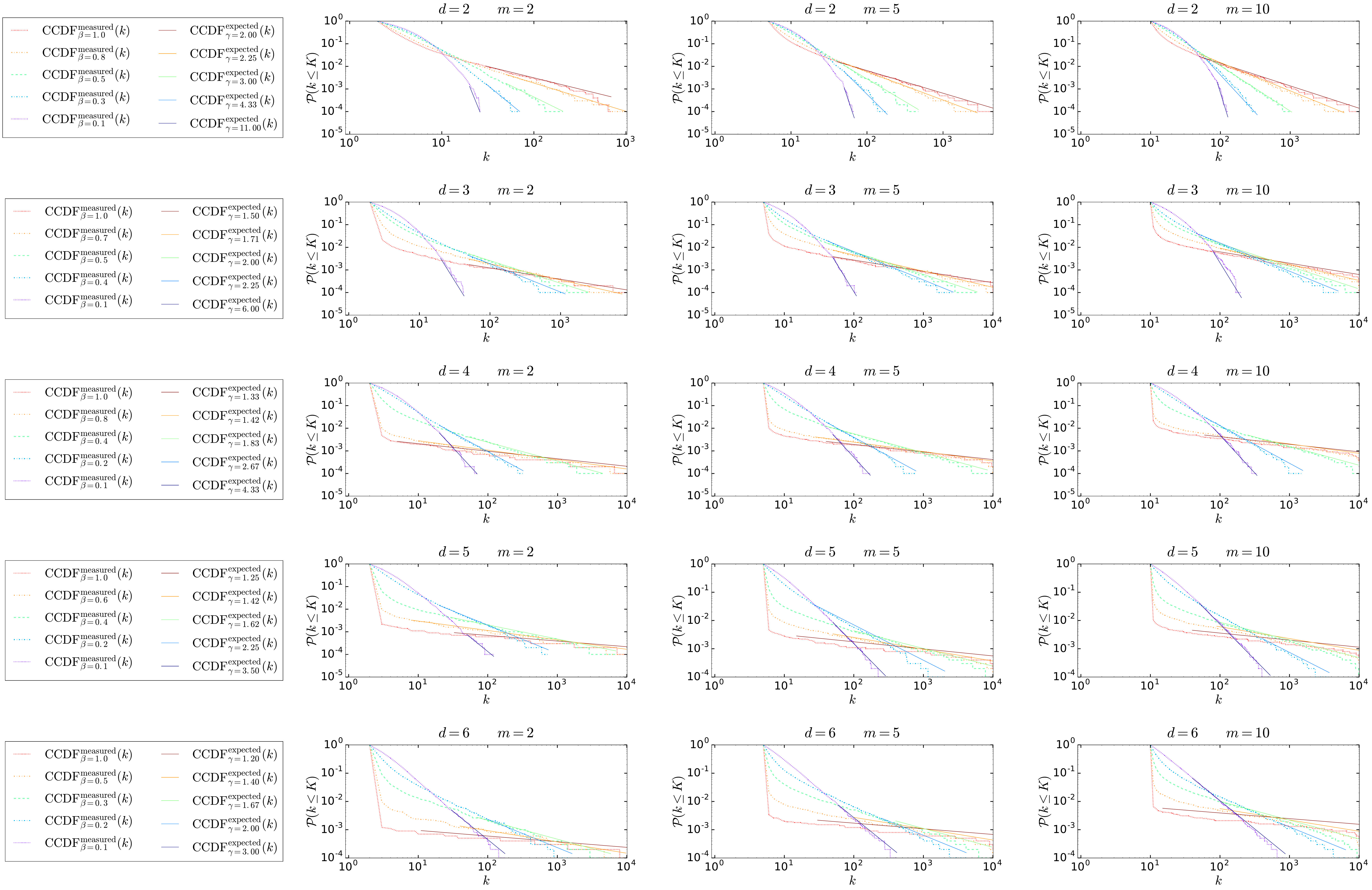}
    \caption{ {\bf Complementary cumulative distribution function (CCDF) of the node degrees for networks generated by the \textit{d}PSO model using different parametrisations.} Each row of panels was created using a given dimension $d$, while each column of subplots presents the results obtained with a given value of $m$, as indicated in the title of the subplots. The popularity fading parameters and the corresponding degree decay exponents tested at each value of $m$ for a given dimension $d$ are listed in the leftmost panel of each row. The curvature of the hyperbolic space, the number of nodes and the temperature were the same for all networks, namely $K=-\zeta^2=-1$, $N=10,000$ and $T=0$. One network was generated with each of the parameter sets.}
    \label{fig:degreeDist_supp}
\end{figure}

This section presents simulation results additional to the figures of the main text. First, we confirm by Fig.~\ref{fig:degreeDist_supp} that in networks generated by the \textit{d}PSO model of different expected average degree $2m$, the tail of the complementary cumulative distribution function (CCDF) of the node degrees follows a power-law written as ${\pazocal{P}(k\leq K) \sim k^{-(\gamma-1)}}$, where the degree decay exponent $\gamma$ can be expressed with the dimension $d$ of the hyperbolic space and the popularity fading parameter $\beta$ as $1+\frac{1}{(d-1)\cdot\beta}$. As it is demonstrated by Fig.~2 of the main text, the temperature $T$ does not have any substantial impact on the degree distribution; therefore, here we study only the $T=0$ case.

\begin{figure}[h!]
    \centering
    \captionsetup{width=1.0\textwidth}
    \includegraphics[width=1.0\textwidth]{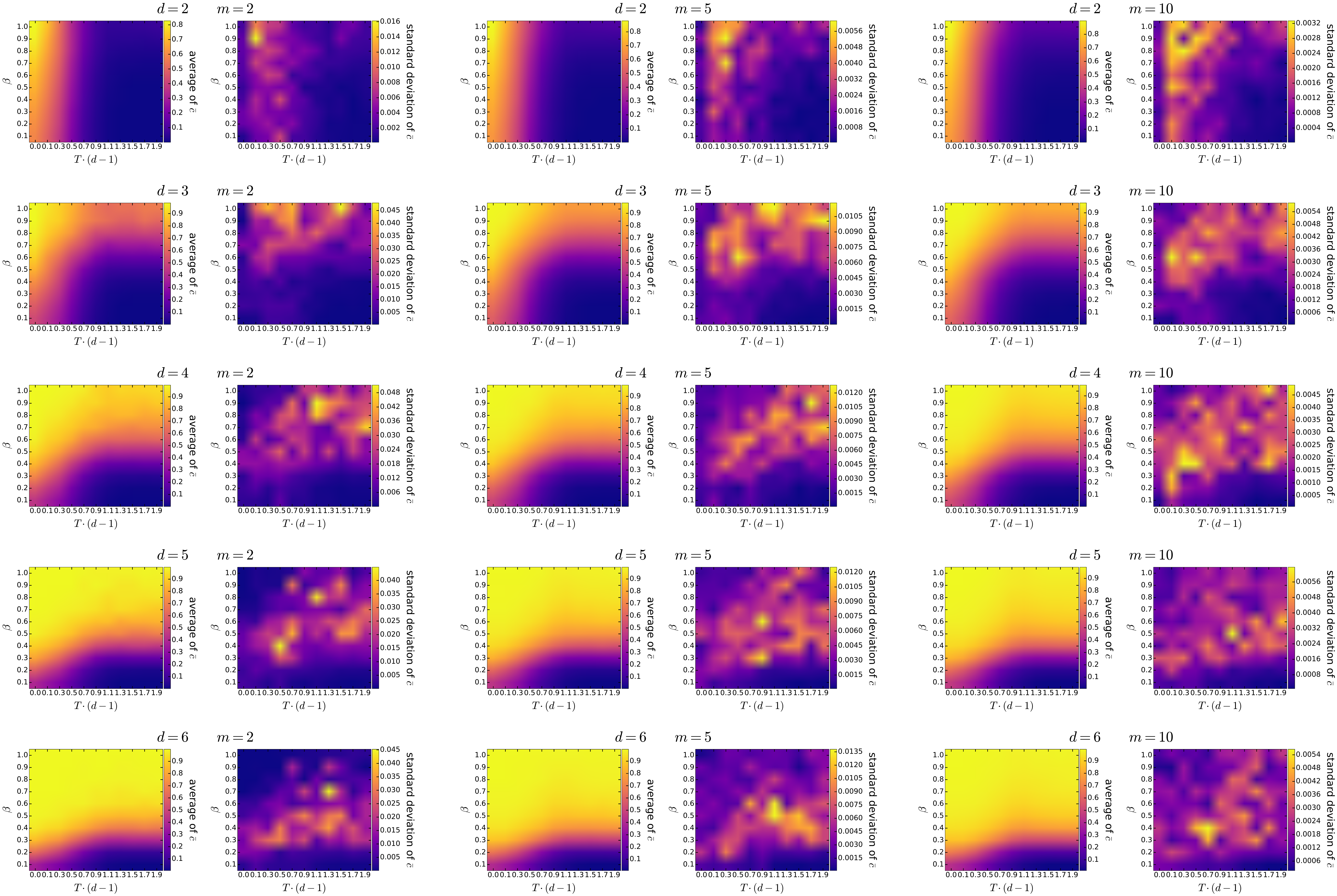}
    \caption{{\bf The mean and the standard deviation of the average clustering coefficient $\bar{c}$ measured in 5 \textit{d}PSO networks in the case of different parametrisations.} Each pair of subplots depicts the effect of changing the popularity fading parameter $\beta$ and the rescaled temperature $T\cdot(d-1)$, with the number of dimensions $d$ and the half $m$ of the expected average degree $\bar{k}$ given in the title of the subplot pair. The number of nodes was $N=10,000$ in each network. The curvature $K$ of the hyperbolic space was always set to $-1$, i.e. we used $\zeta=1$.}
    \label{fig:clustCoeff_colormap}
\end{figure}

\begin{figure}[h!]
    \centering
    \captionsetup{width=1.0\textwidth}
    \includegraphics[width=1.0\textwidth]{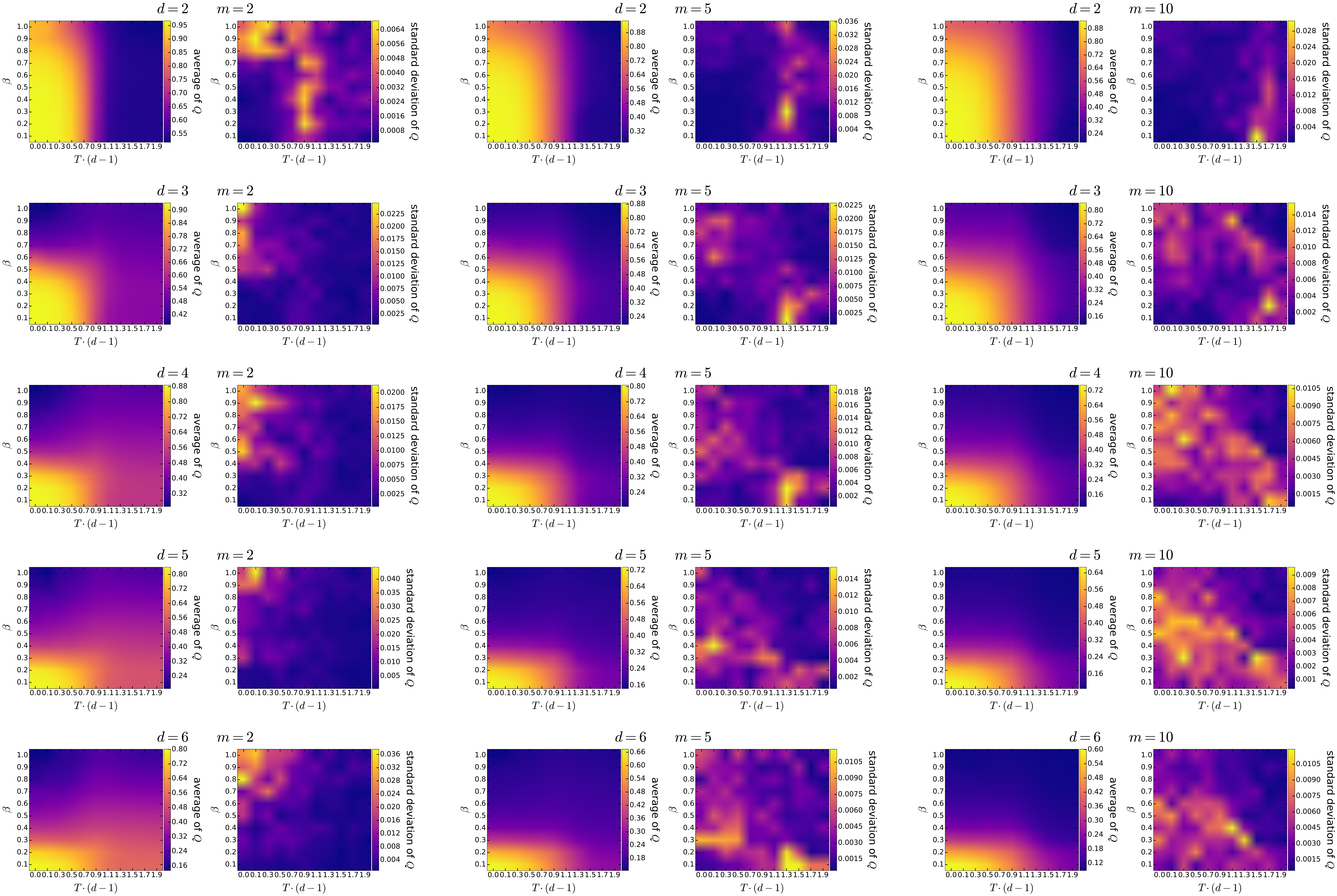}
    \caption{{\bf The mean and the standard deviation of the modularity $Q$ of the community structure detected by the \textit{Louvain} algorithm in 5 \textit{d}PSO networks in the case of different parametrisations.} Each pair of subplots depicts the effect of changing the popularity fading parameter $\beta$ and the rescaled temperature $T\cdot(d-1)$, with the number of dimensions $d$ and the half $m$ of the expected average degree $\bar{k}$ given in the title of the subplot pair. The number of nodes was $N=10,000$ in each network. The curvature $K$ of the hyperbolic space was always set to $-1$, i.e. we used $\zeta=1$.}
    \label{fig:modularity_Louvain_uw}
\end{figure}

\begin{figure}[h!]
    \centering
    \captionsetup{width=1.0\textwidth}
    \includegraphics[width=1.0\textwidth]{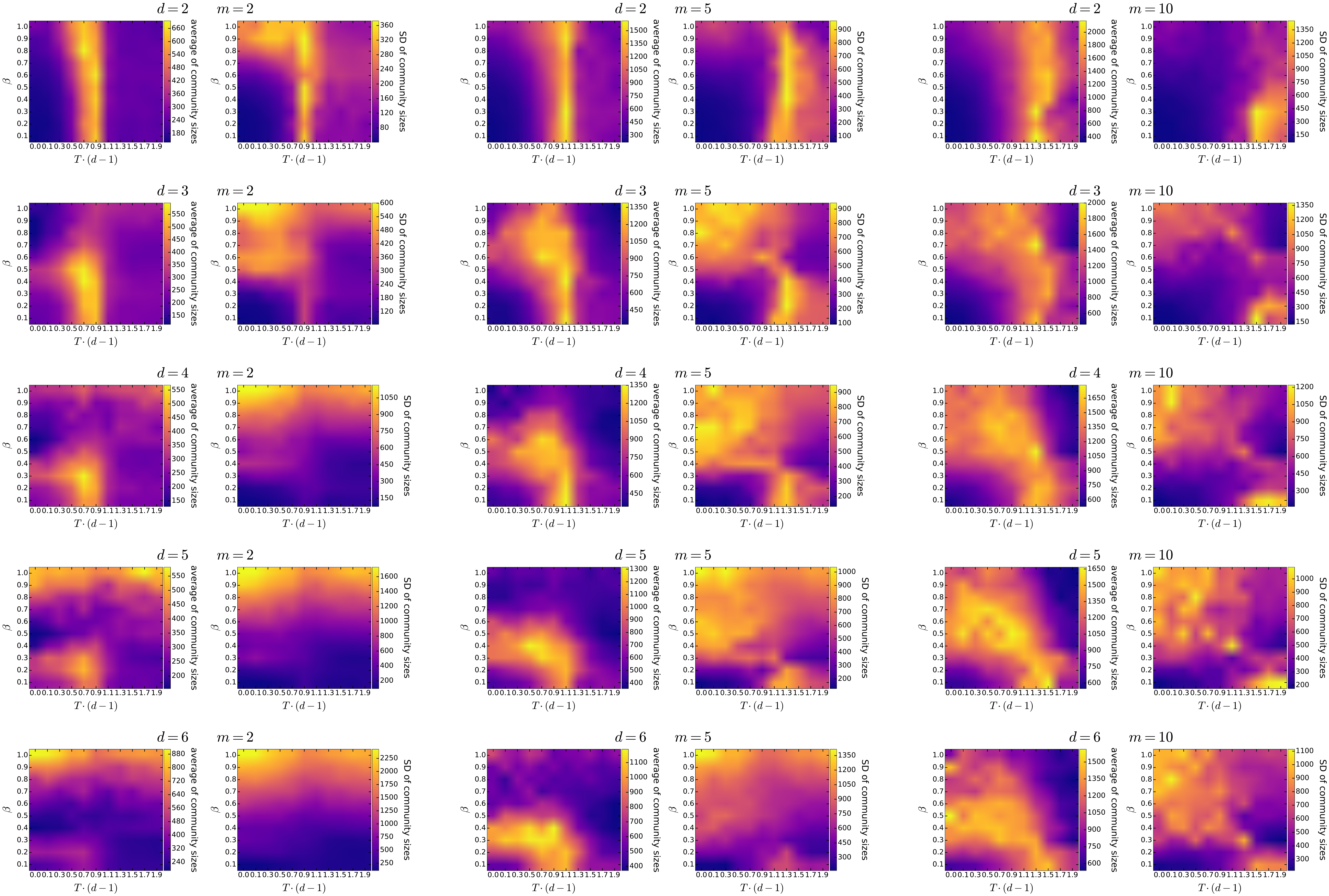}
    \caption{ {\bf The mean and the standard deviation of the size of the communities detected by the \textit{Louvain} algorithm in 5 \textit{d}PSO networks in the case of different parametrisations.} Each pair of subplots depicts the effect of changing the popularity fading parameter $\beta$ and the rescaled temperature $T\cdot(d-1)$, with the number of dimensions $d$ and the half $m$ of the expected average degree $\bar{k}$ given in the title of the subplot pair. The number of nodes was $N=10,000$ in each network. The curvature $K$ of the hyperbolic plane was always set to $-1$, i.e. we used $\zeta=1$.}
    \label{fig:groupSizes_Louvain_uw}
\end{figure}

\begin{figure}[h!]
    \centering
    \captionsetup{width=1.0\textwidth}
    \includegraphics[width=1.0\textwidth]{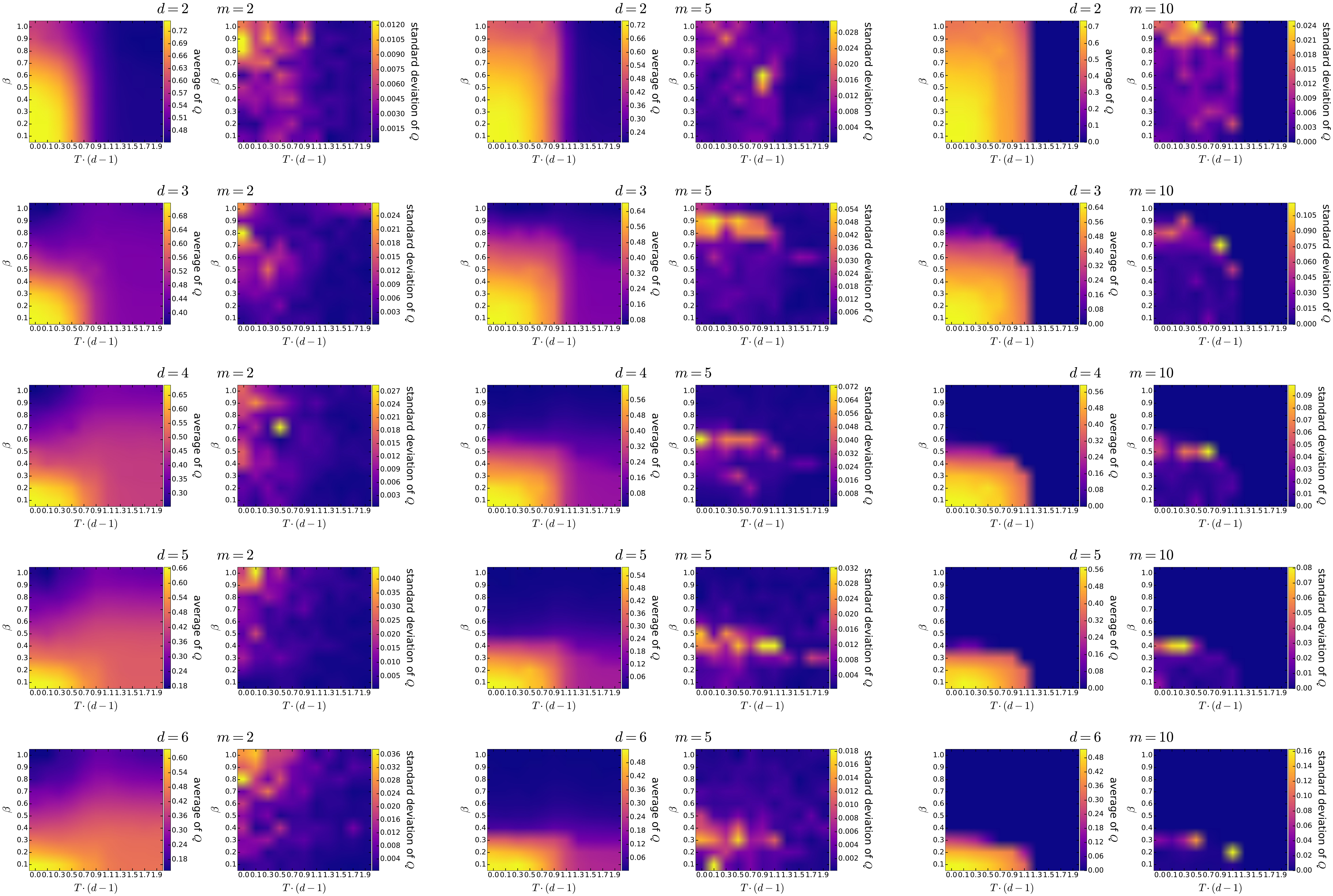}
    \caption{{\bf The mean and the standard deviation of the modularity $Q$ of the community structure detected by the \textit{Infomap} algorithm in 5 \textit{d}PSO networks in the case of different parametrisations.} Each pair of subplots depicts the effect of changing the popularity fading parameter $\beta$ and the rescaled temperature $T\cdot(d-1)$, with the number of dimensions $d$ and the half $m$ of the expected average degree $\bar{k}$ given in the title of the subplot pair. The number of nodes was $N=10,000$ in each network. The curvature $K$ of the hyperbolic space was always set to $-1$, i.e. we used $\zeta=1$.}
    \label{fig:modularity_Infomap_uw}
\end{figure}

\begin{figure}[h!]
    \centering
    \captionsetup{width=1.0\textwidth}
    \includegraphics[width=1.0\textwidth]{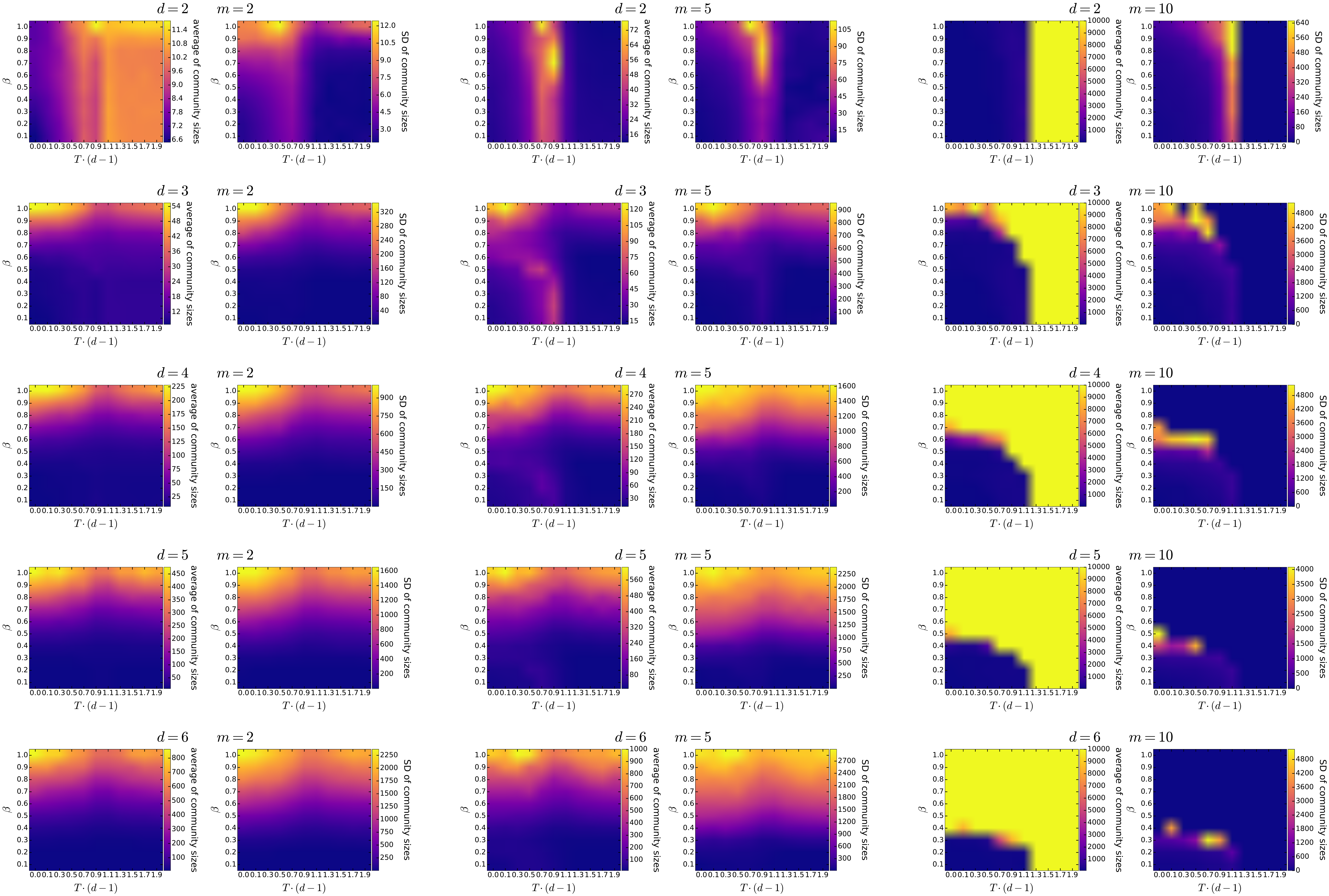}
    \caption{ {\bf The mean and the standard deviation of the size of the communities detected by the \textit{Infomap} algorithm in 5 \textit{d}PSO networks in the case of different parametrisations.} Each pair of subplots depicts the effect of changing the popularity fading parameter $\beta$ and the rescaled temperature $T\cdot(d-1)$, with the number of dimensions $d$ and the half $m$ of the expected average degree $\bar{k}$ given in the title of the subplot pair. The number of nodes was $N=10,000$ in each network. The curvature $K$ of the hyperbolic plane was always set to $-1$, i.e. we used $\zeta=1$.}
    \label{fig:groupSizes_Infomap_uw}
\end{figure}

\begin{figure}[h!]
    \centering
    \captionsetup{width=1.0\textwidth}
    \includegraphics[width=1.0\textwidth]{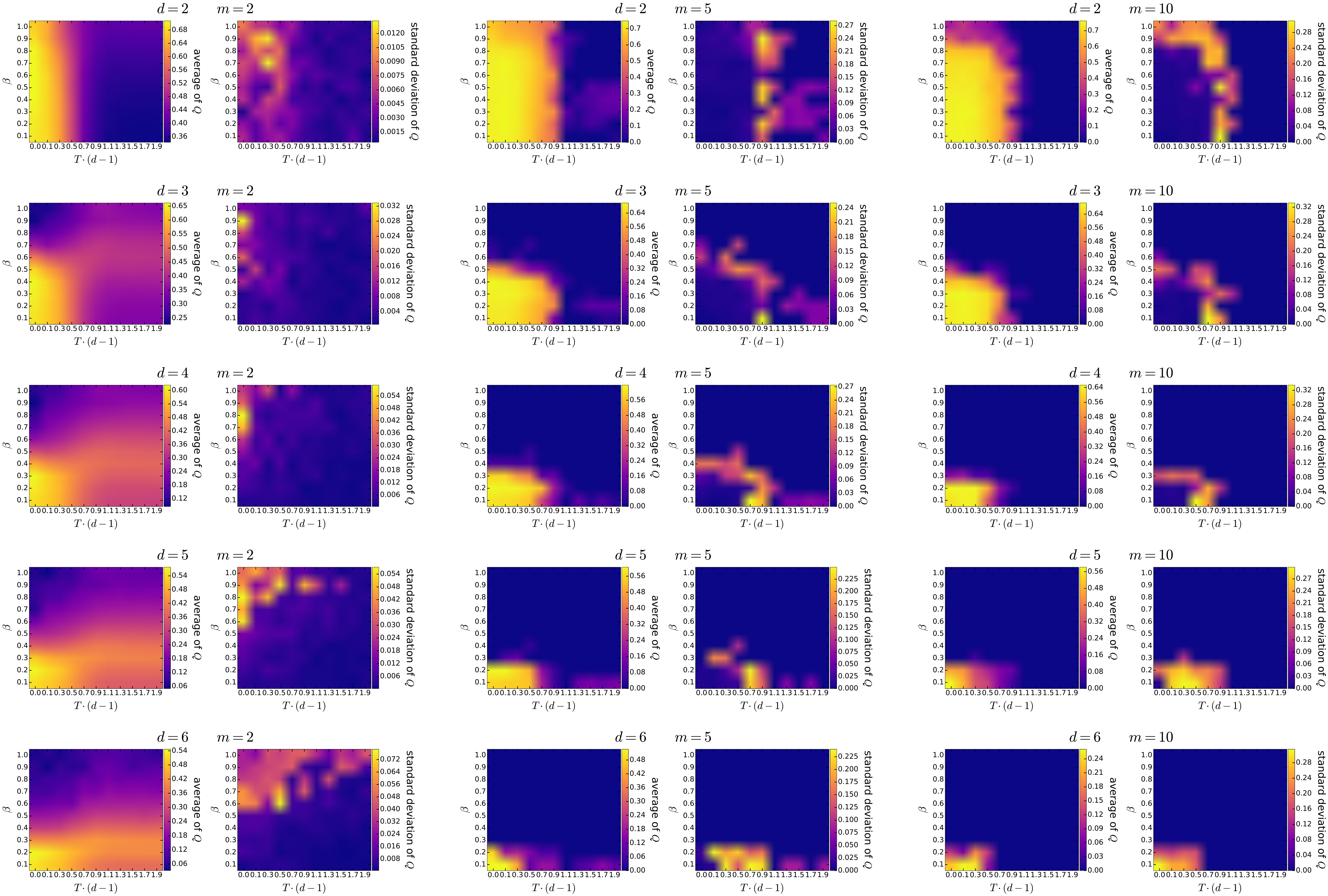}
    \caption{{\bf The mean and the standard deviation of the modularity $Q$ of the community structure detected by the \textit{asynchronous label propagation} algorithm in 5 \textit{d}PSO networks in the case of different parametrisations.} Each pair of subplots depicts the effect of changing the popularity fading parameter $\beta$ and the rescaled temperature $T\cdot(d-1)$, with the number of dimensions $d$ and the half $m$ of the expected average degree $\bar{k}$ given in the title of the subplot pair. The number of nodes was $N=10,000$ in each network. The curvature $K$ of the hyperbolic space was always set to $-1$, i.e. we used $\zeta=1$.}
    \label{fig:modularity_alabprop_uw}
\end{figure}

\begin{figure}[h!]
    \centering
    \captionsetup{width=1.0\textwidth}
    \includegraphics[width=1.0\textwidth]{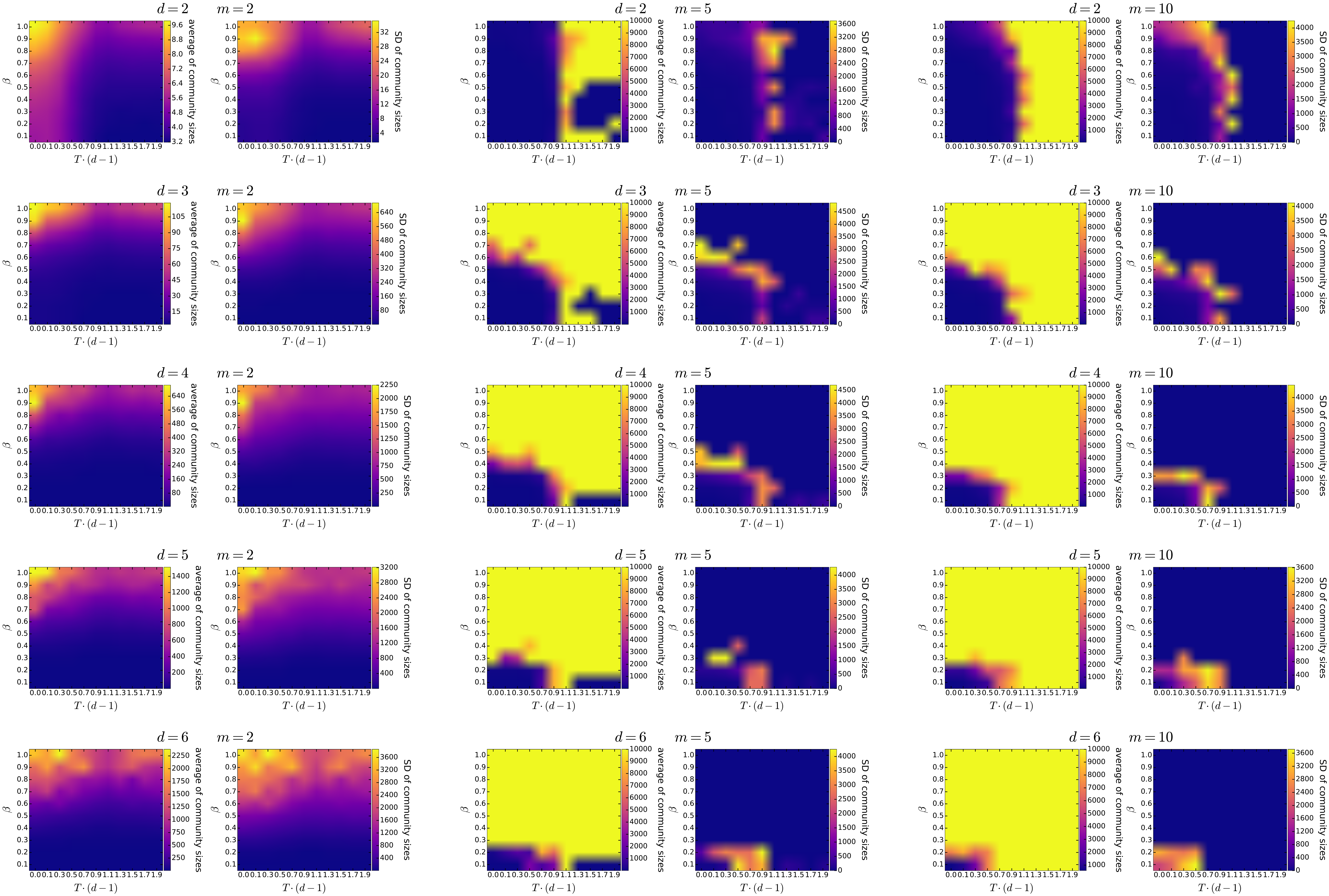}
    \caption{ {\bf The mean and the standard deviation of the size of the communities detected by the \textit{asynchronous label propagation} algorithm in 5 \textit{d}PSO networks in the case of different parametrisations.} Each pair of subplots depicts the effect of changing the popularity fading parameter $\beta$ and the rescaled temperature $T\cdot(d-1)$, with the number of dimensions $d$ and the half $m$ of the expected average degree $\bar{k}$ given in the title of the subplot pair. The number of nodes was $N=10,000$ in each network. The curvature $K$ of the hyperbolic plane was always set to $-1$, i.e. we used $\zeta=1$.}
    \label{fig:groupSizes_alabprop_uw}
\end{figure}

\clearpage
Next, to supplement Fig.~3 of the main text, in Fig.~\ref{fig:clustCoeff_colormap} we depict on a $10\times10$ grid in the $T-\beta$ parameter plane the average clustering coefficient $\bar{c}$ measured in \textit{d}PSO networks of different number of dimensions $d$ and expected average degree $2m$. Similarly, we add more detail to Fig.~5 of the main text by Figs.~\ref{fig:modularity_Louvain_uw}, \ref{fig:modularity_Infomap_uw} and \ref{fig:modularity_alabprop_uw}, where we present for several different parameter settings the modularity $Q$ (described in Sect.~\ref{sect:comms} of the main text) achieved by each of the examined community detection algorithms, namely Louvain~\cite{Louvain,Louvain_code}, Infomap~\cite{Infomap,Infomap_code} and asynchronous label propagation~\cite{alabprop,alabprop_code}. In Figs.~\ref{fig:groupSizes_Louvain_uw}, \ref{fig:groupSizes_Infomap_uw} and \ref{fig:groupSizes_alabprop_uw} we also plot the average and the standard deviation of the size of the communities found by the given algorithms.

\begin{figure}[h!]
    \centering
    \captionsetup{width=1.0\textwidth}
    \includegraphics[width=1.0\textwidth]{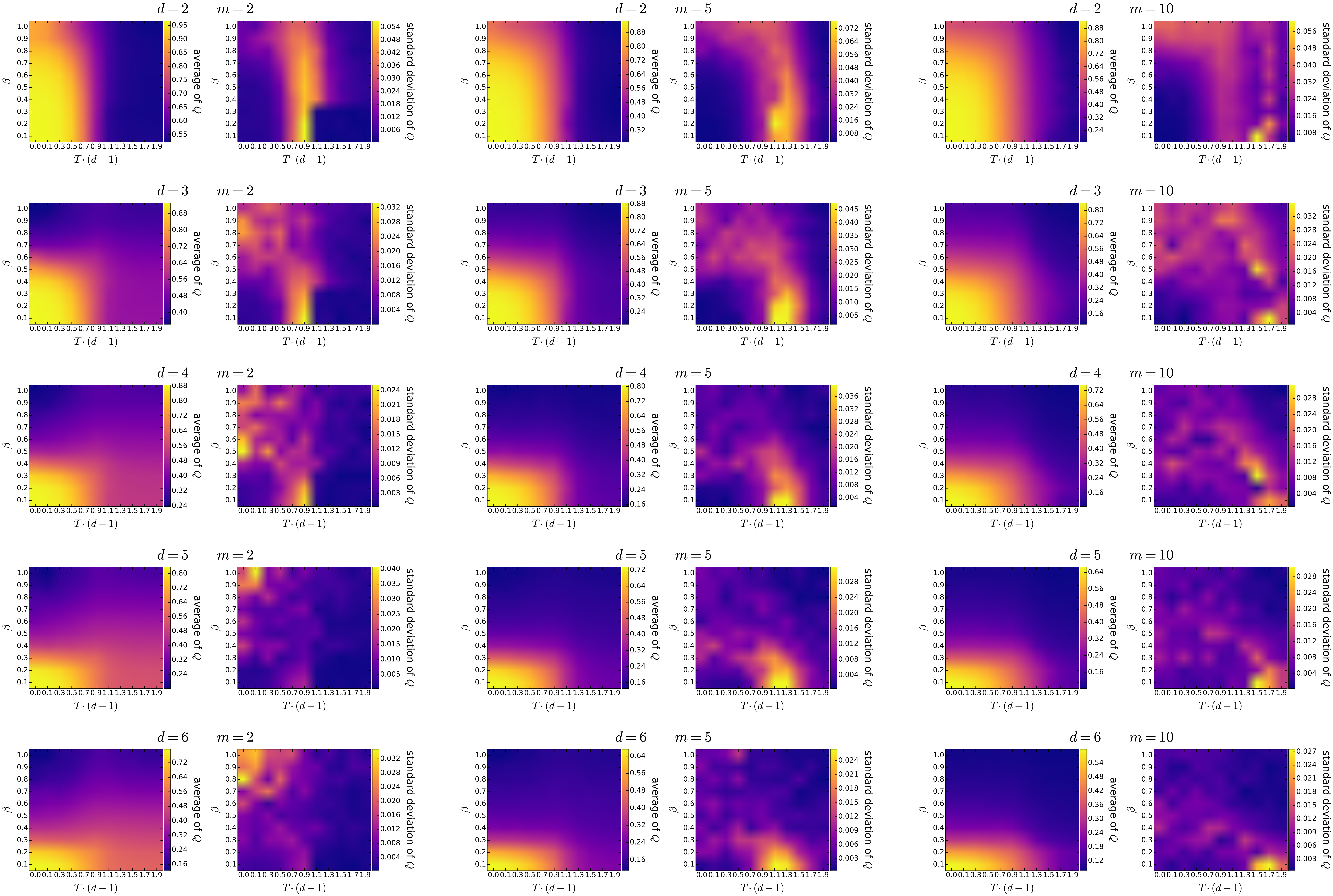}
    \caption{{\bf The mean and the standard deviation of the \textit{weighted} modularity $Q$ of the community structure detected by the \textit{Louvain} algorithm in 5 \textit{d}PSO networks in the case of different parametrisations, using link weights calculated from the hyperbolic distances between the connected nodes according to Eq.~(\ref{eq:linkWeights}).} Each pair of subplots depicts the effect of changing the popularity fading parameter $\beta$ and the rescaled temperature $T\cdot(d-1)$, with the number of dimensions $d$ and the half $m$ of the expected average degree $\bar{k}$ given in the title of the subplot pair. The number of nodes was $N=10,000$ in each network. The curvature $K$ of the hyperbolic space was always set to $-1$, i.e. we used $\zeta=1$.}
    \label{fig:modularity_Louvain_w}
\end{figure}

\begin{figure}[h!]
    \centering
    \captionsetup{width=1.0\textwidth}
    \includegraphics[width=1.0\textwidth]{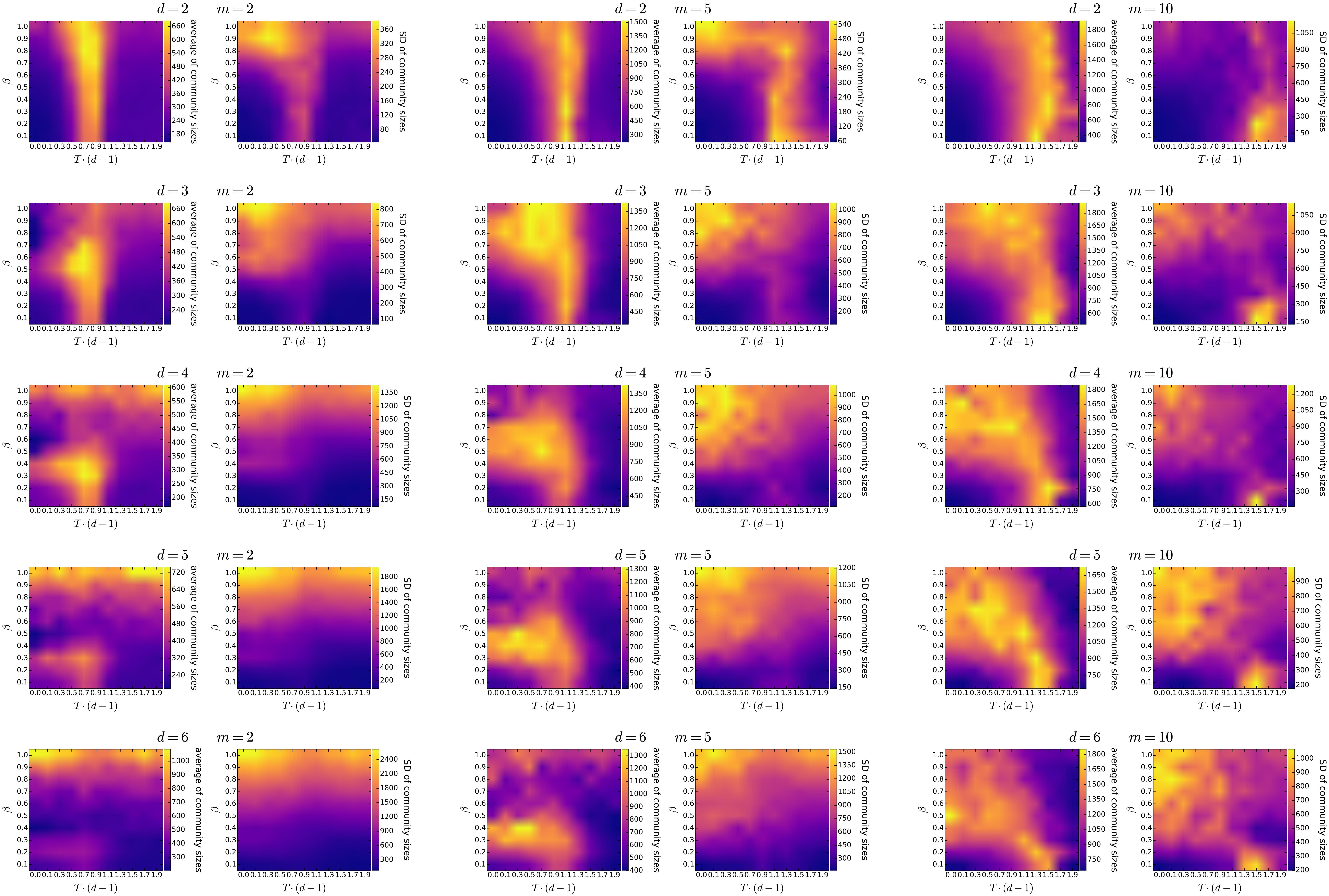}
    \caption{ {\bf The mean and the standard deviation of the size of the communities detected by the \textit{Louvain} algorithm in 5 \textit{weighted} \textit{d}PSO networks in the case of different parametrisations.} Each pair of subplots depicts the effect of changing the popularity fading parameter $\beta$ and the rescaled temperature $T\cdot(d-1)$, with the number of dimensions $d$ and the half $m$ of the expected average degree $\bar{k}$ given in the title of the subplot pair. The number of nodes was $N=10,000$ in each network. The curvature $K$ of the hyperbolic plane was always set to $-1$, i.e. we used $\zeta=1$.}
    \label{fig:groupSizes_Louvain_w}
\end{figure}

\begin{figure}[h!]
    \centering
    \captionsetup{width=1.0\textwidth}
    \includegraphics[width=1.0\textwidth]{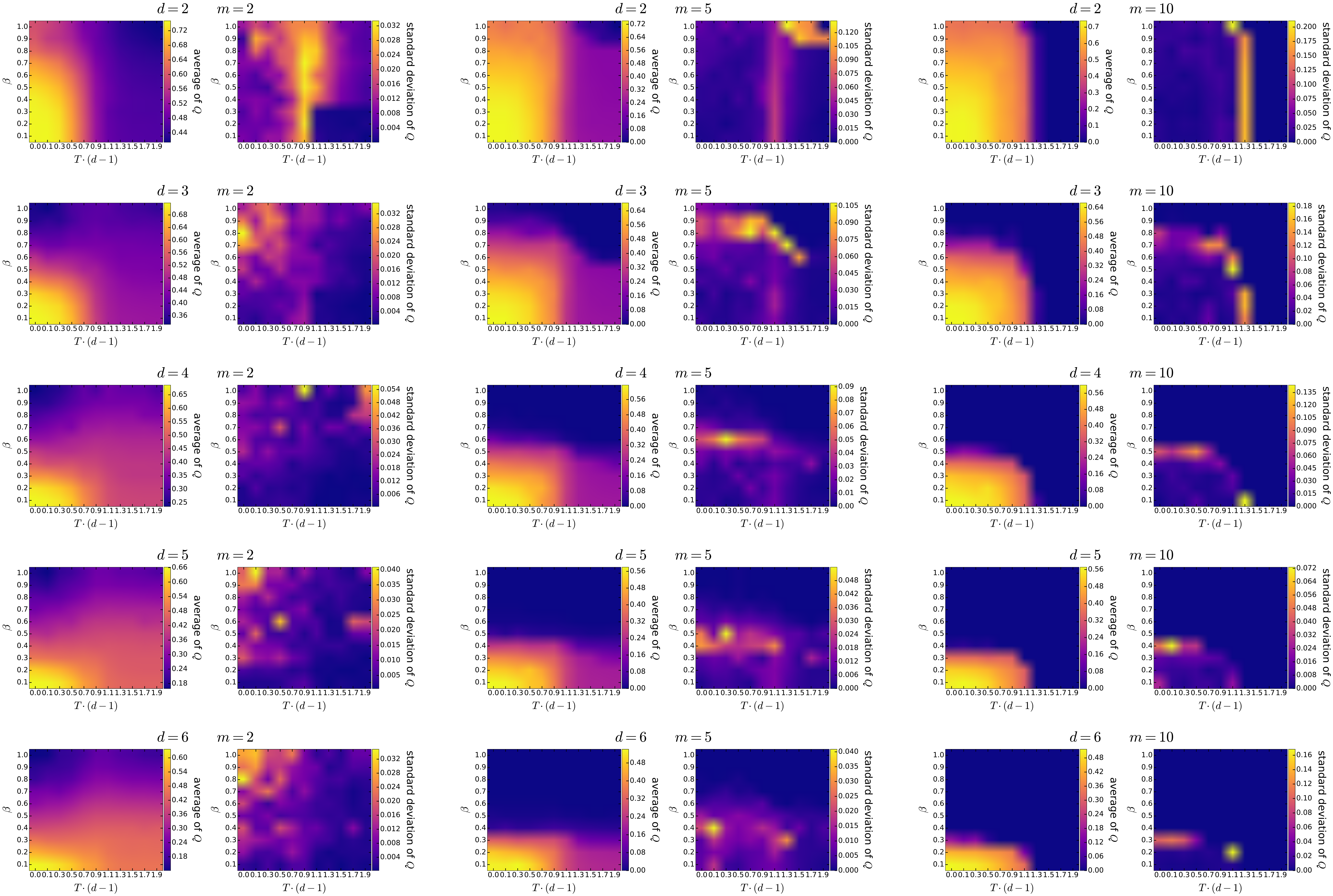}
    \caption{{\bf The mean and the standard deviation of the \textit{weighted} modularity $Q$ of the community structure detected by the \textit{Infomap} algorithm in 5 \textit{d}PSO networks in the case of different parametrisations, using link weights calculated from the hyperbolic distances between the connected nodes according to Eq.~(\ref{eq:linkWeights}).} Each pair of subplots depicts the effect of changing the popularity fading parameter $\beta$ and the rescaled temperature $T\cdot(d-1)$, with the number of dimensions $d$ and the half $m$ of the expected average degree $\bar{k}$ given in the title of the subplot pair. The number of nodes was $N=10,000$ in each network. The curvature $K$ of the hyperbolic space was always set to $-1$, i.e. we used $\zeta=1$.}
    \label{fig:modularity_Infomap_w}
\end{figure}

\begin{figure}[h!]
    \centering
    \captionsetup{width=1.0\textwidth}
    \includegraphics[width=1.0\textwidth]{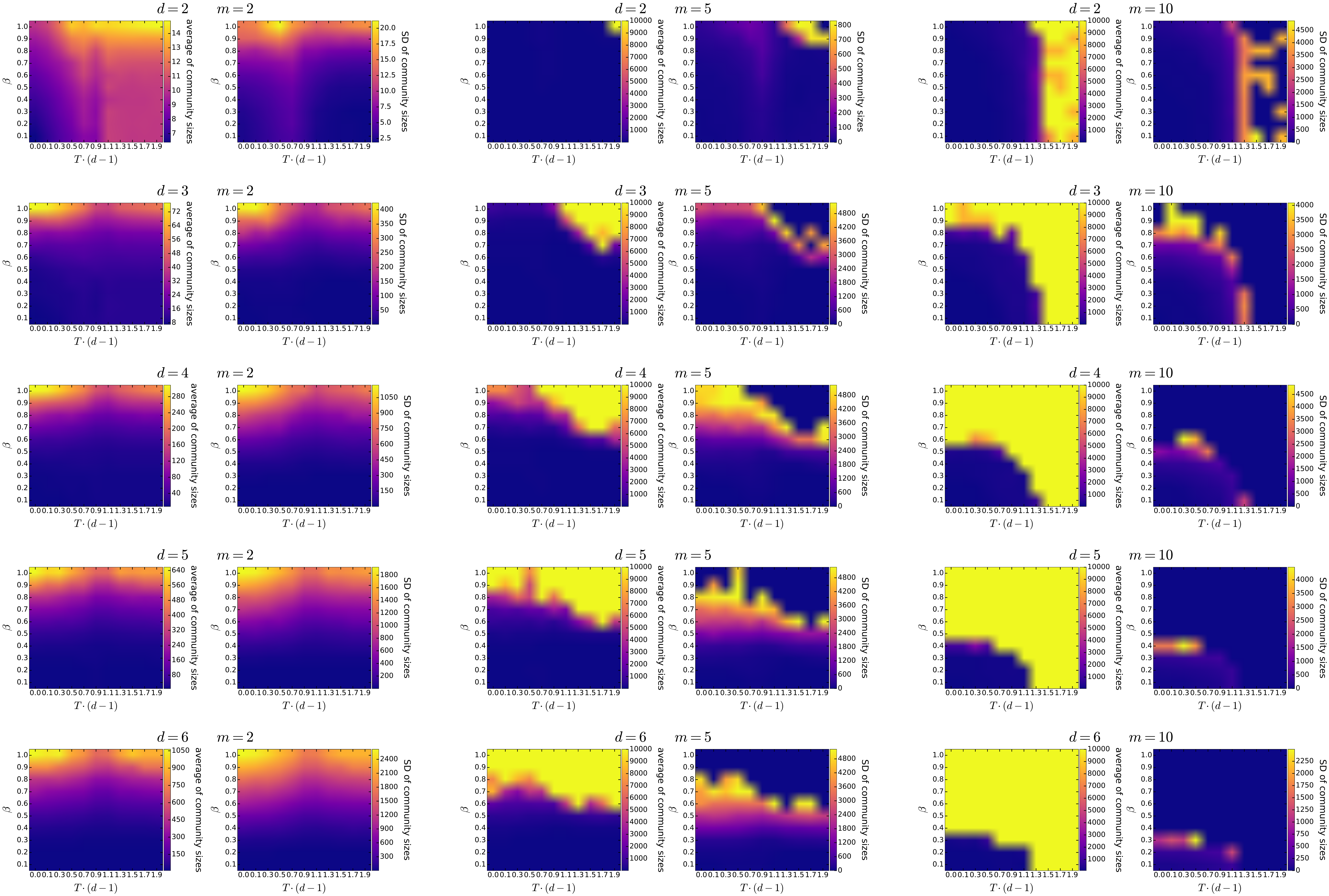}
    \caption{ {\bf The mean and the standard deviation of the size of the communities detected by the \textit{Infomap} algorithm in 5 \textit{weighted} \textit{d}PSO networks in the case of different parametrisations.} Each pair of subplots depicts the effect of changing the popularity fading parameter $\beta$ and the rescaled temperature $T\cdot(d-1)$, with the number of dimensions $d$ and the half $m$ of the expected average degree $\bar{k}$ given in the title of the subplot pair. The number of nodes was $N=10,000$ in each network. The curvature $K$ of the hyperbolic plane was always set to $-1$, i.e. we used $\zeta=1$.}
    \label{fig:groupSizes_Infomap_w}
\end{figure}

\begin{figure}[h!]
    \centering
    \captionsetup{width=1.0\textwidth}
    \includegraphics[width=1.0\textwidth]{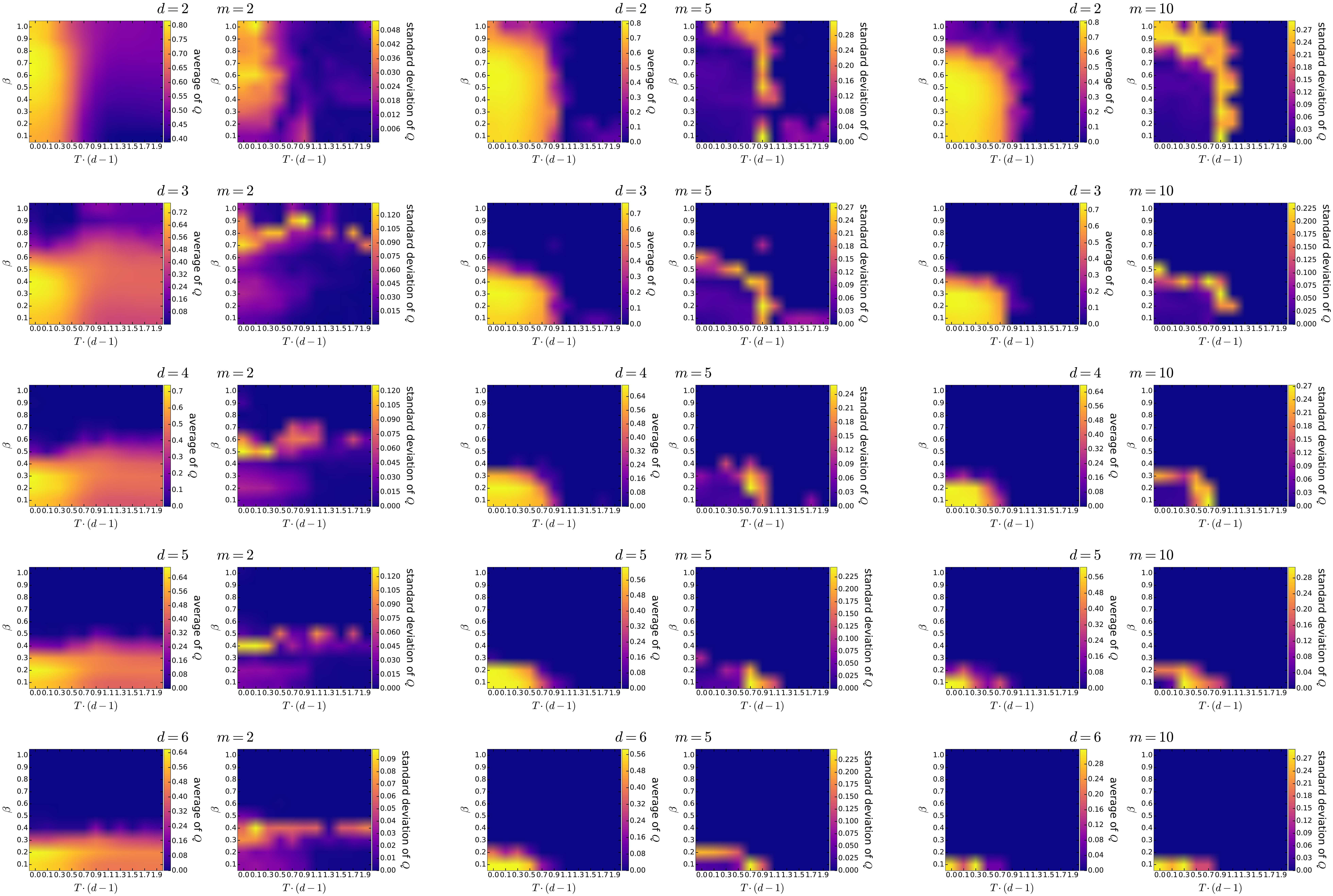}
    \caption{{\bf The mean and the standard deviation of the \textit{weighted} modularity $Q$ of the community structure detected by the \textit{asynchronous label propagation} algorithm in 5 \textit{d}PSO networks in the case of different parametrisations, using link weights calculated from the hyperbolic distances between the connected nodes according to Eq.~(\ref{eq:linkWeights}).} Each pair of subplots depicts the effect of changing the popularity fading parameter $\beta$ and the rescaled temperature $T\cdot(d-1)$, with the number of dimensions $d$ and the half $m$ of the expected average degree $\bar{k}$ given in the title of the subplot pair. The number of nodes was $N=10,000$ in each network. The curvature $K$ of the hyperbolic space was always set to $-1$, i.e. we used $\zeta=1$.}
    \label{fig:modularity_alabprop_w}
\end{figure}

\begin{figure}[h!]
    \centering
    \captionsetup{width=1.0\textwidth}
    \includegraphics[width=1.0\textwidth]{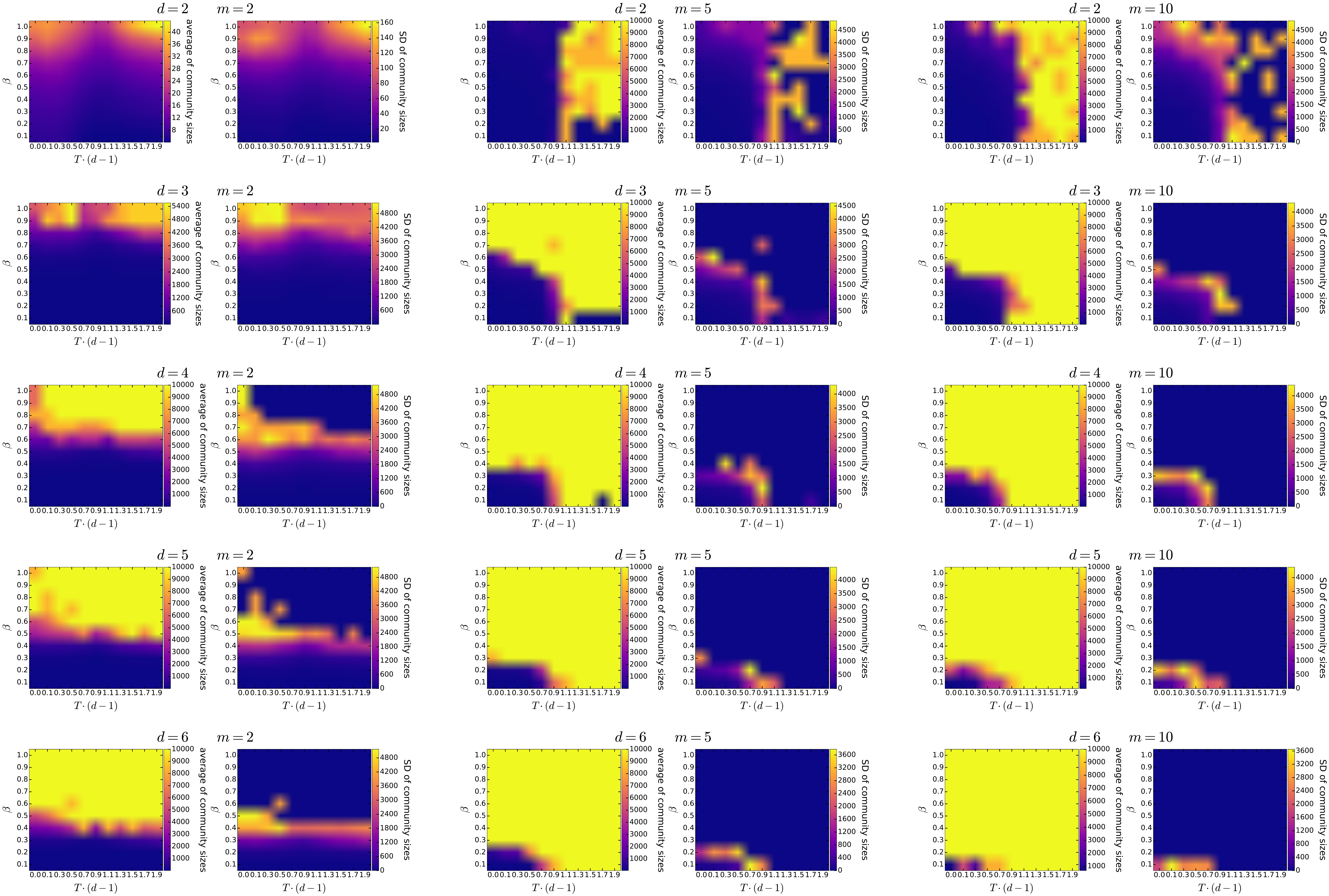}
    \caption{{\bf The mean and the standard deviation of the size of the communities detected by the \textit{asynchronous label propagation} algorithm in 5 \textit{weighted} \textit{d}PSO networks in the case of different parametrisations.} Each pair of subplots depicts the effect of changing the popularity fading parameter $\beta$ and the rescaled temperature $T\cdot(d-1)$, with the number of dimensions $d$ and the half $m$ of the expected average degree $\bar{k}$ given in the title of the subplot pair. The number of nodes was $N=10,000$ in each network. The curvature $K$ of the hyperbolic plane was always set to $-1$, i.e. we used $\zeta=1$.}
    \label{fig:groupSizes_alabprop_w}
\end{figure}

\clearpage
We also repeated our community analysis with a slight modification, taking into account the hyperbolic distances along the links. For this, we adopted the practice suggested in Ref.~\cite{linkWeights_coalescentEmbedding} and assigned weights to the links calculated from the hyperbolic distances between adjacent nodes as 
\begin{equation}
    w_{ij}\equiv w_{ji}=\frac{1}{1+x_{ij}}.
    \label{eq:linkWeights}
\end{equation}
Then, we searched for the communities of the obtained weighted graphs with the Louvain, the Infomap and the asynchronous label propagation methods (where all algorithms allow link weights to be taken into account). As before, for characterising the strength of the detected community structures we used modularity. However, instead of its original version described in Sect.~\ref{sect:comms} of the main text, here we used an extended form defined for weighted networks~\cite{weightedModularityDef}, where the total number of links $E$ is replaced by $M=\frac{1}{2}\cdot\sum\limits_{i=1}^N\sum\limits_{j=1}^N w_{ij}$ (with $w_{ij}$ denoting the link weight between nodes $i$ and $j$), and the node degrees $k_i$ and $k_j$ are replaced by the node strengths $s_i$ and $s_j$, defined e.g. for node $i$ as $s_i = \sum_{\ell=1}^N w_{i\ell}$, resulting in the formula
\begin{equation}
    Q=\frac{1}{2M}\cdot\sum\limits_{i=1}^N\sum\limits_{j=1}^N \left[w_{ij}-\frac{s_is_j}{2M}\right]\delta_{c_i,c_j}.
    \label{eq:weightedModularityDef}
\end{equation}
Similarly to Figs.~\ref{fig:modularity_Louvain_uw}-\ref{fig:groupSizes_alabprop_uw} dealing with the results of the community detection on unweighted networks, we present the averages and the standard deviations of the obtained modularities and community sizes in Figs.~\ref{fig:modularity_Louvain_w}-\ref{fig:groupSizes_alabprop_w} for the weighted case.

\begin{figure}[h!]
    \centering
    \captionsetup{width=1.0\textwidth}
    \includegraphics[width=1.0\textwidth]{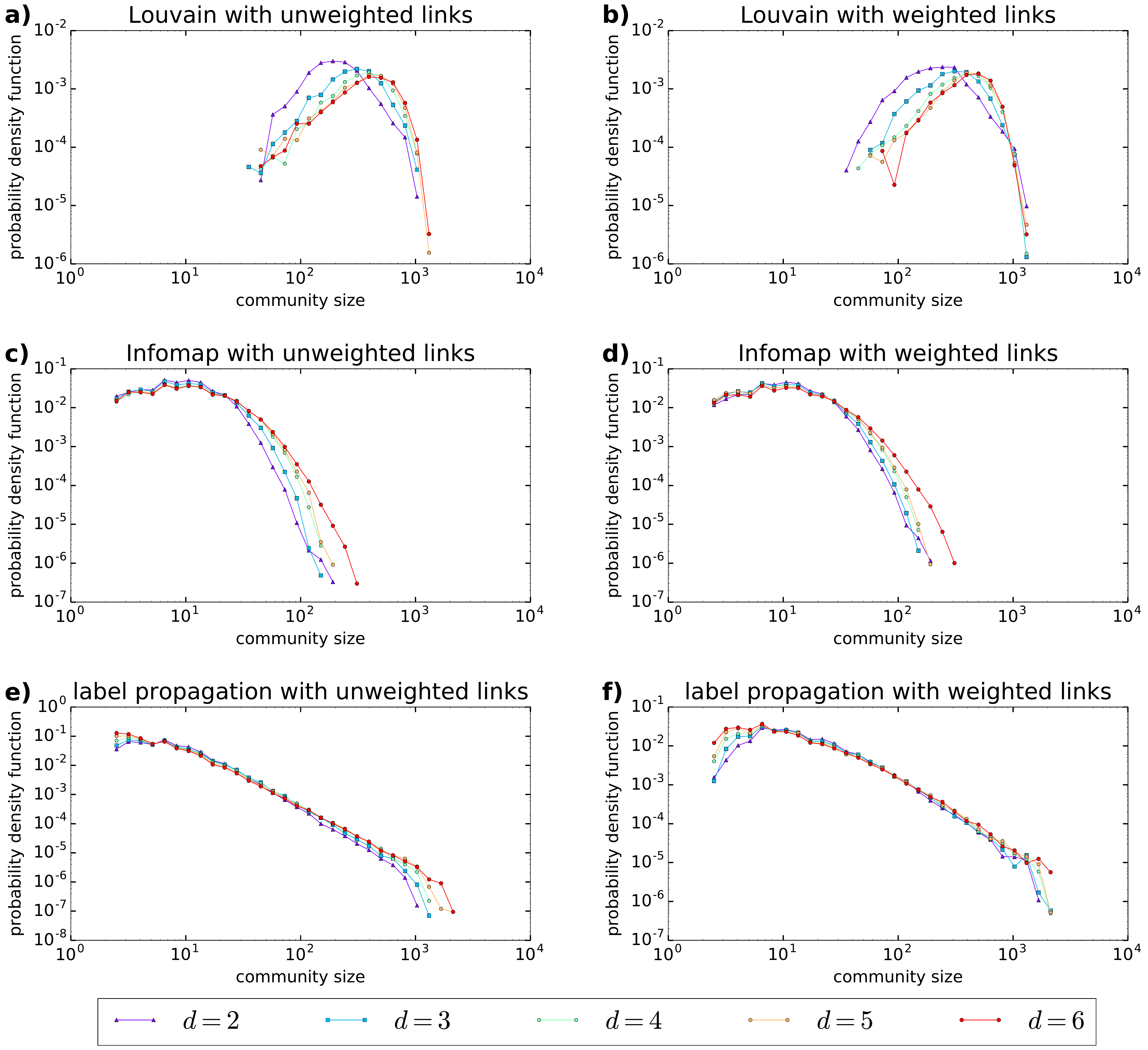}
    \caption{{\bf The size distribution of the communities detected by different community detection algorithms in 100 \textit{d}PSO networks in the case of different values of the dimension $d$.} Each panel shows the results of a community detection method specified in the panel title. The panels in the left column were obtained setting all the link weights in the networks to 1, whereas the panels in the right column were created using link weights calculated from the hyperbolic distances between the connected nodes according to Eq.~(\ref{eq:linkWeights}). The different colours of the curves correspond to the different values of the dimension $d$, listed in the legend. The curvature of the hyperbolic plane $K$ was always set to $-1$, i.e. we used $\zeta=1$. The number of nodes $N$ was 10,000, the expected average degree $\bar{k}=2m$ was 10, the temperature $T$ was 0 and the degree decay exponent $\gamma$ was set to $2.5$ by using the popularity fading parameter $\beta=\frac{1}{(d-1)\cdot(2.5-1)}$ in each case.}
    \label{fig:groupSizeHists}
\end{figure}

\clearpage
Finally, we show some examples of the community size distributions obtained with the applied community finding methods for \textit{d}PSO networks of different number of dimensions in Fig.~\ref{fig:groupSizeHists}. With regard to the shape of the curves, the same conclusion can be drawn for each dimension $d$ as in Ref.~\cite{our_hyp_coms} for $d=2$, namely that Louvain yields relatively narrow, bell-shaped community size distributions concentrated at higher community sizes, while the community size distributions provided by asynchronous label propagation and Infomap are rather skewed, following more or less a power law in the former case and decaying somewhat faster towards the larger sizes in the latter case. As indicated by the slight right shift of the community size distributions in Fig.~\ref{fig:groupSizeHists}, when the dimension $d$ of the hyperbolic space is increased while keeping the curvature $K$, the number of nodes $N$, the expected average degree $2m$, the degree decay exponent $\gamma$ and the temperature $T$ unaltered, all the examined community detection methods tend to find larger modules in the networks. 

\clearpage
\section{Nonuniform angular distribution}
\label{sect:nPSO}
\setcounter{figure}{0}
\setcounter{table}{0}
\setcounter{equation}{0}
\renewcommand{\thefigure}{S4.\arabic{figure}}
\renewcommand{\thetable}{S4.\arabic{table}}
\renewcommand{\theequation}{S4.\arabic{equation}}

We have shown for the two-dimensional popularity-similarity optimisation model~\cite{PSO} in Ref.~\cite{our_hyp_coms} and then for higher-dimensional cases in the present article that despite the absence of any intentional community formation mechanisms built into the model construction, $d$PSO networks possess an inherent, relevant community structure for a wide range of parameter settings. However, the $d$PSO model does not allow control over the number and the size of the communities. To deal with this problem, one can introduce heterogeneity in the angular node arrangement and generate networks using a nonuniform angular distribution of the network nodes, where denser angular regions can serve as built-in communities. 
This nonuniform popularity-similarity optimisation (nPSO) model has been studied in details in the two-dimensional case in Refs.~\cite{nPSO,nPSO_2}. Here, as an example, we examine a simple three-dimensional case, where the $N$ number of network nodes are distributed in equal proportions among $C$ number of planted (angular) groups that are created by setting the angular node distribution to a mixture of $C$ number of von Mises–Fisher distributions~\cite{vMFdistr_book} of the same concentration parameter $\kappa$, with their mean directions $\underline{\mu}_i\,(i=1,2,...,C)$ distributed equidistantly over the unit 2-sphere. That is, we determined the angular coordinates of each network node by first choosing randomly one of the $C$ number of mean directions $\underline{\mu}_i$ (representing the central locations of the angular sectors) and then sampling~\cite{vMFdistr_code} a 3-dimensional unit vector \underline{y} from the corresponding von Mises–Fisher distribution for which the probability density function is given by Eq.~(\ref{eq:vMFdistr}). Note that a higher value of the concentration parameter $\kappa$ means a higher concentration of the distributions around their mean direction, i.e. less spread and overlapping angular sectors, while $\kappa=0$ pertains to the uniform angular distribution of the nodes. Our implementation of this three-dimensional nPSO model is available from Ref.~\cite{our_code}.

\begin{equation} 
    f(\underline{y};\underline{\mu}_i,\kappa)=\frac{\kappa}{2\pi(e^{\kappa}-e^{-\kappa})}\cdot e^{\kappa\underline{\mu}_i^{\mathrm{T}}\underline{y}}
    \label{eq:vMFdistr}
\end{equation}

\begin{figure}[h!]
    \centering
    \captionsetup{width=1.0\textwidth}
    \includegraphics[width=0.85\textwidth]{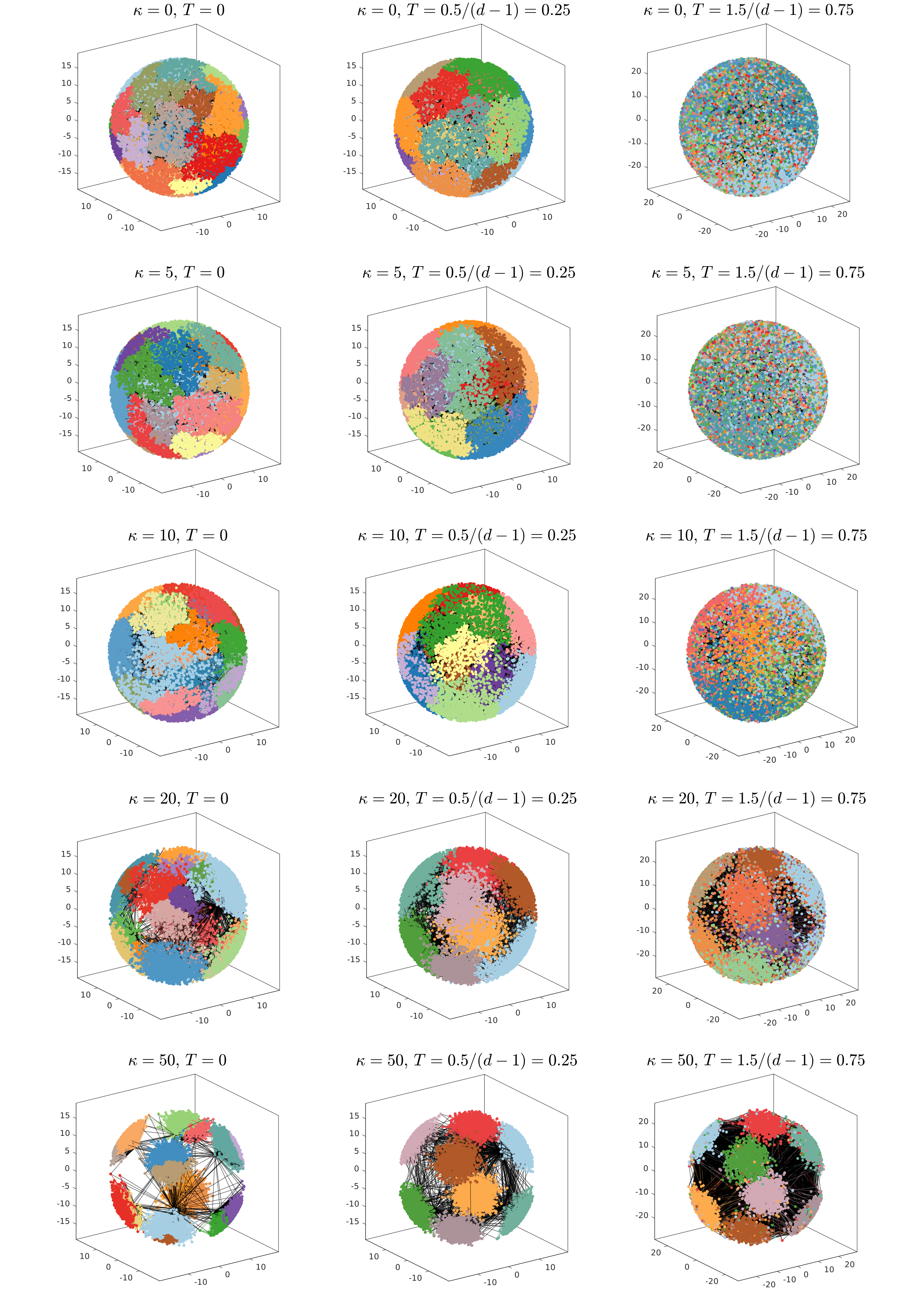}
    \caption{ {\bf Layout of three-dimensional nPSO networks in the native representation of the hyperbolic space of curvature $K=-1$ at different values of the concentration parameter $\kappa$ and the temperature $T$.} Each row of panels was created using a given concentration parameter $\kappa$, and each column of subplots presents the results obtained with a given value of the temperature $T$, as written in the panel titles. Each network was generated setting the number $N$ of nodes to 10,000, the half $m$ of the expected average degree $\bar{k}$ to 5, the popularity fading parameter $\beta$ to 1/3 (yielding the degree decay exponent $\gamma=2.5$) and the number $C$ of components of the mixture distribution describing the angular arrangement of the network nodes to 8. Note that the critical temperature was $T_{\mathrm{c}}=1/(d-1)=0.5$.}
    \label{fig:layout_kappaTdependence}
\end{figure}

The degree of separation among the planted groups is primarily determined by the concentration parameter $\kappa$ and the temperature $T$, on the level of the angular node arrangement and the links, respectively. The effect of these two model parameters is demonstrated via some layouts in Fig.~\ref{fig:layout_kappaTdependence}. Although e.g. $\kappa=20$ already yields a noticeable separation among the patches of the network nodes, when $T$ is increased to $1.5\cdot T_{\mathrm{c}}$, then so many interconnections emerge between the patches that the planted modules eventually do not form actual communities. On the other hand, if the temperature is set to 0, the links become so localised that the planted groups split according to narrower angular regions. Nevertheless, for high enough values of the concentration parameter $\kappa$ and moderate temperatures $T$, the planted groups were successfully identified by the community detection algorithm Louvain~\cite{Louvain,Louvain_code} based on the edge list, without inputting any information about the network geometry.

\begin{figure}[h!]
    \centering
    \captionsetup{width=1.0\textwidth}
    \includegraphics[width=1.0\textwidth]{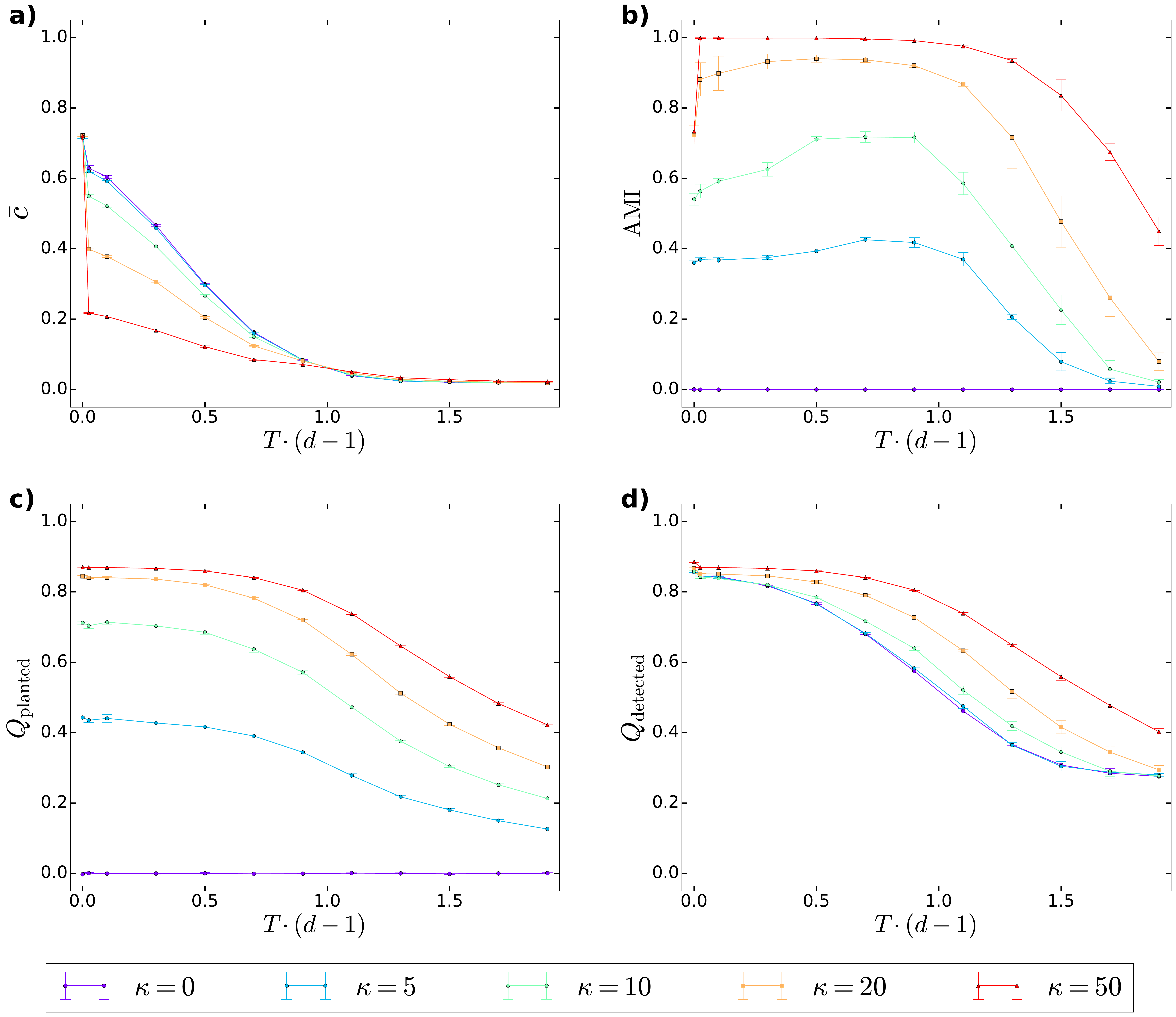}
    \caption{ {\bf Average clustering coefficient $\bar{c}$, adjusted mutual information $\mathrm{AMI}$ of the planted modules and the ones detected by the Louvain algorithm and modularity $Q$ of the planted and the detected network partitions as a function of the rescaled temperature $T\cdot(d-1)$ in 3-dimensional nPSO networks of different values of the concentration parameter $\kappa$.} The data points show the measured values averaged over 5 networks of the given parameter setting, and the error bars indicate the standard deviations among the 5 networks. Each network was generated in the 3-dimensional hyperbolic space of curvature $K=-1$, setting the number $N$ of nodes to 10,000, the half $m$ of the expected average degree $\bar{k}$ to 5, the popularity fading parameter $\beta$ to $1/3$ (i.e., the degree decay exponent $\gamma$ to $2.5$), and the number $C$ of components of the mixture distribution describing the angular arrangement of the network nodes to 8.} 
    \label{fig:clustCoeff_mod_AMI_Tdependence_nPSO}
\end{figure}

In Fig.~\ref{fig:clustCoeff_mod_AMI_Tdependence_nPSO}, we show how the average clustering coefficient $\bar{c}$, the adjusted mutual information $\mathrm{AMI}$~\cite{AMI,AMI_code} of the planted community structure and the one detected by the Louvain algorithm (setting all the link weights to 1), and the modularity~\cite{Newman_modularity_original,modularity_code} of the planted and the detected network partitions depend on the temperature for different concentration parameters. Similarly to what has been shown in Fig.~3 of the main text for the uniform $d$PSO model, the average clustering coefficient $\bar{c}$ measured in three-dimensional nPSO networks gradually decreases with the increase in the temperature $T$ before settling to a more or less constant value just above the critical temperature $T_{\mathrm{c}}=1/(d-1)=0.5$. One can also observe that if $\kappa$ is higher, i.e. the angular patches of the network nodes are more separated from each other, then the drop in $\bar{c}$ occurring when switching from the deterministic connection rule ($T=0$) to the probabilistic one ($0<T$) is larger. 

As expected, the adjusted mutual information of the planted and the detected partitions is 0 for $\kappa=0$, when the angular position of the nodes is sampled from a mixture of uniform distributions and the nodes are assigned to the planted groups randomly, regardless of their angular coordinates. As we increase $\kappa$ and create hereby larger angular gaps between the regions occupied by the network nodes, the $\mathrm{AMI}$ tends to become higher. 
At small temperatures, as exemplified by Fig.~\ref{fig:layout_kappaTdependence}, the links are strongly localised, i.e. crowded into narrower sectors within the region of each planted group. Similarly to how the nodes of the whole hyperbolic space become grouped according to different angular regions in the case of the uniform PSO model~\cite{our_hyp_coms}, this angular confinement of the links splits the planted modules into smaller parts, yielding weaker agreement between the planted and the detected partitions. As the temperature begins to increase, at first the boundaries within the planted groups blur, and thus the $\mathrm{AMI}$ increases. Note that at large enough $\kappa$ and moderate values of $T$, even $\mathrm{AMI}=1$ was achieved, meaning that the planted and the detected partitions were identical. However, even higher temperatures and farther-reaching connections already raise the interconnectedness of the adjacent planted groups too, until eventually the nodes lose their preference for primarily connecting to the members of their own planted group. Obviously, the larger the gaps between the occupied patches, the higher temperature is needed to enable the neighbouring patches to reach each other; thus, as $\kappa$ increases, the $\mathrm{AMI}$ decreases more slowly to 0 as a function of $T$. 

The behaviour of the modularities $Q_{\mathrm{planted}}$ and $Q_{\mathrm{detected}}$ as a function of the temperature $T$ agrees with the results shown in Fig.~5 of the main text for the uniform $d$PSO model: according to this measure, both the planted and the detected community structure of three-dimensional nPSO networks lose from their strength as the temperature increases, and eventually both modularities settle to a constant value. It is important to note that the modularity of the planted network partitions practically does not exceed the modularity of the detected community structures, 
meaning that whenever we measure low modularity values, these arise due to the network structure and not because of some failure in the applied community finding algorithm. 
Since the presence of wider angular gaps between the patches of the nodes decreases the probability of interconnections, larger values of $\kappa$ lead to higher modularities. Nevertheless, at high enough temperatures the probability for the emergence of connections can become non-negligible even for the members of angularly well-separated patches, diminishing the modularity differences among the networks characterised by different concentration parameters.

\begin{figure}[h!]
    \centering
    \captionsetup{width=1.0\textwidth}
    \includegraphics[width=0.78\textwidth]{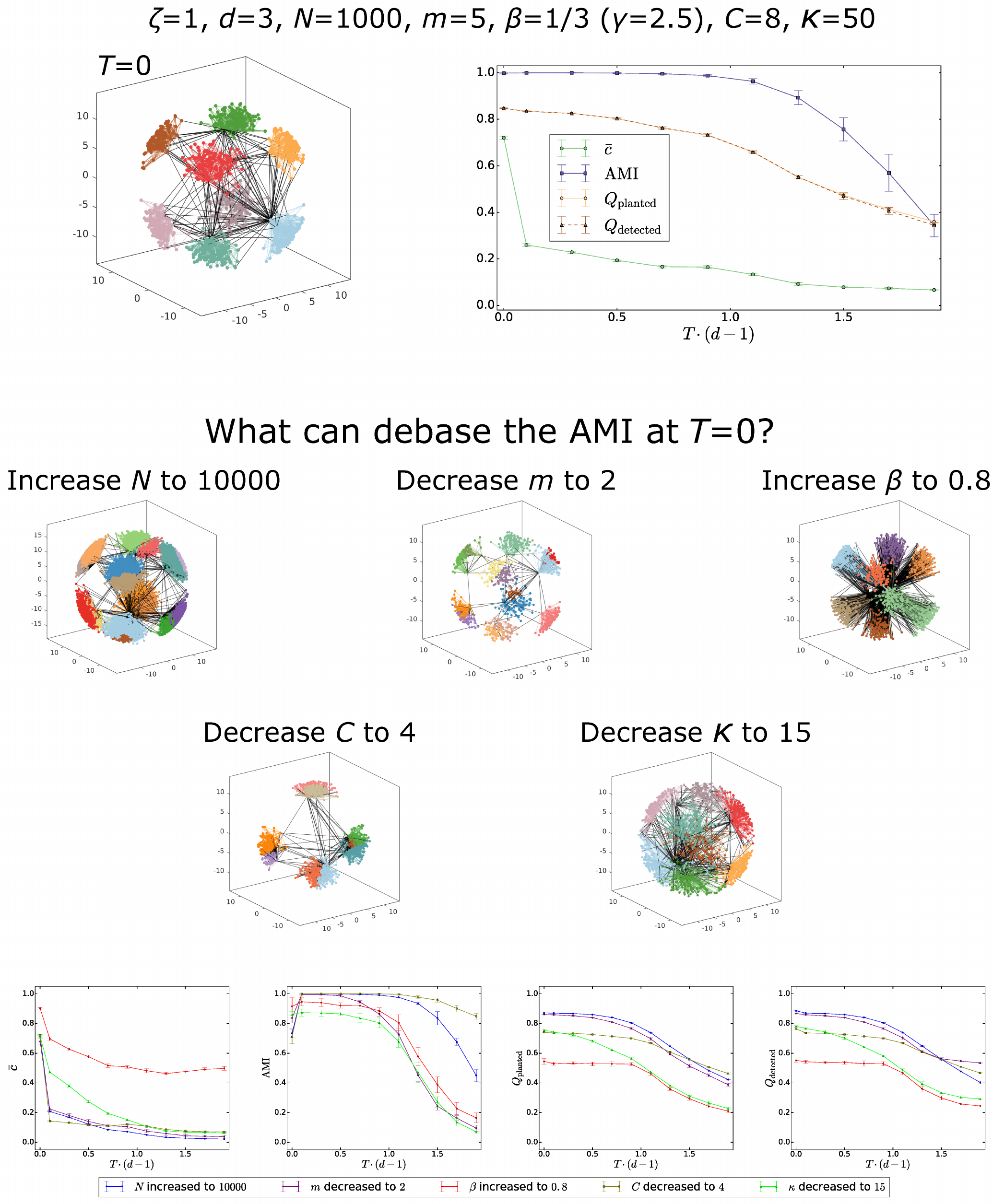}
    \caption{ {\bf With proper settings of the parameters of the three-dimensional nPSO model, large adjusted mutual information of the planted and the detected modules can be achieved even at the temperature minimum $T=0$, where the average clustering coefficient and the modularities are maximised for the given parametrisation. However, increasing the number $N$ of nodes, decreasing the expected average degree $2\cdot m$, increasing the popularity fading parameter $\beta$, decreasing the number $C$ of components of the mixture distribution describing the angular arrangement of the network nodes, or decreasing the concentration parameter $\kappa$ may reduce the $\mathrm{AMI}$ at $T=0$ compared to the values obtained at slightly higher temperatures.} In the right uppermost chart we show an example where -- in contrast with Fig.~\ref{fig:clustCoeff_mod_AMI_Tdependence_nPSO} -- the decay of the $\mathrm{AMI}$ towards the smallest possible temperatures does not appear. The layouts below exemplify how the change of the different model parameters compared to the settings of the uppermost figures can debase the high $\mathrm{AMI}$ value achieved at $T=0$. The lowermost charts depict how the increase in the rescaled temperature $T\cdot(d-1)$ affects the average clustering coefficient $\bar{c}$, the adjusted mutual information $\mathrm{AMI}$ and the modularity $Q$ of the planted partitions and the ones detected by Louvain in the five cases presented by the layouts above. The plotted data points correspond to the values averaged over 5 networks in the case of $N=10,000$ and 10 networks for $N=1000$. The error bars indicate the standard deviations among the networks of the same parameter setting.} 
    \label{fig:AMIatT0_nPSO}
\end{figure}

According to Fig.~\ref{fig:clustCoeff_mod_AMI_Tdependence_nPSO}, while the maximum point of both the average clustering coefficient $\bar{c}$ and the modularities $Q_{\mathrm{planted}}$ and $Q_{\mathrm{detected}}$ is at $T=0$, with respect to the adjusted mutual information of the planted and the detected partitions it is better to choose higher temperatures. Nonetheless, Fig.~\ref{fig:AMIatT0_nPSO} demonstrates that this is not a general rule, and with proper parameter settings one can generate networks that are simultaneously highly clustered and possess a strong planted community structure that is also well detectable. 

\begin{figure}[h!]
    \centering
    \captionsetup{width=1.0\textwidth}
    \includegraphics[width=1.0\textwidth]{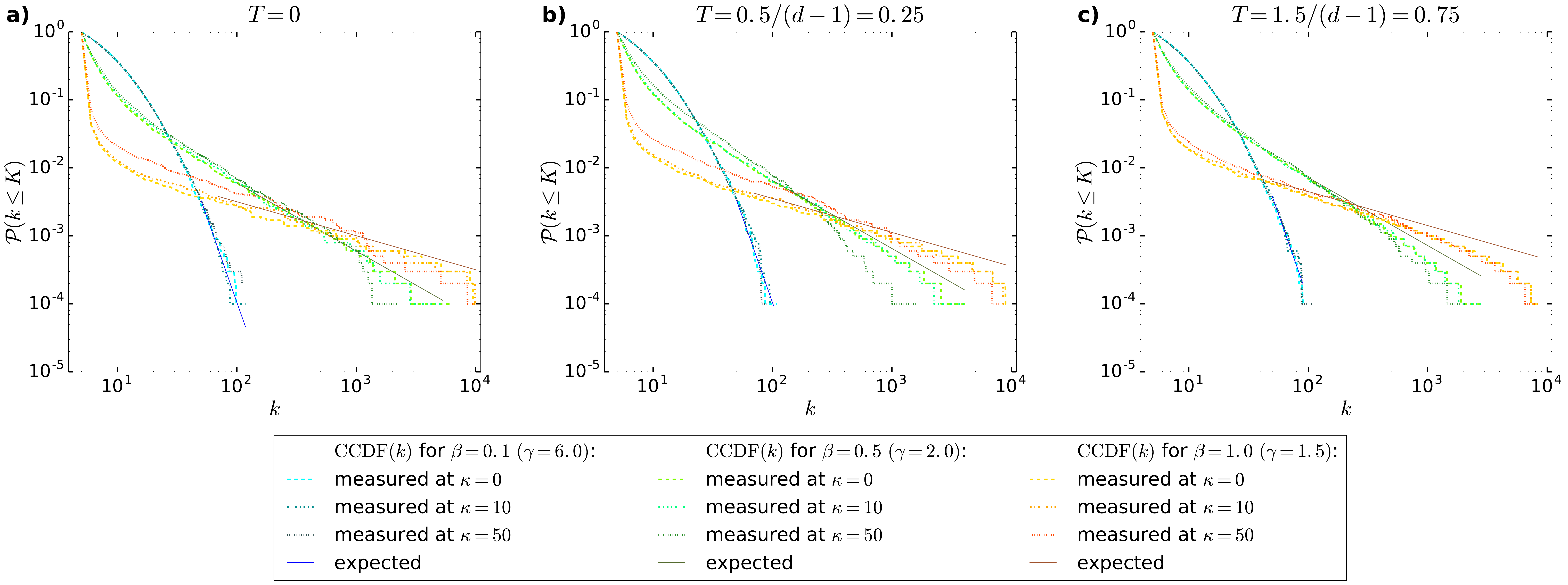}
    \caption{ {\bf Degree distribution of networks generated by the three-dimensional nPSO model using different values of the temperature $T$, the popularity fading parameter $\beta$ and the concentration parameter $\kappa$. As expected, the tail of the complementary cumulative distribution function (CCDF) of the node degrees follows a power-law that can be written in the form of $\pazocal{P}(k\leq K) \sim k^{-(\gamma-1)}$ with $\gamma=1+\frac{1}{(d-1)\cdot\beta}$, independently of the choice of $\kappa$ and $T$.} One network was generated with all the parameter settings. The curvature of the three-dimensional hyperbolic space, the number of nodes, the half of the expected average degree and the number of components of the mixture distribution describing the angular arrangement of the network nodes were the same for each network, namely $K=-\zeta^2=-1$, $N=10,000$, $m=5$ and $C=8$. Panel a) corresponds to the case of the deterministic connection rule ($T=0$), panel b) shows the curves obtained at the half of the critical temperature $T_{\mathrm{c}}=1/(d-1)$, while panel c) presents the results at a higher temperature of $1.5\cdot T_{\mathrm{c}}$. The CCDF curves obtained with the different values of the concentration parameter $\kappa$ are well grouped according to the popularity fading parameters, i.e. the expected degree decay exponents in all panels.}
    \label{fig:degreeDist_nPSO}
\end{figure}

\clearpage
Lastly, we demonstrate in Fig.~\ref{fig:degreeDist_nPSO} that when each network node is assigned randomly to one of the 8 planted groups corresponding to von Mises–Fisher distributions of the same concentration parameter $\kappa$ and mean directions pointing toward the vertices of a cube, then the resulting degree distribution reasonably preserves its form of $\pazocal{P}(K=k)\sim k^{-\gamma}$ with $\gamma=1+\frac{1}{(d-1)\cdot\beta}$, even if the angular distribution of the nodes becomes more and more heterogeneous due to the increasing separation of the occupied angular regions obtained at higher and higher values of $\kappa$. Similarly to what has been shown in Fig.~2 of the main text concerning the uniform $d$PSO model, the temperature $T$ does not have a significant effect on the degree distribution even for the nonuniform three-dimensional PSO model.

\end{document}